\documentclass[a4paper, UTF8, 11pt]{article}

\usepackage[title]{appendix}
\usepackage{float}
\usepackage{geometry}
\usepackage{amsmath}
\usepackage{amssymb}
\usepackage{amsfonts}
\usepackage{amsthm}
\usepackage{mathtools}
\usepackage{extarrows}
\usepackage{setspace}
\usepackage{indentfirst}
\usepackage[bottom]{footmisc}
\usepackage{titlesec}
\usepackage{natbib}
\usepackage{graphicx}
\usepackage{mathrsfs}
\usepackage{diagbox}
\usepackage{caption}
\usepackage{subcaption}
\usepackage{color}
\usepackage{xcolor}
\usepackage{enumitem}
\usepackage{bbm}
\usepackage{fancyhdr}
\usepackage{lastpage}
\usepackage{multirow}
\usepackage{booktabs}
\usepackage{thmtools}
\usepackage{titletoc}

\usepackage{chngcntr}
\counterwithin{table}{section}

\usepackage[colorlinks=true, urlcolor=blue, citecolor=blue, linkcolor=blue, backref=none,breaklinks=true, hypertexnames = false]{hyperref}

\usepackage{chapterbib}

\titleformat*{\section}{\large\bf}
\titleformat*{\subsection}{\normalsize\bf}

\setcitestyle{authoryear}
\allowdisplaybreaks[4]

\makeatletter
\renewcommand\p@enumii{}
\makeatother

\newcommand{\E}{\mathbb{E}}

\newcommand{\p}{\mathbb{P}}
\newcommand{\ddd}{\;\mathrm{d}}

\newcommand{\argmin}{\operatornamewithlimits{arg\,min}}

\newcommand{\mr}{\mathbb{R}}
\newcommand{\mf}{\mathscr{F}}
\newcommand{\convd}{\rightsquigarrow}
\newcommand{\convp}{\xrightarrow{\mathbb{P}}}
\newcommand{\convas}{\xrightarrow{\mathrm{a.s.}}}

\newcommand{\lp}{\left(}
\newcommand{\lbk}{\left[}
\newcommand{\lbr}{\left\{}
\newcommand{\rp}{\right)}
\newcommand{\rbk}{\right]}
\newcommand{\rbr}{\right\}}
\newcommand{\lv}{\left\vert}
\newcommand{\rv}{\right\vert}
\newcommand{\lvv}{\left\Vert}
\newcommand{\rvv}{\right\Vert}
\newcommand{\indicator}{\mathbbm{1}}
\newcommand{\T}{\mathsf{T}}

\newtheoremstyle{thrysty}{10pt}{10pt}{\normalfont}{\parindent}{\bfseries}{:}{0.5em}{}
\theoremstyle{thrysty}

\newtheorem{thry}{Theorem}[section]
\newtheorem{prop}{Proposition}[section]

\newtheorem{lemma}{Lemma}[section]
\newtheorem{definition}{Definition}[section]
\newtheorem{assum}{Assumption}[section]
\newtheorem{exam}{Example}[section]
\newtheorem{remark}{Remark}[section]
\newenvironment{pf}[1]{\textbf{Proof{#1}:}}{\qed\vskip10pt}
\renewcommand\thmcontinues[1]{Cont.}

\title{\Large\bf Unified Inference on Moment Restrictions with Nuisance Parameters\footnote{This paper is a revised version of Chapters 2 and 3 of the first author's doctoral dissertation at Peking University. The authors are grateful to the Editors, two anonymous referees, Zheng Fang, Yixiao Sun, and all seminar and conference participants for their insightful and constructive comments.}
}    
\author{Xingyu Li\\School of Economics\\Zhejiang University\\
	lixyecon@zju.edu.cn
	\and Xiaojun Song\footnote{This work was supported by the National Natural Science Foundation of China [Grant Numbers 72373007 and 72333001]. The author also gratefully acknowledges the research support from the Center for Statistical Science of Peking University, China, and the Key Laboratory of Mathematical Economics and Quantitative Finance (Peking University) of the Ministry of Education, China.}\\
	Guanghua School of Management
    \\ Peking University\\
	sxj@gsm.pku.edu.cn
	\and Zhenting Sun\footnote{This work was supported by the National Natural Science Foundation of China [Grant Number 72103004].}\\Department of Economics\\University of Melbourne\\
	zhenting.sun@unimelb.edu.au}
\date{\today}

\begin{document}	

\maketitle
\begin{abstract}
This paper proposes a simple unified inference approach on moment restrictions in the presence of nuisance parameters. The proposed test is constructed based on a new characterization that avoids the estimation of nuisance parameters and can be broadly applied across diverse settings. Under suitable conditions, the test is shown to be asymptotically size controlled and consistent for both independent and dependent samples. Monte Carlo simulations show that the test performs well in finite samples. Numerical results from the application to conditional moment restriction models with weak instruments demonstrate that the proposed method may improve upon existing approaches in the literature.
\end{abstract}

\thispagestyle{fancy}

\setlength\abovedisplayskip{2pt plus 0pt minus 2pt}
\setlength\belowdisplayskip{2pt plus 0pt minus 2pt}


\newcommand{\borel}{\mathscr{B}}
\newcommand{\mL}{\mathcal{L}}
\newcommand{\mLhat}{\widehat{\mathcal{L}}''_n}
\newcommand{\Phat}{\widehat{P}_n}
\newcommand{\Fhat}{\widehat{F}_{n}}
\newcommand{\Ghat}{\widehat{G}_{n}}
\newcommand{\Hhat}{\widehat{H}_{n}}
\newcommand{\phihat}{\widehat{\phi}_n}
\newcommand{\bZn}{\mathbf{Z}_n}
\newcommand{\HH}{\mathrm{H}}
\newcommand{\bG}{\mathbb{G}}
\newcommand{\bD}{\mathbb{D}}
\newcommand{\bW}{\mathbb{W}}
\newcommand{\bx}{\mathbb{X}}
\newcommand{\cF}{\mathcal{F}}
\newcommand{\cx}{\mathcal{X}}
\newcommand{\cxhat}{\widehat{\mathcal{X}}_{n_1}}
\newcommand{\by}{\mathbb{Y}}
\newcommand{\cy}{\mathcal{Y}}
\newcommand{\cyhat}{\widehat{\mathcal{Y}}_{n_2}}
\newcommand{\bv}{\mathbb{V}}
\newcommand{\cvv}{\mathcal{V}}
\newcommand{\cvhat}{\widehat{\mathcal{V}}_{n_1}}
\newcommand{\mS}{\mathcal{S}}
\newcommand{\mR}{\mathcal{R}}
\newcommand{\mC}{\mathcal{C}}
\newcommand{\mCb}{\mathcal{C}_\mathrm{b}}
\newcommand{\mD}{\mathcal{D}}
\newcommand{\mE}{\mathcal{E}}
\newcommand{\mQ}{\mathcal{Q}}
\newcommand{\sg}{\mathscr{G}}
\newcommand{\fg}{\mathfrak{g}}
\newcommand{\bF}{\mathbb{F}}
\newcommand{\bM}{\mathbb{M}}
\newcommand{\Qhat}{\widehat{\mQ}_n}
\newcommand{\chat}{\widehat{c}}
\newcommand{\bhat}{\widehat{b}}
\newcommand{\Pihat}{\widehat{\Pi}_n}
\newcommand{\lsv}{{L^2(\nu)}}
\newcommand{\sh}{\mathscr{H}}
\newcommand{\bH}{\mathbb{H}}
\newcommand{\bK}{\mathbb{K}}
\newcommand{\thetalb}{\underline{\theta}}
\newcommand{\varepsilonbar}{\overline{\varepsilon}}
\newcommand{\gbar}{\overline{g}}
\newcommand{\That}{\widehat{T}}
\newcommand{\thetahat}{\widehat{\theta}}

\newpage
\section{Introduction}\label{sec:intro}

The analysis of moment restriction models plays a central role in econometric theory and applications. 
Considerable efforts have been devoted to estimating unknown key parameters and to testing hypotheses related to these parameters in the moment restrictions; see, e.g., \citet{chamberlain1987asymptotic}, \citet{NEWEY1993419}, \citet{dominguez2004consistent}, \citet{kitamura2004empirical}, \citet{smith2007efficient}, and \citet{lavergne2013smooth}, among many others; see also \citet{kunitomo2011moment} for an overview of the moment restriction-based econometric methods. 

Valid statistical inference on these parameters relies crucially on the correct specification of the postulated moment restriction models. Assessing the suitability of the moment restrictions has therefore generated an extensive literature; see, e.g., \citet{bierens1982consistent}, \citet{tauchen1985diagnostic}, \citet{newey1985generalized}, and  \citet{donald2003empirical}. In testing the moment restrictions, the unknown parameters may not be of primary interest under the null hypothesis and can be regarded as nuisance parameters. Handling nuisance parameters in the considered testing procedures is an important theoretical issue. Existing specification tests for moment restrictions typically employ procedures that first estimate the nuisance parameters and then test the moment restrictions using the estimators; see, e.g., \citet{tripathi2003testing}, \citet{delgado2006consistent}, and \citet{muandet2020kernel}. As a result, classical approaches are generally model- or estimator-dependent, requiring different theories and implementation procedures for different cases. 
In addition, these approaches may encounter theoretical difficulties due to the estimation of nuisance parameters. For example, obtaining reliable estimates of nuisance parameters may be nontrivial in conditional moment restriction models when instruments are weak. 

In this paper, we propose a unified testing framework for moment restrictions with nuisance parameters that is broadly applicable to various settings. The critical values of our test are constructed using the numerical delta methods developed by \citet{hong2018numerical} and \citet{chen2019inference} who provide novel methodologies for addressing nonstandard testing issues.\footnote{More discussions on this topic can be found in \citet{dumbgen1993nondifferentiable}, \citet{andrews2000inconsistency}, \citet{hirano2012impossibility}, \citet{hansen2017regression}, and \citet{fang2019inference}. Other discussions and applications of related bootstrap methods can be found in \citet{beare2015nonparametric}, \citet{Beare2016global}, \citet{Seo2016tests}, \citet{Beare2015improved}, \citet{chen2019improved}, \citet{hong2020numerical}, \citet{Beare2017improved}, and \citet{sun2018ivvalidity}.} The proposed method in the paper effectively circumvents the estimation of nuisance parameters, thus providing a general and robust inferential tool for different settings where nuisance parameters are present.  A comparison between the proposed test and existing approaches in conditional moment restriction models with weak instruments demonstrates that the test can achieve performance improvement.

We summarize the main features of the proposed test as follows: (i) It is case-independent; (ii) it is free of the estimation of nuisance parameters, and is particularly appealing in cases where desirable estimation is challenging; (iii) it is asymptotically size controlled and consistent against a broad class of alternatives to the null; (iv) it works for both independent and dependent samples; (v) the bootstrap test procedure is simple.

Now we introduce our testing framework. Let $d_{\theta}\in\mathbb{Z}_+$ and $d_z\in\mathbb{Z}_+$. Let $\Theta\subset\mathbb{R}^{d_\theta}$ be a parameter space. Let $\Psi=\{\psi_{x,\theta}:\mr^{d_z}\to \mr:x\in\mr, \theta\in\Theta\}$ be a class of functions indexed by $(x,\theta)\in\mr\times\Theta$ such that $\psi_{x,\theta}$ is measurable for all $(x,\theta)\in\mr\times\Theta$. Throughout the paper, all random elements are defined on a probability space $\lp \Omega, \mf, \p \rp$. Let $P$ be an unknown probability distribution on $(\mr^{d_z}, \borel(\mr^{d_z}))$ and $Z\sim P$ be a random vector such that for every Borel set $B\subset \mathbb{R}^{d_z}$, $P(B)=\mathbb{P}(Z\in B)$. We are interested in the null hypothesis
\begin{align}\label{eq:para-original-null-0}
\HH_0:  \text{For some } \theta\in \Theta,  \E_P[\psi_{x,\theta}(Z)]=0 \text{ for all } x\in \mr.
\end{align}
This can be viewed as a specification test on a set of moment restrictions. The parameter $\theta$ in \eqref{eq:para-original-null-0} is the nuisance parameter we need to take into account. Let $\phi_P:\mr\times\Theta\to\mr$ be a function depending on $P$ such that $\phi_P(x,\theta)=P(\psi_{x,\theta})=\E_P[\psi_{x,\theta}(Z)]$ for every $(x,\theta)\in\mr\times\Theta$. Clearly, the null hypothesis in \eqref{eq:para-original-null-0} is equivalent to
\begin{align}\label{eq:para original null}
\HH_0:  \text{For some } \theta\in \Theta, \phi_P(x,\theta)=0  \text{ for all } x\in \mr.
\end{align}
The above formulation can easily be extended to cases where $x\in\mathbb{R}^k$ for some $k>1$. To simplify exposition, we present the results for scalar $x$ in the main text.

The testing approach provided in the paper can be readily applied in a wide range of empirical studies. In the following, we present several important examples where the hypothesis of interest can be formulated into \eqref{eq:para original null}.

\subsection{Examples}

\begin{exam}[label=CMR](Conditional Moment Restrictions)
\label{exam:CMR.prob}
Let $Z=(X,Y)$ be a $d_z$-dimensional random vector with scalar $X$ and $d_y$-dimensional vector $Y$, where $d_z=d_y+1\ge 2$. Let $g:\mr^{d_y}\times \Theta\to\mr$ be a known function. The null hypothesis of interest is 
\begin{align*}
\HH_0:  \text{For some } \theta\in \Theta, \E_P[g(Y,\theta)|X]=0 \text{ almost surely}.
\end{align*}
This null hypothesis is equivalent to \begin{align*}
\HH_0:  \text{For some } \theta\in \Theta, \E_P [g(Y,\theta)\indicator\{X\le x\}]=0 \text{ for all } x\in \mr.
\end{align*}
In this case, $\psi_{x,\theta}(z)=g(y,\theta)\indicator\{w\le x\}$ for every $z=(w,y)\in\mr\times\mr^{d_y}$ and every $(x,\theta)\in\mr\times\Theta$, and $\phi_P(x,\theta)=\E_P [g(Y,\theta)\indicator\{X\le x\}]$ for every $(x,\theta)\in\mr\times\Theta$. \citet{tripathi2003testing} construct a smoothed empirical likelihood-based test for the conditional moment restrictions, \citet{escanciano2014specification} use a projected empirical process to eliminate the estimation effect of nuisance parameters, \citet{dominguez2015simple} introduce an omnibus test statistic as the minimized value of the objective function considered in \citet{dominguez2004consistent}, and \citet{berger2022testing} proposes a new empirical likelihood test for parameters of conditional moment restriction models. 

\citet{jun2009semiparametric} propose semi-parametric tests of conditional moment restrictions with weak instruments. The null rejection probabilities of their tests are shown to be asymptotically no greater than the nominal significance level, suggesting possible conservativeness. Under suitable conditions, the test proposed in this paper has an exact asymptotic size, which allows for dependent data as well. The performance improvement of our method over existing approaches is illustrated through simulation studies in Section \ref{sec: performance improvement}, where the data generating processes (DGPs) are tailored to conditional moment restriction models with weak instruments.
\end{exam}

\begin{exam}[label=symmetry](Symmetry)
\label{exam:symmetry.prob}
Let $G$ be the cumulative distribution function of the random variable $Z$. The null hypothesis of symmetry about center $\theta$ is
\begin{align*}
\HH_0: \text{For some } \theta\in \Theta, G(x)=1-G(2\theta-x) \text{ for all } x\in \mr.
\end{align*}
In this case, $\psi_{x,\theta}(z)=\indicator\{z\le x\}+\indicator\{z\le 2\theta-x\}-1$ for every $z\in\mr$ and every $(x,\theta)\in\mr\times\Theta$, and $\phi_P(x,\theta)=G(x)+G(2\theta-x)-1$ for every $(x,\theta)\in\mr\times\Theta$. \citet{psaradakis2015quantile} use a quantile-based measure of skewness to test symmetry about an unspecified center, \citet{psaradakis2016using} considers the autoregressive sieve bootstrap to obtain critical values for tests of symmetry, and \citet{psaradakis2022using} employ a U-statistic involving triples of observations to assess symmetry. \citet{psaradakis2019bootstrap} provide an overview of symmetry tests.
\end{exam}

\begin{exam}[label=fit](Goodness of Fit)
\label{exam:fit.prob}
Let $G$ be the cumulative distribution function of the random variable $Z$. Suppose there is a given class of distribution functions $\{G_0 (\cdot, \theta):\theta\in\Theta\}$ so that $x\mapsto G_0(x,\theta)$ is a distribution function on $\mr$ for every $\theta\in\Theta$. We assume the identifiability of $\theta$ in the sense that for all $\theta_1,\theta_2\in\Theta$ with $\theta_1\ne\theta_2$, there exists $x_0\in\mr$ such that $G_0(x_0,\theta_1)\ne G_0(x_0,\theta_2)$. The null hypothesis of correct specification is 
\begin{align*}
\HH_0: \text{For some } \theta\in \Theta, G(x)=G_0(x,\theta) \text{ for all } x\in \mr.
\end{align*}
In this case, $\psi_{x,\theta}(z)=\indicator\{z\le x\}-G_0(x,\theta)$ for every $z\in\mr$ and every $(x,\theta)\in\mr\times\Theta$, and $\phi_P(x,\theta)=G(x)-G_0(x,\theta)$. Goodness-of-fit tests based on parametric empirical processes have been extensively studied since \citet{durbin1973weak}. For example, the martingale approach proposed by \citet{khmaladze1982martingale} is applied to the problem of testing goodness of fit with estimated parameters, and \citet{genest2008validity} consider goodness-of-fit tests using a parametric bootstrap approach. A more recent work is \citet{parker2013comparison}, which recommends conducting sup-norm inference for tests based on \citet{durbin1985first}’s approximations.
\end{exam}

\begin{exam}[label=LST](Location-scale Transformation)
\label{exam:LST.prob}
We wish to test the null hypothesis of equal distributions up to some location-scale transformation. This is a generalization of the classical two-sample problem. Let $Z=(X,Y)$ be a two-dimensional random vector and $H$ be the joint cumulative distribution function of $Z$ with marginal distribution functions $F$ (for $X$) and $G$ (for $Y$). The null hypothesis is 
\begin{align}\label{eq: LST}
\HH_0:  \text{For some } \theta=(\theta_1,\theta_2)\in \Theta, F(x)=
G\left(\frac{x-\theta_1}{\theta_2}\right) \text{ for all } x\in \mr.
\end{align}
In this case, $\psi_{x,\theta}(z)=\indicator\{z_1\le x\}-\indicator\{z_2\le(x-\theta_1)/\theta_2\}$ for every $z=(z_1,z_2)\in\mr^2$ and every $(x,\theta)\in\mr\times\Theta$, and $\phi_P(x,\theta)=F(x)-G((x-\theta_1)/\theta_2)$.
A substantial number of tests exist for comparing two or multiple distributions. See, for example, \citet{lehmann2005testing} and \citet{chen2018modern} for extensive reviews. 
\citet{hall2013new} propose an extension of the Cram{\'e}r--von Mises type test based on empirical characteristic functions to examine whether the two samples come from the same location-scale family of distributions. \citet{henze2005checking} and \citet{jimenez2017fast} deal with the two-sample problem using similar test statistics. 

An important special case of Example \ref{exam:LST.prob} is testing for heterogeneous treatment effects.
We follow \citet{ding2016randomization} and \citet{chung2021permutation} and consider a randomized experiment model. 
Let $Y$ denote the observable outcome of interest, and $D$ denote the binary treatment variable. If an individual is randomly assigned to the treatment group and receives treatment, then $D=1$; otherwise, the individual is randomly assigned to the control group and does not receive treatment, with $D=0$. 
Suppose that $Y(1)$ is the potential outcome of an individual if treated, and $Y(0)$ is the potential outcome if not treated. 
The treatment effect is constant if $Y(1)-Y(0)=\theta$ almost surely for some fixed constant $\theta$; otherwise, the treatment effect is said to be heterogeneous. The null hypothesis of constant treatment effect is 
\begin{align}\label{eq: HTE}
\HH_0^{s}:  \text{For some } \theta\in \Theta, Y(1)-Y(0)=\theta \text{ almost surely}.
\end{align}
Hypothesis \eqref{eq: HTE} is a more restrictive sharp null and is usually not directly testable. A necessary and weaker condition of this sharp null hypothesis, which is considered by \citet{ding2016randomization} and \citet{chung2021permutation}, is 
\begin{align*}
\HH_0:  \text{For some } \theta\in \Theta, F(x)=G(x-\theta) \text{ for all }x\in\mr,
\end{align*}
where $F$ and $G$ are the CDFs of $Y(1)$ and $Y(0)$, respectively. Clearly, this condition can be incorporated into \eqref{eq: LST}. 
\end{exam}

\textbf{{Organization of the Paper}:}
Section \ref{sec:Parametric Transformations} introduces the framework and develops theoretical results for testing general moment restrictions in the presence of nuisance parameters. 
Section \ref{sec:Extension to Dependent Data} extends the results to dependent data.
Section \ref{sec:simulation} provides Monte Carlo simulation evidence to show the performance of the test in finite samples. Section \ref{sec:conclusion} concludes the paper. Auxiliary lemmas, analyses and extensions of examples, all mathematical proofs, and additional simulation results are collected in the Online Supplementary Appendix.

\textbf{{Notation:}} Notation follows common conventions \citep[e.g.,][]{van1996weak,kosorok2008introduction}. 
We use $M^\T$ to denote the transpose of a matrix $M$. For $a,b\in\mr$, we define $a\wedge b=\min\{a,b\}$ and $a\vee b=\max\{a,b\}$. We use two forms of indicator functions: $\indicator\{S\}=1$ if the statement $S$ is true, and $\indicator\{S\}=0$ otherwise; $\indicator_A(x)=1$ if $x\in A$, and $\indicator_A(x)=0$ if $x\notin A$.  For an arbitrary set $A$, let $\ell^\infty(A)$ be the set of bounded real-valued functions on $A$. Equip $\ell^\infty(A)$ with the supremum norm $\lvv \cdot \rvv_{\infty}$ such that $\lvv f \rvv_{\infty}=\sup_{x\in A} \lv f(x) \rv$ for every $f\in \ell^\infty(A)$. For a subset $B$ of a metric space, let $\mC(B)$ be the set of continuous real-valued functions on $B$, and $\mCb(B)$ be the set of bounded continuous functions on $B$, that is, $\mCb(B)=\mC(B)\cap \ell^\infty(B)$. Following the notation of \citet{van1996weak}, for every normed space $\mathbb{B}$ with a norm $\lvv \cdot \rvv_\mathbb{B}$, we define \begin{align*}
\mathrm{BL}_1(\mathbb{B})=\lbr  \Gamma: \mathbb{B} \to \mr : 	\lv \Gamma(a) \rv\le 1 \; \text{and}\; \lv \Gamma(a)-\Gamma(b) \rv \le 	\lvv a-b \rvv_\mathbb{B}\; \text{for all}\; a, b \in \mathbb{B} \rbr.
\end{align*}
Let $\mathbb{F}$ be an arbitrary vector space equipped with a norm $\Vert\cdot\Vert_{\mathbb{F}}$. For every $C\subset \bF$ and every $\varepsilon>0$, define the $\varepsilon$-neighborhood of $C$ to be \begin{align*}
C^\varepsilon=\lbr g\in \bF : \inf_{f\in C} \lvv f- g \rvv_\bF	\le \varepsilon \rbr.
\end{align*}
For every measure $\nu$ on $\lp \mr, \borel (\mr) \rp$,
let $L^p(\nu)$ be the set of functions such that 
\begin{align*}
L^p(\nu)=\lbr f: \mr\to \mr :  \int_\mr | f(x) |^p \ddd \nu(x)
<\infty  \rbr
\end{align*}
with $p\ge 1$.
Equip $L^p(\nu)$ with the norm $\lvv \cdot \rvv_{L^p(\nu)}$ such that 
\begin{align*}
\lvv f \rvv_{L^p(\nu)}= \left\{\int_\mr | f(x) |^p \ddd \nu(x)\right\}^{1/p}
\end{align*}
for every $f\in L^p(\nu)$. 

Let $\mu$ be the Lebesgue measure on $\lp \mr, \borel(\mr) \rp$. 
For an arbitrary space $\cF$, we say $\bW$ is a $P$-Brownian bridge in $\ell^\infty(\cF)$ if and only if $\bW$ is a tight Borel measurable Gaussian process with $\E_P[\bW(f_1)]=0$ and $\E_P[\bW(f_1)\bW(f_2)]=P(f_1 f_2)-P(f_1)P(f_2)$ for all $f_1,f_2\in\cF$. Let $\leadsto$ denote the weak convergence defined in \citet[p.~4]{van1996weak}. Let $\overset{\mathbb{P}}{\leadsto}$ and $\overset{\text{a.s.}}{\leadsto}$ denote the weak convergence in probability conditional on the sample and almost sure weak convergence conditional on the sample, respectively, as defined in \citet[pp.~19--20]{kosorok2008introduction}. 

\section{Test Formulation} \label{sec:Parametric Transformations}

\subsection{Setup}

Let $\nu$ be a probability measure on $\lp \mr, \borel(\mr) \rp$.
We first introduce the following assumptions.  

\begin{assum}\label{ass:para properties of g}
For every $\theta\in\Theta$,
the function $x\mapsto \phi_P(x,\theta)$ is continuous.
\end{assum}

\begin{assum}\label{ass:para nu and mu}
The probability measure $\nu$ on $\lp \mr, \borel(\mr) \rp$ satisfies $\mu \ll \nu$, that is, if $\nu(B)=0$ for some $B\in\mathscr{B}(\mathbb{R})$, then $\mu(B)=0$.
\end{assum}

\begin{assum}\label{ass:para Theta is compact}
The set $\Theta$ is compact in $\mr^{d_\theta}$.
\end{assum}

\begin{assum}\label{ass:para g mapsto f(g) is continuous}
For every $\theta_0\in\Theta$ and every $\varepsilon>0$,
there exists $\delta>0$ such that \begin{align*}
\sup_{x\in\mr} P\lbk (\psi_{x,\theta}-\psi_{x,\theta_0})^2 \rbk <\varepsilon
\end{align*}
for all $\theta\in\Theta$ with $\lvv \theta-\theta_0 \rvv_2 <\delta$.
\end{assum}

Assumption \ref{ass:para properties of g} shows that we focus on moment restrictions that are continuous in $x$ for every $\theta\in\Theta$. Assumption \ref{ass:para nu and mu} requires the absolute continuity of the Lebesgue measure $\mu$ with respect to the probability measure $\nu$. For example, $\nu$ could be set as the probability measure corresponding to a normal distribution with a large variance.\footnote{See the discussion and simulation results in Section \ref{sec:simulation}.} Assumption \ref{ass:para Theta is compact} is a common condition on the compactness of $\Theta$. Assumption \ref{ass:para g mapsto f(g) is continuous} can be understood as the continuity of $\psi_{x,\theta}$ with respect to $\theta$ under a certain metric.

Define a function space
\begin{align*}
\bD_{\mL0} = \lbr \varphi\in\ell^\infty(\mr\times\Theta):
\theta\mapsto \varphi(\cdot,\theta), \text{ as a map from } \Theta
\text{ to } \lsv, \text{ is continuous} \rbr.
\end{align*}
In the definition of $\bD_{\mL0}$, the continuity of the map 
$\theta\mapsto \varphi(\cdot,\theta)$
is understood in the sense that for every $\theta_0\in \Theta$ and every $\varepsilon>0$,
there exists $\delta>0$ such that 
\begin{align*}
\int_\mr \lbk \varphi(x,\theta)-\varphi \lp x, \theta_0\rp \rbk^2 \ddd \nu(x)<\varepsilon
\end{align*}
for all $\theta\in\Theta$ with $\lvv \theta-\theta_0 \rvv_2 <\delta$. Note that for every $x\in\mr$ and all $\theta,\theta_0\in\Theta$, by Jensen's inequality,
\begin{align*}
[\phi_P(x,\theta)-\phi_P(x,\theta_{0})]^2 \le P\lbk (\psi_{x,\theta}-\psi_{x,\theta_0})^2 \rbk.
\end{align*}
Since $\nu$ is a probability measure, Assumption \ref{ass:para g mapsto f(g) is continuous} implies that $\phi_P\in\bD_{\mL 0}$.

The proposition below provides an equivalent characterization
of the null hypothesis in \eqref{eq:para original null}. We construct the test based on this equivalent characterization to avoid estimating the nuisance parameter $\theta$ under the null.

\begin{prop}\label{prop:para equivalent null}
If Assumptions 
\ref{ass:para properties of g}--\ref{ass:para g mapsto f(g) is continuous}
hold, then the null hypothesis in \eqref{eq:para original null} is equivalent to 
\begin{align}
\HH_0: \inf_{\theta\in\Theta}\int_{\mr} \lbk\phi_P(x,\theta) \rbk^2
\ddd \nu(x)=0.
\label{eq:para working null}
\end{align}
\end{prop}

It is worth noting that different measures $\nu$ may deliver
different power properties of the test. However, searching for the optimal $\nu$ to maximize power is challenging, as it may depend in a complicated manner on the DGP.

The measure $\nu(\mathbb{R})$ is assumed to be finite (Assumption \ref{ass:para nu and mu}) to obtain the theoretical results in the paper. In practice, we suggest setting $\nu$ to a normal probability measure with a large variance so that it does not heavily concentrate on some region of the real line, given no prior information about the DGP. Other probability measures satisfying Assumption \ref{ass:para nu and mu} also work asymptotically for the proposed method. For finite samples, the simulation results in Section \ref{sec:simulation} show that normal probability measures with different variances ($\mathcal{N}(0,1)$, $\mathcal{N}(0,5^2)$, $\mathcal{N}(0,10^2)$) perform well.

\subsection{Test Statistic}\label{sec:para Test Statistic}

We first restrict our attention to independent and identically distributed (i.i.d.)\ samples, and will extend the results to dependent data in Section \ref{sec:Extension to Dependent Data}. Let $\Phat$ be the empirical probability measure of the sample $\bZn$, which assigns weight $1/n$ to each observation $Z_i$ with $i\in\{1,\ldots,n\}$. Then the sample analogue of $\phi_P$ is defined as
\begin{align*}
\phihat(x,\theta)=\Phat(\psi_{x,\theta})=\frac{1}{n}\sum_{i=1}^n \psi_{x,\theta}(Z_i)
\end{align*}
for every $(x,\theta)\in\mr\times\Theta$. We present the exact function form of $\widehat{\phi}_n$ in every example.

\begin{exam}[continues=CMR]
\label{exam:CMR.est}
With the known function $g$, it follows by definition that \begin{align*}
\phihat(x,\theta)=\Phat(\psi_{x,\theta})=\frac{1}{n}\sum_{i=1}^n \psi_{x,\theta}(Z_i)=\frac{1}{n}\sum_{i=1}^n g(Y_i,\theta)\indicator\{X_i\le x\}
\end{align*}
for every $(x,\theta)\in\mr\times\Theta$.
\end{exam}

\begin{exam}[continues=symmetry]
\label{exam:symmetry.est}
The cumulative distribution function $G$ can be estimated by the empirical distribution function $\widehat{G}_{n}$ such that for every $x\in \mr$,
\begin{align*}
\widehat{G}_{n}(x)=\Phat(\indicator_{(-\infty,x]})=\frac{1}{n}\sum_{i=1}^{n} \indicator_{(-\infty,x]}\lp Z_i \rp .
\end{align*} 
Then \begin{align*}
\phihat(x,\theta)=\Phat(\psi_{x,\theta})=\Phat(\indicator_{(-\infty,x]})+\Phat(\indicator_{(-\infty,2\theta-x]})-1= \widehat{G}_{n}(x)+\widehat{G}_{n}(2\theta-x)-1
\end{align*}
for every $(x,\theta)\in\mr\times\Theta$.
\end{exam}

\begin{exam}[continues=fit]
\label{exam:fit.est}
The cumulative distribution function $G$ can be estimated by the empirical distribution function $\widehat{G}_{n}$ such that for every $x\in \mr$,
\begin{align*}
\widehat{G}_{n}(x)=\Phat(\indicator_{(-\infty,x]})=\frac{1}{n}\sum_{i=1}^{n} \indicator_{(-\infty,x]}\lp Z_i \rp .
\end{align*} 
Then $\phihat(x,\theta)=\Phat(\psi_{x,\theta})=\Phat(\indicator_{(-\infty,x]})-G_0(x,\theta)=\widehat{G}_{n}(x)-G_0(x,\theta)$ for every $(x,\theta)\in\mr\times\Theta$.
\end{exam}

\begin{exam}[continues=LST]
\label{exam:LST.est}
For every $i\in\{1,\ldots, n\}$, the observation $Z_i=(X_i,Y_i)$. Let $\Phat$ be the empirical distribution of $\{Z_i\}_{i=1}^n$, and $\widehat{H}_{n}$ be its empirical distribution function so that \begin{align*}
\widehat{H}_{n}(x,y)=\frac{1}{n}\sum_{i=1}^{n} \indicator_{(-\infty,x]\times(-\infty,y]}\lp X_i,Y_i \rp 
\end{align*} 
for all $(x,y)\in\mathbb{R}^2$. Let $\widehat{P}_{X,n}$ and $\widehat{P}_{Y,n}$ be the marginal distributions of $\Phat$, i.e., the (marginal) empirical distributions of $\lbr X_i \rbr_{i=1}^{n}$ and $\lbr Y_i \rbr_{i=1}^{n}$, respectively. It follows that \begin{align*}
\phihat(x,\theta)=\Phat(\psi_{x,\theta})=\widehat{P}_{X,n}(\indicator_{(-\infty,x]})-\widehat{P}_{Y,n}(\indicator_{(-\infty,(x-\theta_1)/\theta_2]})
\end{align*}
for every $(x,\theta)\in\mr\times\Theta$. The marginal distribution functions $F$ and $G$ can be estimated by the empirical distribution functions $\widehat{F}_n$ and $\widehat{G}_n$, respectively, where for every $x\in\mr$, \begin{align*}
&\Fhat(x)=\widehat{P}_{X,n}(\indicator_{(-\infty,x]})=\frac{1}{n}\sum_{i=1}^{n} \indicator_{(-\infty,x]}\lp X_i \rp \text{ and }\\
&\Ghat(x)=\widehat{P}_{Y,n}(\indicator_{(-\infty,x]})=\frac{1}{n}\sum_{i=1}^{n} \indicator_{(-\infty,x]}\lp Y_i \rp.
\end{align*}
This implies that $\phihat(x,\theta)=\Fhat (x)-\Ghat [ (x-\theta_1)/\theta_2 ]$ for every $(x,\theta)\in\mr\times\Theta$.

We may also test this null hypothesis with two independent samples of different sizes. This case will be discussed in Appendix \ref{sec:multiple CDFs}, where we present the results for comparing multiple samples.
\end{exam}

To obtain the asymptotic law of the stochastic process $\phihat$, we need the following assumption on the function class $\Psi$.

\begin{assum}\label{ass:para Donsker class}
The function class $\Psi=\{\psi_{x,\theta}:(x,\theta)\in\mr\times\Theta\}$ satisfies that
\begin{align}
\sup_{f\in\Psi} |P(f)|<\infty \text{ and } \sup_{f\in\Psi} \lv f(z)-P(f) \rv <\infty	 \label{eq:uniform boundedness of Psi}
\end{align}
for all $z\in\mr^{d_z}$, and is $P$-Donsker in the sense that 
\begin{align}
\sqrt{n}(\Phat-P)\convd \bW \text{ in } \ell^\infty(\Psi)  \label{eq:weak convergence of Psi}
\end{align}
as $n\to\infty$, where $\bW$ is a $P$-Brownian bridge in $\ell^\infty(\Psi)$.
\end{assum}

Lemma \ref{prop:para Donsker of phihat} establishes the consistency of $\phihat$ and the weak convergence of $\sqrt{n}( \phihat-\phi_P )$ in $\ell^\infty(\mr\times\Theta)$ as $n\to\infty$.

\begin{lemma}\label{prop:para Donsker of phihat}
If Assumptions \ref{ass:para g mapsto f(g) is continuous} and \ref{ass:para Donsker class} hold, then $(\phihat-\phi_P)\in\ell^\infty(\mr\times\Theta)$ for all $n\in\mathbb{Z}_+$. In addition, 
\begin{align*}
\sup_{(x,\theta)\in\mr\times\Theta} \lv \phihat(x,\theta)-\phi_P(x,\theta) \rv \convp 0 \text{ and }
\sqrt{n}( \phihat-\phi_P )\convd \bG_0 \text{ in } \ell^\infty(\mr\times\Theta)
\end{align*}
as $n\to\infty$, where $\bG_{0}$ is some tight random element which almost surely takes values in $\bD_{\mL0}$.
\end{lemma}

Define a function space
\begin{align*}
\bD_{\mL}=\left\{  \varphi\in\ell^{\infty}(\mathbb{R}\times\Theta): \int_{\mr} \lbk \varphi (x,\theta) \rbk^2 \ddd \nu(x)<\infty \text{ for all } \theta\in\Theta \right\}.
\end{align*}
Define a map $\mathcal{L}$ on $\bD_{\mL}$ such that $\mL(\varphi)= \inf_{\theta\in\Theta} \int_{\mr} \lbk \varphi (x,\theta) \rbk^2 \ddd \nu(x)$
for every $\varphi\in \bD_{\mL}$.
Then under Assumptions 
\ref{ass:para properties of g}--\ref{ass:para Donsker class}, the null and the alternative hypotheses can be expressed as 
\begin{align}\label{eq:para null L}
\HH_0: \mL (\phi_P)=0 \text{ and } \HH_1: \mL (\phi_P)>0.
\end{align}
To test the null hypothesis in \eqref{eq:para null L}, we set the test statistic to $n \mL ( \phihat )$.

Next, we show that the map $\mL$ is Hadamard directionally differentiable,
but its Hadamard directional derivative is degenerate under $\HH_0$.\footnote{See Definition \ref{def:first order Hadamard} for Hadamard directional differentiability.} 
Define $$\bD_0=\lbr \varphi\in \bD_{\mL0}:\mL(\varphi)=0 \rbr.$$ 
The following lemma provides the Hadamard directional derivative of $\mathcal{L}$ and its first order degeneracy under $\mathrm{H}_0$.

\begin{lemma}\label{lemma:para-HDDL}
If Assumptions \ref{ass:para Theta is compact} and \ref{ass:para g mapsto f(g) is continuous}
hold, then $\mL$ is Hadamard directionally differentiable at $\phi_P\in \bD_{\mL}$
tangentially to $\bD_{\mL0}$ with the Hadamard directional
derivative \begin{align*}
\mL'_{\phi_P}(h)=2\inf_{\theta\in \Theta_0(\phi_P)} \int_{\mr} \phi_P(x,\theta)h(x,\theta)
\ddd \nu(x) \text{ for all } h\in \bD_{\mL0},
\end{align*}
where $\Theta_0(\phi_P)=\argmin_{\theta\in\Theta}\int_{\mr} \lbk \phi_P (x,\theta) \rbk^2 \ddd \nu(x)$.
Moreover, if $\phi_P\in\bD_0$, then the derivative $\mL'_{\phi_P}$ is
well defined on the whole of $\ell^\infty(\mr\times\Theta)$ with 
$\mL'_{\phi_P}(h)=0$ for every $h\in\ell^\infty(\mr\times\Theta)$.
\end{lemma}

The first order degeneracy of $\mL$ under $\HH_0$ implies that we may need to find the second order Hadamard directional derivative of $\mL$.\footnote{See Definition \ref{def:second order Hadamard CF} for second order Hadamard directional differentiability.} We assume the following conditions to guarantee the existence of the second order Hadamard directional derivative of $\mL$. 

\begin{assum}\label{ass:para G is 2nd differentiable}
The function $\phi_P$ is twice differentiable with respect to $\theta$, and the second partial derivative satisfies
\begin{align}\label{eq:second order derivative bound}
\int_{\mathbb{R}}	\sup_{\theta\in\Theta} \lvv \left.
\frac{\partial^2 \phi_P (z,\vartheta)}{\partial \vartheta \partial \vartheta^\T} 
\rv_{(z,\vartheta)=(x,\theta)} \rvv_{2}^2 \,\mathrm{d}\nu(x)
<\infty,
\end{align}
where $\lvv \cdot \rvv_{2}$ denotes the $\ell^2$ operator norm of a matrix.
\end{assum}

\begin{assum}\label{ass:para strong identifiability}
The set $\Theta_0\equiv\{\theta\in\Theta:\int_{\mr} \lbk \phi_P (x,\theta) \rbk^2 \ddd \nu(x)=0\}\subset\mathrm{int}(\Theta)$,
and there exist $\kappa\in(0,1]$, $\varepsilonbar>0$, and $C>0$ such 
that for all $\varepsilon\in (0,\varepsilonbar)$, 
\begin{align}\label{eq.identification para}
\inf_{\theta\in\Theta\setminus\Theta_0^\varepsilon}
\left\{\int_\mr \lbk \phi_P(x,\theta) \rbk^2 \ddd \nu(x)\right\}^{1/2} \ge C \varepsilon^\kappa.
\end{align}
\end{assum}

We provide Assumptions \ref{ass:para G is 2nd differentiable}
and \ref{ass:para strong identifiability} following the basic idea of \citet{chen2019inference}. Assumption \ref{ass:para G is 2nd differentiable} requires the boundedness of the second partial derivative of $\phi_P$ in the sense of \eqref{eq:second order derivative bound}. Assumption \ref{ass:para strong identifiability} requires that the set $\Theta_0$ is in the interior of $\Theta$ and it is well separated. The condition in \eqref{eq.identification para} is similar to the partial identification assumption used in \citet[p.~1265]{chernozhukov2007estimation}. It is worth noting that these conditions are sufficient but not necessary for our results, as also mentioned by \citet{chen2019inference}.  We impose such high level conditions for theoretical completeness.
In Section \ref{sec:simulation}, we verify these assumptions for a conditional moment restriction model.

\begin{lemma}\label{prop:para second order Hadamard of L}
If Assumptions \ref{ass:para Theta is compact},
\ref{ass:para g mapsto f(g) is continuous}, \ref{ass:para G is 2nd differentiable},
and \ref{ass:para strong identifiability}
hold, and $\phi_P\in\bD_0$, then the function $\mL$ is second order Hadamard directionally differentiable
at $\phi_P$ tangentially to $\bD_{\mL0}$ with the second order Hadamard
directional derivative \begin{align*}
\mL''_{\phi_P}(h)=\inf_{\theta\in\Theta_0(\phi_P)} \inf_{v\in \mathbb{R}^{d_{\theta}}}
\lvv \lbk \Phi'(\theta) \rbk^\T v +\sh (\theta) \rvv_\lsv^2 \text{ for all } h\in \bD_{\mL0},
\end{align*}	
where $\Phi'(\theta): \mr\to\mr^{d_{\theta}}$ with \begin{align*}
\Phi'(\theta)(x)=\left.
\frac{\partial \phi_P (z,\vartheta)}{\partial \vartheta} 
\rv_{(z,\vartheta)=(x,\theta)}
\quad \text{for every } (x,\theta)\in\mr\times\Theta,
\end{align*}
and $\sh:\Theta\to \ell^\infty(\mathbb{R})$ with $\sh(\theta)(x)=h(x,\theta)$ for every
$(x,\theta)\in\mr\times\Theta$.
\end{lemma}

\begin{remark}
Lemma \ref{prop:para second order Hadamard of L} provides the explicit expression of the complicated second order Hadamard directional derivative of $\mathcal{L}$. We employ a numerical method that does not require exploring this function form. 
\end{remark}

With Lemma \ref{prop:para second order Hadamard of L}, the asymptotic null distribution of the test statistic $n\mL( \phihat )$
is obtained by applying the second order delta method.

\begin{prop}\label{thry:para asymptotic distribution of test stat}
If Assumptions \ref{ass:para properties of g}--\ref{ass:para strong identifiability} hold and $\HH_0$ is true ($\phi_P\in\bD_0$), then \begin{align*}
n \mL(\phihat )	\convd \mL''_{\phi_P}\lp \bG_0 \rp \text{ as } n\to \infty.
\end{align*}
\end{prop}

\subsection{Bootstrap Procedure}

The distribution of $\mL''_{\phi_P}\lp \bG_0 \rp$ in Proposition \ref{thry:para asymptotic distribution of test stat} is unknown because
both the function $\mL''_{\phi_P}$ and the stochastic process $\bG_0$ depend
on the unknown underlying distribution $P$.
Motivated by \citet{hong2018numerical} and
\citet{chen2019inference}, we propose to approximate $\mL''_{\phi_P}$
by a consistent estimator
and approximate the distribution of $\bG_0$ by bootstrap.\footnote{Bootstrap may not be the only method to approximate the distribution of $\bG_0$ in our framework. Other consistent estimators of $\bG_0$ might also suffice for the proposed approach.}
We use the numerical second order Hadamard
directional derivative $\mLhat$ to approximate $\mL''_{\phi_P}$, which is defined as
\begin{align*}
\mLhat(h)=\frac{\mL( \phihat+\tau_n h )-\mL( \phihat )}{\tau_n^2}
\end{align*}
for all $h\in \ell^\infty(\mr\times\Theta)$, where $\lbr \tau_n \rbr$ is a sequence of tuning parameters
satisfying the assumption below.\footnote{As discussed in \citet{chen2019inference}, the modified bootstrap in \citet{babu1984bootstrapping} (Babu correction) is inappropriate when $\mathcal{L}$ is only second order Hadamard directionally differentiable but $\mathcal{L}''_{\phi_P}$ is not ``continuous'' in $\phi_P$. To ensure that our method can accommodate more general cases, we employ the bootstrap method of \citet{hong2018numerical} and \citet{chen2019inference}.}

\begin{assum}\label{ass:para rate of tau}
$\lbr \tau_n \rbr\subset\mr_+$ is a sequence of scalars 
such that $\tau_n\downarrow 0$
and $\tau_n \sqrt{n}\to\infty$ as $n\to\infty$.
\end{assum}

Assumption \ref{ass:para rate of tau} provides the rate at which $\tau_n\downarrow0$. Under this condition, we show that $\mLhat$ approximates $\mL''_{\phi_P}$ well in the following lemma. 
\begin{lemma}\label{prop:para consistency of second derivative estimator}
If Assumptions \ref{ass:para properties of g}--\ref{ass:para rate of tau}
hold and $\HH_0$ is true ($\phi_P\in\bD_0$),
then for every sequence $\lbr h_n \rbr\subset \ell^\infty
(\mr\times\Theta)$ and every $h\in\bD_{\mL0}$ such that $h_n\to h$
in $\ell^\infty	(\mr\times\Theta)$ as $n\to\infty$, we have \begin{align*}
\mLhat\lp h_n \rp \convp \mL''_{\phi_P} (h)\text{ as }	n\to\infty.
\end{align*}
\end{lemma}

We next approximate the distribution of $\bG_0$ via bootstrap. The bootstrap sample $\bZn^*=\{Z_i^*\}_{i=1}^n$ is i.i.d.\ drawn from the empirical distribution $\Phat$ of the original sample $\bZn$. Equivalently, $\bZn^*$ is a random sample of size $n$, drawn from the set $\bZn$ with replacement. Let $\Phat^*$ be the empirical distribution of $\bZn^*$. Then the bootstrap version of $\phihat$ is $\phihat^*$ such that \begin{align*}
\phihat^*(x,\theta)=\Phat^*(\psi_{x,\theta})=\frac{1}{n}\sum_{i=1}^n \psi_{x,\theta}(Z_i^*)
\end{align*}
for every $(x,\theta)\in\mr\times\Theta$.

\begin{exam}[continues=CMR] \label{exam:CMR.boot}
It follows by definition that \begin{align*}
\phihat^*(x,\theta)=\Phat^*(\psi_{x,\theta})=\frac{1}{n}\sum_{i=1}^n \psi_{x,\theta}(Z_i^*)=\frac{1}{n}\sum_{i=1}^n g(Y_i^*,\theta)\indicator\{X_i^*\le x\}
\end{align*}
for every $(x,\theta)\in\mr\times\Theta$, where $Z_i^*=(X_i^*,Y_i^*)$.
\end{exam}

\begin{exam}[continues=symmetry]\label{exam:symmetry.boot}
Define \begin{align*}
\widehat{G}_n^*(x)=\Phat^*(\indicator_{(-\infty,x]})=\frac{1}{n}\sum_{i=1}^{n} \indicator_{(-\infty,x]}\lp Z^*_i \rp
\end{align*}
for every $x\in\mr$. Then \begin{align*}
\phihat^*(x,\theta)=\Phat^*(\psi_{x,\theta})=\Phat^*(\indicator_{(-\infty,x]})+\Phat^*(\indicator_{(-\infty,2\theta-x]})-1= \widehat{G}_n^*(x)+\widehat{G}_n^*(2\theta-x)-1
\end{align*}
for every $(x,\theta)\in \mr\times\Theta$. 
\end{exam}

\begin{exam}[continues=fit]\label{exam:fit.boot}
Define \begin{align*}
\widehat{G}_n^*(x)=\Phat^*(\indicator_{(-\infty,x]})=\frac{1}{n}\sum_{i=1}^{n} \indicator_{(-\infty,x]}\lp Z^*_i \rp
\end{align*}
for every $x\in\mr$. Then \begin{align*}
\phihat^*(x,\theta)=\Phat^*(\psi_{x,\theta})=\Phat^*(\indicator_{(-\infty,x]})-G_0(x,\theta)=\widehat{G}_n^*(x)-G_0(x,\theta)
\end{align*} 
for every $(x,\theta)\in \mr\times\Theta$. 
\end{exam}

\begin{exam}[continues=LST]\label{exam:LST.boot}
Define $\Phat^*$ as the empirical distribution of $\{Z_i^*\}_{i=1}^n$ with $Z_i^*=(X_i^*,Y_i^*)$. Let $\widehat{P}^*_{X,n}$ and $\widehat{P}^*_{Y,n}$ be the marginal distributions of $\Phat^*$, i.e., the (marginal) empirical distributions of $\lbr X_i^* \rbr_{i=1}^{n}$ and $\lbr Y_i^* \rbr_{i=1}^{n}$, respectively. It follows that \begin{align*}
\phihat^*(x,\theta)=\Phat^*(\psi_{x,\theta})=\widehat{P}_{X,n}^*(\indicator_{(-\infty,x]})-\widehat{P}_{Y,n}^*(\indicator_{(-\infty,(x-\theta_1)/\theta_2]})
\end{align*}
for every $(x,\theta)\in\mr\times\Theta$. Define $\Fhat^*$ and $\Ghat^*$ to be the (marginal) empirical distribution functions of $\lbr X_i^* \rbr_{i=1}^{n}$ and $\lbr Y_i^* \rbr_{i=1}^{n}$, respectively, such that for every $x\in\mr$, \begin{align*}
&\Fhat^*(x)=\widehat{P}_{X,n}^*(\indicator_{(-\infty,x]})=\frac{1}{n}\sum_{i=1}^{n} \indicator_{(-\infty,x]}\lp X_i^* \rp\text{ and } \\
&\Ghat^*(x)=\widehat{P}_{Y,n}^*(\indicator_{(-\infty,x]})=\frac{1}{n}\sum_{i=1}^{n} \indicator_{(-\infty,x]}\lp Y_i^* \rp.
\end{align*}
This implies that $\phihat^*(x,\theta)=\Fhat^* (x)-\Ghat^*\lp (x-\theta_1)/\theta_2 \rp$ for every $(x,\theta)\in\mr\times\Theta$.
\end{exam}

The following lemma establishes the conditional weak convergence of $\sqrt{n}(\phihat^*-\phihat)$ in probability as $n\to\infty$.

\begin{lemma}\label{prop:para weak convergence of bootstrap}
If Assumption \ref{ass:para Donsker class} holds, then as $n\to\infty$, \begin{align*}
\sup_{\Gamma\in\mathrm{BL}_1\lp \ell^\infty(\mr\times\Theta) \rp}\lv \E \lbk \left. \Gamma \lp \sqrt{n} \lp \phihat^*-\phihat \rp \rp	\rv \bZn \rbk-\E \lbk \Gamma \lp \bG_0 \rp \rbk \rv \convp 0,
\end{align*}
and $\sqrt{n}( \phihat^*-\phihat )$ is asymptotically measurable, where $\bG_0$ is defined as in Lemma \ref{prop:para Donsker of phihat}.
\end{lemma}

With the numerical estimator $\mLhat$ for $\mL''_{\phi_P}$ and a suitable bootstrap approximation $\sqrt{n}( \phihat^*-\phihat )$ for $\bG_0$ at hand, we can naturally approximate the distribution
of $\mL''_{\phi_P}\lp \bG_0 \rp$ by the conditional distribution of the bootstrap test statistic $\mLhat \{ 
\sqrt{n}( \phihat^*-\phihat ) \}$ given the original sample. This is justified by the following proposition.

\begin{prop}\label{prop:para consistency of approximation of test statistic distribution}
If Assumptions \ref{ass:para properties of g}--\ref{ass:para rate of tau} hold and $\HH_0$ is true ($\phi_P\in\bD_0$), then \begin{align*}
\sup_{\Gamma\in\mathrm{BL}_1\lp \mr \rp}
\lv \E \lbk \left. \Gamma \lp \mLhat \lbk \sqrt{n} \lp \phihat^*
-\phihat \rp \rbk \rp \rv \bZn \rbk-\E \lbk \Gamma \lp \mL''_{\phi_P} \lp \bG_0 \rp
\rp \rbk \rv \convp 0
\end{align*}
as $n\to\infty$.
\end{prop}

\subsection{Asymptotic Properties}

Now we construct the test for the null hypothesis $\HH_0$.	
For a given level of significance $\alpha\in(0,1)$, define the bootstrap critical value \begin{align*}
\chat_{1-\alpha,n}=\inf\lbr c\in \mr: \p\lp\left. 
\mLhat \lbk \sqrt{n} \lp \phihat^*-\phihat \rp \rbk
\le c \rv \bZn\rp \ge 1-\alpha \rbr.
\end{align*}
In practice, $\chat_{1-\alpha,n}$ may be approximated by the $1-\alpha$ empirical 
quantile of the $n_B$ independently	generated bootstrap test statistics, with $n_B$ set to be as large as computationally feasible. We reject $\HH_0$ if and only if $n\mL ( \phihat )>\chat_{1-\alpha,n}$.
The following theorem shows that the proposed test is asymptotically size controlled and consistent.

\begin{thry}\label{thry:para size and power}
Suppose that Assumptions \ref{ass:para properties of g}--\ref{ass:para rate of tau} hold. \begin{enumerate}[label=(\roman*),nosep]
\item If $\HH_0$ is true and the CDF of $\mL''_{\phi_P}\lp \bG_0 \rp$
is strictly increasing and continuous at its $1-\alpha$ quantile,
then \begin{align*}
\lim_{n\to\infty}
\p\lp n \mL( \phihat )>\chat_{1-\alpha,n} \rp = \alpha.
\end{align*}
\item If $\HH_0$ is false, then \begin{align*}
\lim_{n\to\infty}
\p\lp n \mL( \phihat )>\chat_{1-\alpha,n} \rp=1.
\end{align*}
\end{enumerate}
\end{thry}

\subsection{Local Power}\label{sec.local power}

In this section, we consider the local power of the test following the discussion in \citet{chen2019inference}. For each $n\in\mathbb{Z}_+$, let the sample $\bZn=\{Z_i\}_{i=1}^n$ be distributed according to the joint law $P_n^n=\prod_{i=1}^n P_n$, where $P_n$ is a probability distribution on $(\mr^{d_z}, \borel(\mr^{d_z}))$ with $P_n(B)=\mathbb{P}(Z_i\in B)$ for every Borel set $B$. 
That is, for each $n\in\mathbb{Z}_+$, the observations $Z_1,\ldots,Z_n$ are i.i.d.\ with distribution $P_n$. We suppose that the null hypothesis $\mathrm{H}_0$ is false for each $P_n$, that is, for all $\theta\in\Theta$, $P_n(\psi_{x,\theta})\ne 0$ for some $x\in\mr$. Suppose that $P_n$ converges (in a way as described in the following assumption) to the probability measure $P$, and that $P$ satisfies $\mathrm{H}_0$, that is, for some $\theta\in\Theta$, $P(\psi_{x,\theta})= 0$ for all $x\in\mathbb{R}$.

\begin{assum}\label{ass:local sequence matched pairs}
The probability distributions $P_n$ and $P$ satisfy that 
\begin{align}\label{eq:Pn convergence matched}
\lim_{n\to\infty}\int \left[ \sqrt{n}\left( \mathrm{d}P_n^{1/2} - \mathrm{d}P^{1/2} \right)    -\frac{1}{2}v_0\,\mathrm{d}P^{1/2} \right]^2=0
\end{align}
for some measurable function $v_0:\mr^{d_z}\to\mr$, where $\mathrm{d}P_n^{1/2}$ and $\mathrm{d}P^{1/2}$ denote the square roots of the densities of $P_n$ and $P$, respectively.
\end{assum}

Our local power results rely on Assumption \ref{ass:local sequence matched pairs}, which is similar to (3.10.10) of \citet{van1996weak}.
The following proposition states formally the local power property of the test. 

\begin{prop}\label{prop:local power}
Suppose that Assumptions \ref{ass:para properties of g}--\ref{ass:local sequence matched pairs} hold and  $\sup_{f\in\Psi} |P_n(f^2)|=O(1)$. Then $\sqrt{n}( \phihat-\phi_P )\leadsto \mathbb{G}_0+\zeta_P$, where $\mathbb{G}_0$ is some tight random element, and $\zeta_P(x,\theta)= P(\psi_{x,\theta}v_0)$ for every $(x,\theta)\in\mathbb{R}\times\Theta$. Furthermore, if the CDF of $\mathcal{L}''_{\phi_P}(\mathbb{G}_0)$ is strictly increasing and continuous at its $1-\alpha$ quantile $c_{1-\alpha}$, then it follows that
\begin{align*}
\liminf_{n\to\infty}\p\lp n \mL( \phihat )>\chat_{1-\alpha,n} \rp \ge \mathbb{P}(\mathcal{L}''_{\phi_P}(\mathbb{G}_0+\zeta_P)>c_{1-\alpha}).
\end{align*}
\end{prop}

Proposition \ref{prop:local power} follows from Lemma C.1 of \citet{chen2019inference} and provides lower bounds for the power of the test under local perturbations to the null. 

\section{Dependent Data}
\label{sec:Extension to Dependent Data}

This section considers cases where the observations $\{Z_i\}_{i=1}^n$ may be dependent. For results established in Section \ref{sec:Parametric Transformations}, it is worth noting that Lemmas \ref{lemma:para-HDDL}--\ref{prop:para consistency of second derivative estimator}, Propositions \ref{prop:para equivalent null}--\ref{prop:para consistency of approximation of test statistic distribution}, and Theorem \ref{thry:para size and power} do not directly rely on the i.i.d.\ nature of the data observations, possibly given the consistency and weak convergence of $\phihat$ (Lemma \ref{prop:para Donsker of phihat}) and the conditional weak convergence of $\phihat^*$ in probability (Lemma \ref{prop:para weak convergence of bootstrap}). Thus, to obtain the asymptotic properties of the proposed test in dependent samples, it suffices to establish the consistency and weak convergence of $\phihat$ and the conditional weak convergence of $\phihat^*$ in probability under dependency.

A sequence of $d_z$-dimensional random vectors, $\{Z_i:i\in\mathbb{Z}\}$, is said to be strictly stationary, if for all $\{i_1,\ldots,i_n\}\subset\mathbb{Z}$ and all $n\in\mathbb{Z}_+$, the joint distribution of $(Z_{i_1+k},\ldots,Z_{i_n+k})$ does not depend on $k$. For $-\infty\le s\le t \le \infty$, let $\mathscr{S}_{s}^{t}$ be the $\sigma$-field generated by $\{Z_s,\ldots, Z_t\}$. Following Equation (II) of \citet{volkonskii1959some} and (1.1) of \citet{arcones1994central}, the $\beta$-mixing coefficient $\beta_k$ of the sequence $\{Z_i:i\in\mathbb{Z}\}$ is defined as \begin{align*}
\beta_k=\sup_{t\in\mathbb{Z}} \E \lbk \sup_{A\in \mathscr{S}_{t+k}^\infty} \lv \p \lp A \lv \mathscr{S}_{-\infty}^t \right. \rp-\p(A) \rv \rbk,
\end{align*}
and $\{Z_i:i\in\mathbb{Z}\}$ is said to be $\beta$-mixing if and only if $\beta_k\to 0$ as $k\to\infty$.

Throughout our discussion of cases with dependent data, we assume that the sample $\bZn=\{Z_i:i=1,\ldots, n\}$ is a finite segment of the strictly stationary sequence $\{Z_i:i\in\mathbb{Z}\}$ in which the common marginal distribution of $Z_i$ is $P$. We impose the following assumptions.

\begin{assum}\label{ass:dep VC class}
The class $\Psi=\{\psi_{x,\theta}:(x,\theta)\in\mr\times\Theta\}$ is a VC-subgraph class of functions satisfying \eqref{eq:uniform boundedness of Psi} with $P(\overline{\psi}^p)<\infty$ for some $p\in(2,\infty)$, where $\overline{\psi}(z)=\sup_{f\in\Psi}|f(z)|$ for every $z\in\mr^{d_z}$, and $\Psi$ is totally bounded under $\Vert\cdot\Vert_{L^2(P)}$.\footnote{See the definition of VC-subgraph class of functions in Section 2.6 of \citet[p.~141]{van1996weak}.}
\end{assum}

With $p$ specified as in Assumption \ref{ass:dep VC class}, we introduce the following condition for $\beta_k$.

\begin{assum}\label{ass:dep beta mixing}
The sequence $\{Z_i:i\in\mathbb{Z}\}$ is $\beta$-mixing with coefficient $\beta_k=O(k^{-q})$ as $k\to\infty$ for some $q>p/(p-2)$.
\end{assum}

Assumption \ref{ass:dep VC class} emerges as one of the conditions in Theorem 2.1 of \citet{arcones1994central} and Theorem 1 of \citet{radulovic1996bootstrap}. Assumption \ref{ass:dep beta mixing} corresponds to one of the conditions in Theorem 1 of \citet{radulovic1996bootstrap}.

Let $\phihat$ and $\phi_P$ be defined as in Section \ref{sec:Parametric Transformations}. The lemma below establishes the consistency and weak convergence of $\phihat$ as $n\to\infty$.

\begin{lemma}\label{lemma:dep weak convergence}
If Assumptions \ref{ass:para Theta is compact}, \ref{ass:para g mapsto f(g) is continuous}, \ref{ass:dep VC class}, and \ref{ass:dep beta mixing} hold, then $(\phihat-\phi_P)\in\ell^\infty(\mr\times\Theta)$ for all $n\in\mathbb{Z}_+$. In addition, \begin{align*}
\sup_{(x,\theta)\in\mr\times\Theta} \lv \phihat(x,\theta)-\phi_P(x,\theta) \rv \convp 0 \text{ and }
\sqrt{n}( \phihat-\phi_P )\convd \bG_0  \text{ in }  \ell^\infty(\mr\times\Theta)
\end{align*}
as $n\to\infty$, where $\bG_{0}$ is tight and almost surely takes values in $\bD_{\mL0}$.
\end{lemma}

To construct the bootstrap sample $\bZn^*=\{Z_i^*\}_{i=1}^n$, we follow \citet{radulovic1996bootstrap} and use the moving blocks bootstrap (MBB) procedure. Recall that the original sample is $\{Z_i\}_{i=1}^n$. Let $b\in\mathbb{Z}_+$ be the block size satisfying $b\to\infty$ and $b/n\to0$, 
and $k\in\mathbb{Z}_+$ be the number of blocks. Without loss of generality, we may assume that $k$ and $b$ satisfy $kb=n$.\footnote{In practice, $n/b$ may not always be an integer. In this case, we set $k=\lceil n/b \rceil$ and generate $kb>n$ bootstrap observations according to the algorithm described in the main text, and then keep the first $n$ observations as the bootstrap sample.\label{footnote:non integer k}} For $i\in\{1,\ldots, b-1\}$, we set $Z_{n+i}=Z_i$. Let the random variables $I_1,\ldots, I_k$ be i.i.d.\ from $\mathrm{Unif}\{1,\ldots, n\}$ and independent of the original sample. For all $\ell\in\{1,\ldots,k\}$ and $j\in\{1,\ldots,b\}$, set the bootstrap observation $Z_{(\ell-1)b+j}^*=Z_{I_\ell+j-1}$. That is, the bootstrap sample is
\begin{align*}
\bZn^*=\{Z_{I_1}, Z_{I_1+1}, \ldots, Z_{I_1+b-1}, Z_{I_2}, Z_{I_2+1}, \ldots, Z_{I_2+b-1}, \ldots , Z_{I_k}, Z_{I_k+1}, \ldots, Z_{I_k+b-1}\}.
\end{align*}
Let $\Phat^*$ be the empirical distribution of $\bZn^*$. The bootstrap version of $\phihat$ is defined as \begin{align*}
\phihat^*(x,\theta)=\Phat^*(\psi_{x,\theta})=\frac{1}{n}\sum_{i=1}^n \psi_{x,\theta}(Z_i^*)
\end{align*}
for every $(x,\theta)\in\mr\times\Theta$.

We impose the assumption below on the block size $b$, which treats $b$ as a function of $n$, that is, $b=b(n)$. 
This assumption corresponds to one of the conditions in Theorem 1 of \citet{radulovic1996bootstrap}.

\begin{assum}\label{ass:dep block size b}
The block size $b$ is a function of the sample size $n$ such that $b=b(n)=O(n^r)$ as $n\to\infty$ for some $0<r<(p-2)/(2p-2)$.
\end{assum}

The following lemma establishes the conditional weak convergence of $\sqrt{n}(\phihat^*-\phihat)$ in probability.

\begin{lemma}\label{lemma:dep conditional weak convergence of bootstrap}
If Assumptions \ref{ass:dep VC class}--\ref{ass:dep block size b} hold, then as $n\to\infty$, \begin{align*}
\sup_{\Gamma\in\mathrm{BL}_1\lp \ell^\infty(\mr\times\Theta) \rp}\lv \E \lbk \left.  \Gamma \lp \sqrt{n} \lp \phihat^*-\phihat \rp \rp	\rv \bZn \rbk-\E \lbk \Gamma \lp \bG_0 \rp \rbk \rv \convp 0,
\end{align*}
where $\bG_0$ is defined as in Lemma \ref{lemma:dep weak convergence}.
\end{lemma}

Given the modification to the construction of the bootstrap sample, the remaining steps of the test follow the procedure in Section \ref{sec:Parametric Transformations}. For dependent data, the test is also asymptotically size controlled and consistent, as shown in Theorem \ref{thry:dep size and power}.

\begin{thry}\label{thry:dep size and power}
Suppose that Assumptions \ref{ass:para properties of g}--\ref{ass:para g mapsto f(g) is continuous}, \ref{ass:para G is 2nd differentiable}--\ref{ass:para rate of tau}, and \ref{ass:dep VC class}--\ref{ass:dep block size b} hold, and that $\sqrt{n}(\phihat^*-\phihat)$ is asymptotically measurable. \begin{enumerate}[label=(\roman*),nosep]
\item If $\HH_0$ is true and the CDF of $\mL''_{\phi_P}\lp \bG_0 \rp$
is strictly increasing and continuous at its $1-\alpha$ quantile,
then \begin{align*}
\lim_{n\to\infty}
\p\lp n \mL( \phihat )>\chat_{1-\alpha,n} \rp = \alpha.
\end{align*}
\item If $\HH_0$ is false, then \begin{align*}
\lim_{n\to\infty}
\p\lp n \mL( \phihat )>\chat_{1-\alpha,n} \rp=1.
\end{align*}
\end{enumerate}
\end{thry}

\section{Monte Carlo Experiments}\label{sec:simulation}

In this section, we construct the Monte Carlo experiments based on the conditional moment restriction models with weak instrumental variables (IVs) in \citet[Example II]{jun2009semiparametric}. Let $y_i$ be a scalar outcome variable, $Y_i$ be a scalar endogenous variable, and $z_i$ be a scalar instrumental variable. The model of interest is \begin{align}
\E_P[y_i-Y_i\theta_0|z_i]=0 \text{ almost surely}  \label{eq:IV conditional moment restriction}
\end{align}
for a true structural parameter $\theta_0\in\Theta\subset\mr$. We consider the null hypothesis 
\begin{align*}
\HH_0: \text{For some } \theta\in\Theta, \; \E_P[y_i-Y_i\theta|z_i]=0 \text{ almost surely},
\end{align*}
which is equivalent to 
\begin{align*}
\HH_0: \text{For some } \theta\in\Theta, \; \E_P \lbk (y_i-Y_i\theta) \indicator\{z_i\le x\} \rbk=0 \text{ for all } x\in\mr.
\end{align*}

As noted by \citet{jun2009semiparametric}, there are several specification tests for \eqref{eq:IV conditional moment restriction} under strong
point identification and the assumption that $\theta_0$ can be $\sqrt{n}$-consistently estimated under the null  \citep[e.g.,][]{bierens1990consistent,zheng1996consistent,fan1996consistent,fan2000consistent}. Since $\E_P[y_i-Y_i\theta_0|z_i]=0$ almost surely for some $\theta_0$ under the null, typical estimators of $\theta_0$ include two-stage least squares (2SLS) and semi-parametric methods. However, when instruments are weak, these estimators of $\theta_0$ may be undesirable \citep[e.g.,][]{staiger1997instrumental,stock2000gmm,jun2012testing}, and thus two-step tests plugging in preliminary estimators of $\theta_0$ may not perform well.

\citet{jun2009semiparametric} propose semi-parametric specification tests of conditional moment restrictions with weak instruments, which do not require a consistent first-step estimator. As shown in Theorems 1 and 2 of \citet{jun2009semiparametric}, their tests yield limiting rejection probabilities no greater than the nominal significance level under the null. Based on their Example II, \citet{jun2009semiparametric} study the finite sample performance of their tests with weak instruments via Monte Carlo experiments. We first follow \citet{jun2009semiparametric} and consider two cases under the null. In the first case (Case 1), $\E_P[z_iY_i]=0$, that is, the rank condition fails and $z_i$ is not a valid instrument for $Y_i$ when estimating $\theta_0$ by 2SLS in two-step tests, which may be seen as an extreme case of weak instruments. Thus, the 2SLS estimator of $\theta_0$ that uses $z_i$ as the instrument for $Y_i$ is unreliable. In the second case (Case 2), $\E_P[Y_i|z_i]\to 0$ almost surely as $n\to\infty$, that is, all measurable functions $f$ of $z_i$ with $\E_P[|f(z_i)Y_i|]<\infty$ may be weak instruments for $Y_i$ when estimating $\theta_0$ by 2SLS in two-step tests because $\E_P[f(z_i)Y_i]$ may converge to $0$. As discussed in \citet[]{jun2012testing}, semi-parametric estimators of $\theta_0$ may also break down when $\E_P[Y_i|z_i]$ decays too fast in $n$.
In addition, we consider a third case (Case 3), which is an extreme case of Case 2: $\E_P[Y_i|z_i]= 0$ almost surely.

As demonstrated in Tables 1 and 2 of \citet{jun2009semiparametric}, their tests improve greatly upon two-step plug-in methods in the presence of weak instruments, while they are often conservative, which is in line with their theoretical results. 
The proposed test in this paper is asymptotically exactly size controlled and consistent under certain conditions, regardless of the strength of instruments. We numerically present these properties through Monte Carlo experiments, where the DGPs are designed for conditional moment restriction models with weak instruments as in the above cases. 

Now we introduce the designs of our simulations. For i.i.d.\ samples, we follow the design of \citet{jun2009semiparametric}: \begin{align*}
y_i&=Y_i+\delta \ln \lp Y_i^2+1 \rp+u_i ,\\
Y_i&=\lambda g(z_i)+v_i,
\end{align*}
where $\{(u_i,v_i,z_i):i=1,\ldots,n\}$ are i.i.d.\ with \begin{align*}
\begin{bmatrix}
u_i \\ v_i \\ z_i
\end{bmatrix}\sim \mathcal{N} \lp \begin{bmatrix}
0 \\ 0 \\ 0
\end{bmatrix}, \; \begin{bmatrix}
1 & \rho & 0 \\ \rho & 1 & 0 \\ 0 & 0 & 1
\end{bmatrix} \rp.
\end{align*}
The aforementioned three cases are realized in the following manner:
\begin{itemize}
\item Case 1: $\rho=0.5$, $\lambda=1$, and $g(z)=z^2-1$. The moment $\E_P[z_iY_i]=0$.

\item Case 2: $\rho=-0.99$, $\lambda=0.07\sqrt{200/n}$, and $g(z)=z$. The conditional moment $\E_P[Y_i|z_i]\to 0$ almost surely as $n\to\infty$. 

\item Case 3: $\rho=-0.5$ and $\lambda=0$. The conditional moment $\E_P[Y_i|z_i]=0$ almost surely. 
\end{itemize}
For each case, we consider four DGPs characterized by the values of $\delta$:
\begin{itemize}
\item DGP (0): $\delta=0$. The null is true.
\item DGP (1): $\delta=0.2$. The null is false.
\item DGP (2): $\delta=0.6$. The null is false.
\item DGP (3): $\delta=1$. The null is false.
\end{itemize}

We also consider dependent data. For every DGP introduced above, we construct the dependent-data counterpart by generating $\{z_i:i=1,\ldots,n\}$ as \begin{align*}
z_0=0,z_i=0.5 z_{i-1}+\varepsilon_{i},
\end{align*}
where $\{\varepsilon_{i}:i=1,\ldots,n\}$ are i.i.d.\ $\mathcal{N}(0,1)$ and independent of $\{(u_i,v_i):i=1,\ldots,n\}$.

\begin{remark}
\label{remark:verify ass for CMR}
For illustration of the high level assumptions in Section \ref{sec:para Test Statistic}, we consider Case 1 with $\delta=0$. 
With $\theta_{0}=1$, $\mathrm{H}_{0}$ is true and thus
\[
\mathbb{E}_{P}\left[  \left(  y_{i}-Y_{i}\theta_{0}\right)  \mathbbm{1}\left\{
z_{i}\leq x\right\}  \right]  =0
\]
for all $x$. We have that for all $\theta$,
\begin{align*}
\phi_{P}\left(  x,\theta\right)   &  =\mathbb{E}_{P}\left[  \left(
y_{i}-Y_{i}\theta\right)  \mathbbm{1}\left\{  z_{i}\leq x\right\}  \right]  \\
&  =\mathbb{E}_{P}\left[  y_{i}\mathbbm{1}\left\{  z_{i}\leq x\right\}  \right]
-\mathbb{E}_{P}\left[  Y_{i}\mathbbm{1}\left\{  z_{i}\leq x\right\}  \right]  \theta.
\end{align*}
It follows that for all $\theta$,
\begin{align*}
&  \int_{\mathbb{R}}\phi_{P}\left(  x,\theta\right)  ^{2}\mathrm{d}\nu\left(
x\right)  \\
=&\,\int_{\mathbb{R}}(\mathbb{E}_{P}\left[  y_{i}\mathbbm{1}\left\{  z_{i}\leq x\right\}
\right])  ^{2}\mathrm{d}\nu\left(  x\right)  -2\int_{\mathbb{R}}\mathbb{E}%
_{P}\left[  y_{i}\mathbbm{1}\left\{  z_{i}\leq x\right\}  \right]  \mathbb{E}_{P}\left[
Y_{i}\mathbbm{1}\left\{  z_{i}\leq x\right\}  \right]  \mathrm{d}\nu\left(  x\right)
\theta\\
&+\int_{\mathbb{R}}(\mathbb{E}_{P}\left[  Y_{i}\mathbbm{1}\left\{  z_{i}\leq x\right\}
\right])  ^{2}\mathrm{d}\nu\left(  x\right)  \theta^{2}\geq0.
\end{align*}
The value $\theta_{0}=1$ satisfies $\int_{\mathbb{R}}\phi_{P}\left(
x,\theta_{0}\right)  ^{2}\mathrm{d}\nu\left(  x\right)  =0$, so we have
\[
\Theta_{0}=\left\{  \theta_{0}\right\}  ,\theta_{0}=\frac{\int_{\mathbb{R}%
}\mathbb{E}_{P}\left[  y_{i}\mathbbm{1}\left\{  z_{i}\leq x\right\}  \right]
\mathbb{E}_{P}\left[  Y_{i}\mathbbm{1}\left\{  z_{i}\leq x\right\}  \right]
\mathrm{d}\nu\left(  x\right)  }{\int_{\mathbb{R}}(\mathbb{E}_{P}\left[
Y_{i}\mathbbm{1}\left\{  z_{i}\leq x\right\}  \right] ) ^{2}\mathrm{d}\nu\left(
x\right)  }=1.
\]
For every $\varepsilon>0$,
\begin{align*}
&  \int_{\mathbb{R}}\phi_{P}\left(  x,\theta_{0}-\varepsilon\right)
^{2}\mathrm{d}\nu\left(  x\right)  \\
=&\,\int_{\mathbb{R}}(\mathbb{E}_{P}\left[  y_{i}\mathbbm{1}\left\{  z_{i}\leq x\right\}
\right] ) ^{2}\mathrm{d}\nu\left(  x\right)  -2\int_{\mathbb{R}}\mathbb{E}%
_{P}\left[  y_{i}\mathbbm{1}\left\{  z_{i}\leq x\right\}  \right]  \mathbb{E}_{P}\left[
Y_{i}\mathbbm{1}\left\{  z_{i}\leq x\right\}  \right]  \mathrm{d}\nu\left(  x\right)
\left(  \theta_{0}-\varepsilon\right)  \\
&  +\int_{\mathbb{R}}(\mathbb{E}_{P}\left[  Y_{i}\mathbbm{1}\left\{  z_{i}\leq x\right\}
\right] ) ^{2}\mathrm{d}\nu\left(  x\right)  \left(  \theta_{0}-\varepsilon
\right)  ^{2}\\
=&\,2\int_{\mathbb{R}}\mathbb{E}_{P}\left[  y_{i}\mathbbm{1}\left\{  z_{i}\leq
x\right\}  \right]  \mathbb{E}_{P}\left[  Y_{i}\mathbbm{1}\left\{  z_{i}\leq x\right\}
\right]  \mathrm{d}\nu\left(  x\right)  \varepsilon-2\int_{\mathbb{R}%
}(\mathbb{E}_{P}\left[  Y_{i}\mathbbm{1}\left\{  z_{i}\leq x\right\}  \right])
^{2}\mathrm{d}\nu\left(  x\right)  \theta_{0}\varepsilon\\
&  +\int_{\mathbb{R}}(\mathbb{E}_{P}\left[  Y_{i}\mathbbm{1}\left\{  z_{i}\leq x\right\}
\right] ) ^{2}\mathrm{d}\nu\left(  x\right)  \varepsilon^{2}\\
=&\,\int_{\mathbb{R}}(\mathbb{E}_{P}\left[  Y_{i}\mathbbm{1}\left\{  z_{i}\leq x\right\}
\right] ) ^{2}\mathrm{d}\nu\left(  x\right)  \varepsilon^{2}.
\end{align*}
This implies that
\[
\left\{  \int_{\mathbb{R}}\phi_{P}\left(  x,\theta_{0}-\varepsilon\right)
^{2}\mathrm{d}\nu\left(  x\right)  \right\}  ^{1/2}=\left\{  \int_{\mathbb{R}%
}(\mathbb{E}_{P}\left[  Y_{i}\mathbbm{1}\left\{  z_{i}\leq x\right\}  \right])
^{2}\mathrm{d}\nu\left(  x\right)  \right\}  ^{1/2}\varepsilon.
\]
In this case, Assumptions \ref{ass:para G is 2nd differentiable} and \ref{ass:para strong identifiability} hold.
The asymptotic limit of the test statistic is
\begin{align}\label{eq: asymptotic limit CMR}
\mathcal{L}_{\phi_{P}}^{\prime\prime}\left(  \mathbb{G}_{0}\right)
=\inf_{v\in \mathbb{R}  }\int
_{\mathbb{R}}\left(  \mathbb{G}_{0}\left(  x,\theta_{0}\right)  -\mathbb{E}
_{P}\left[  Y_{i}\mathbbm{1}\left\{  z_{i}\leq x\right\}  \right]  v\right)
^{2}\mathrm{d}\nu\left(  x\right)  .
\end{align}
Theorem \ref{thry:para size and power}(i) requires that the CDF of $\mathcal{L}_{\phi_{P}}^{\prime\prime
}\left(  \mathbb{G}_{0}\right)  $ in \eqref{eq: asymptotic limit CMR} is strictly increasing and continuous at its
$1-\alpha$ quantile.
\end{remark}

The sample size is set to $n\in\{100,200,400,800\}$. We set the tuning parameter $\tau_n$ as $\tau_n=\sqrt{\ln(n)/n}$, $n^{-2/5}$, $n^{-1/3}$, $n^{-1/4}$, $n^{-1/5}$, and $n^{-1/6}$, which all satisfy Assumption \ref{ass:para rate of tau}. For dependent data, the moving blocks bootstrap involves  
an additional tuning parameter $b(n)$. We set $b(n)=n^{1/6}$, $n^{1/5}$, $n^{1/4}$, and $n^{1/3}$. Recall that the test statistic involves an integration with respect to a measure $\nu$ and an infimum. The integration is approximated by an equally weighted average on the grid $\{-3,-2.998, -2.996, \ldots, 3\}$ of $x$, and the infimum is achieved by a search on the grid $\{0.7, 0.702, 0.704, \ldots, 1.3\}$ of $\theta$. Furthermore, we apply the warp-speed method \citep{giacomini2013warp} to implement all the Monte Carlo experiments. Specifically, for each DGP and sample size, we generate $1000$ samples and compute one original statistic $n\mL(\phihat)$ and one bootstrap statistic $\mLhat[\sqrt{n}(\phihat^*-\phihat)]$ for each sample. The critical value $\chat_{1-\alpha,n}$ is approximated by the $(1-\alpha)$-empirical quantile of the $1000$ bootstrap statistics, and the rejection rate is computed by comparing the $1000$ original statistics with the critical value $\chat_{1-\alpha,n}$. 

We present some main simulation results in the following and leave the remaining results to Section \ref{sec: additional simulations} of the Online Supplementary Appendix. Tables \ref{tab:case1 null}--\ref{tab:case1dep DGP (3)} and \ref{tab:case2 null}--\ref{tab:case3dep DGP (3)} show the rejection rates for different DGPs, tuning parameters, and nominal significance levels with measure $\nu$ being the probability measure of $\mathcal{N}(0,10^2)$. Tables \ref{tab:case1 sig1}--\ref{tab:case1dep DGP (3) sig5} display the rejection rates for Case 1 with the measure $\nu$ being the probability measure of $\mathcal{N}(0,1)$ or $\mathcal{N}(0,5^2)$. The results are stable for different choices of $\tau_n$, $b(n)$, and $\nu$. Most of the rejection rates under the null are close to the nominal significance levels. The rejection rates under the alternatives increase to one as the sample size $n$ increases. For dependent samples, the rejection rates under the null may exceed the significance level $\alpha$ for some $\tau_n$, $b(n)$, and $\nu$ as shown, for example, in Table \ref{tab:case1dep null}. As we increase the sample sizes, the results become closer to $\alpha$, as shown in Table \ref{tab:case1dep null large samples}.

\begin{table}[H]
\small\centering\setstretch{1.3}
\caption{Size for Case 1 with i.i.d.\ data}
\label{tab:case1 null}
\begin{tabular*}{15cm}{@{\extracolsep{\fill}}cccccccc}
\hline\hline
\multirow{2}{*}{$\alpha$} & \multirow{2}{*}{$n$} & \multicolumn{6}{c}{$\tau_n$} \\
\cline{3-8}
& & $\sqrt{\ln(n)/n}$ & $n^{-2/5}$ & $ n^{-1/3}$ & $ n^{-1/4}$ & $ n^{-1/5}$ & $ n^{-1/6}$ \\
\hline
\multirow{4}{*}{$0.01$}
&$	100	$&$	0.011	$&$	0.007	$&$	0.011	$&$	0.011	$&$	0.011	$&$	0.011	$\\
&$	200	$&$	0.004	$&$	0.003	$&$	0.004	$&$	0.006	$&$	0.010	$&$	0.011	$\\
&$	400	$&$	0.003	$&$	0.003	$&$	0.003	$&$	0.005	$&$	0.006	$&$	0.006	$\\
&$	800	$&$	0.008	$&$	0.008	$&$	0.008	$&$	0.014	$&$	0.014	$&$	0.014	$\\
\hline
\multirow{4}{*}{$0.025$}
&$	100	$&$	0.026	$&$	0.017	$&$	0.026	$&$	0.027	$&$	0.026	$&$	0.027	$\\
&$	200	$&$	0.020	$&$	0.015	$&$	0.020	$&$	0.022	$&$	0.025	$&$	0.026	$\\
&$	400	$&$	0.022	$&$	0.021	$&$	0.022	$&$	0.022	$&$	0.023	$&$	0.023	$\\
&$	800	$&$	0.019	$&$	0.016	$&$	0.023	$&$	0.026	$&$	0.026	$&$	0.026	$\\
\hline
\multirow{4}{*}{$0.05$}
&$	100	$&$	0.043	$&$	0.038	$&$	0.043	$&$	0.051	$&$	0.054	$&$	0.054	$\\
&$	200	$&$	0.040	$&$	0.035	$&$	0.041	$&$	0.046	$&$	0.050	$&$	0.051	$\\
&$	400	$&$	0.058	$&$	0.044	$&$	0.067	$&$	0.069	$&$	0.062	$&$	0.067	$\\
&$	800	$&$	0.052	$&$	0.046	$&$	0.052	$&$	0.069	$&$	0.074	$&$	0.076	$\\
\hline
\multirow{4}{*}{$0.1$}
&$	100	$&$	0.101	$&$	0.090	$&$	0.101	$&$	0.111	$&$	0.111	$&$	0.111	$\\
&$	200	$&$	0.098	$&$	0.091	$&$	0.103	$&$	0.110	$&$	0.110	$&$	0.111	$\\
&$	400	$&$	0.109	$&$	0.101	$&$	0.114	$&$	0.124	$&$	0.130	$&$	0.131	$\\
&$	800	$&$	0.128	$&$	0.112	$&$	0.136	$&$	0.127	$&$	0.133	$&$	0.137	$\\
\hline
\multirow{4}{*}{$0.2$}
&$	100	$&$	0.219	$&$	0.209	$&$	0.219	$&$	0.241	$&$	0.244	$&$	0.244	$\\
&$	200	$&$	0.213	$&$	0.198	$&$	0.213	$&$	0.228	$&$	0.238	$&$	0.247	$\\
&$	400	$&$	0.219	$&$	0.206	$&$	0.229	$&$	0.235	$&$	0.238	$&$	0.241	$\\
&$	800	$&$	0.238	$&$	0.215	$&$	0.240	$&$	0.248	$&$	0.255	$&$	0.255	$\\
\hline\hline
\end{tabular*}
\end{table}

\begin{table}[H]
\small\centering\setstretch{1.3}
\caption{Power for Case 1 with i.i.d.\ data ($\alpha=0.05$)}
\label{tab:case1 alt}
\begin{tabular*}{15cm}{@{\extracolsep{\fill}}cccccccc}
\hline\hline
\multirow{2}{*}{DGP} & \multirow{2}{*}{$n$} & \multicolumn{6}{c}{$\tau_n$} \\
\cline{3-8}
& & $\sqrt{\ln(n)/n}$ & $n^{-2/5}$ & $ n^{-1/3}$ & $ n^{-1/4}$ & $ n^{-1/5}$ & $ n^{-1/6}$ \\
\hline
\multirow{4}{*}{DGP (1)}
&$	100	$&$	0.245	$&$	0.184	$&$	0.246	$&$	0.313	$&$	0.345	$&$	0.373	$\\
&$	200	$&$	0.440	$&$	0.362	$&$	0.460	$&$	0.573	$&$	0.623	$&$	0.638	$\\
&$	400	$&$	0.679	$&$	0.583	$&$	0.709	$&$	0.820	$&$	0.850	$&$	0.860	$\\
&$	800	$&$	0.888	$&$	0.822	$&$	0.924	$&$	0.976	$&$	0.990	$&$	0.992	$\\
\hline
\multirow{4}{*}{DGP (2)}
&$	100	$&$	0.865	$&$	0.797	$&$	0.866	$&$	0.926	$&$	0.949	$&$	0.956	$\\
&$	200	$&$	0.997	$&$	0.986	$&$	0.997	$&$	1.000	$&$	1.000	$&$	1.000	$\\
&$	400	$&$	1.000	$&$	1.000	$&$	1.000	$&$	1.000	$&$	1.000	$&$	1.000	$\\
&$	800	$&$	1.000	$&$	1.000	$&$	1.000	$&$	1.000	$&$	1.000	$&$	1.000	$\\
\hline
\multirow{4}{*}{DGP (3)}
&$	100	$&$	0.992	$&$	0.983	$&$	0.992	$&$	0.999	$&$	0.999	$&$	0.999	$\\
&$	200	$&$	1.000	$&$	1.000	$&$	1.000	$&$	1.000	$&$	1.000	$&$	1.000	$\\
&$	400	$&$	1.000	$&$	1.000	$&$	1.000	$&$	1.000	$&$	1.000	$&$	1.000	$\\
&$	800	$&$	1.000	$&$	1.000	$&$	1.000	$&$	1.000	$&$	1.000	$&$	1.000	$\\
\hline\hline
\end{tabular*}
\end{table}

\begin{table}[H]
\small\centering\setstretch{1.3}
\caption{Size for Case 1 with dependent data ($\alpha=0.05$)}
\label{tab:case1dep null}
\begin{tabular*}{15cm}{@{\extracolsep{\fill}}cccccccc}
\hline\hline
\multirow{2}{*}{$b(n)$} & \multirow{2}{*}{$n$} & \multicolumn{6}{c}{$\tau_n$} \\
\cline{3-8}
& & $\sqrt{\ln(n)/n}$ & $n^{-2/5}$ & $ n^{-1/3}$ & $ n^{-1/4}$ & $ n^{-1/5}$ & $ n^{-1/6}$ \\
\hline
\multirow{4}{*}{$n^{1/6}$}
&$	100	$&$	0.037	$&$	0.028	$&$	0.037	$&$	0.046	$&$	0.050	$&$	0.052	$\\
&$	200	$&$	0.050	$&$	0.037	$&$	0.054	$&$	0.058	$&$	0.067	$&$	0.069	$\\
&$	400	$&$	0.071	$&$	0.065	$&$	0.072	$&$	0.075	$&$	0.080	$&$	0.079	$\\
&$	800	$&$	0.060	$&$	0.050	$&$	0.068	$&$	0.077	$&$	0.082	$&$	0.083	$\\
\hline
\multirow{4}{*}{$n^{1/5}$}
&$	100	$&$	0.037	$&$	0.029	$&$	0.037	$&$	0.045	$&$	0.046	$&$	0.047	$\\
&$	200	$&$	0.035	$&$	0.029	$&$	0.037	$&$	0.040	$&$	0.045	$&$	0.049	$\\
&$	400	$&$	0.071	$&$	0.065	$&$	0.072	$&$	0.075	$&$	0.080	$&$	0.079	$\\
&$	800	$&$	0.047	$&$	0.045	$&$	0.064	$&$	0.078	$&$	0.081	$&$	0.085	$\\
\hline
\multirow{4}{*}{$n^{1/4}$}
&$	100	$&$	0.037	$&$	0.029	$&$	0.037	$&$	0.045	$&$	0.046	$&$	0.047	$\\
&$	200	$&$	0.039	$&$	0.035	$&$	0.044	$&$	0.054	$&$	0.058	$&$	0.061	$\\
&$	400	$&$	0.067	$&$	0.058	$&$	0.068	$&$	0.072	$&$	0.072	$&$	0.072	$\\
&$	800	$&$	0.083	$&$	0.072	$&$	0.088	$&$	0.097	$&$	0.097	$&$	0.100	$\\
\hline
\multirow{4}{*}{$n^{1/3}$}
&$	100	$&$	0.056	$&$	0.046	$&$	0.057	$&$	0.065	$&$	0.070	$&$	0.072	$\\
&$	200	$&$	0.047	$&$	0.037	$&$	0.049	$&$	0.055	$&$	0.059	$&$	0.064	$\\
&$	400	$&$	0.067	$&$	0.058	$&$	0.067	$&$	0.069	$&$	0.074	$&$	0.075	$\\
&$	800	$&$	0.057	$&$	0.036	$&$	0.071	$&$	0.078	$&$	0.085	$&$	0.084	$\\
\hline\hline
\end{tabular*}
\end{table}

\begin{table}[H]
\small\centering\setstretch{1.3}
\caption{Power for DGP (1) of Case 1 with dependent data ($\alpha=0.05$)}
\label{tab:case1dep DGP (1)}
\begin{tabular*}{15cm}{@{\extracolsep{\fill}}cccccccc}
\hline\hline
\multirow{2}{*}{$b(n)$} & \multirow{2}{*}{$n$} & \multicolumn{6}{c}{$\tau_n$} \\
\cline{3-8}
& & $\sqrt{\ln(n)/n}$ & $n^{-2/5}$ & $ n^{-1/3}$ & $ n^{-1/4}$ & $ n^{-1/5}$ & $ n^{-1/6}$ \\
\hline
\multirow{4}{*}{$n^{1/6}$}
&$	100	$&$	0.317	$&$	0.249	$&$	0.318	$&$	0.387	$&$	0.410	$&$	0.433	$\\
&$	200	$&$	0.520	$&$	0.393	$&$	0.547	$&$	0.655	$&$	0.683	$&$	0.697	$\\
&$	400	$&$	0.759	$&$	0.671	$&$	0.804	$&$	0.895	$&$	0.915	$&$	0.924	$\\
&$	800	$&$	0.988	$&$	0.964	$&$	0.992	$&$	1.000	$&$	1.000	$&$	1.000	$\\
\hline
\multirow{4}{*}{$n^{1/5}$}
&$	100	$&$	0.257	$&$	0.206	$&$	0.258	$&$	0.333	$&$	0.356	$&$	0.380	$\\
&$	200	$&$	0.482	$&$	0.368	$&$	0.509	$&$	0.617	$&$	0.673	$&$	0.686	$\\
&$	400	$&$	0.759	$&$	0.671	$&$	0.804	$&$	0.895	$&$	0.915	$&$	0.924	$\\
&$	800	$&$	0.990	$&$	0.969	$&$	0.992	$&$	1.000	$&$	1.000	$&$	1.000	$\\
\hline
\multirow{4}{*}{$n^{1/4}$}
&$	100	$&$	0.257	$&$	0.206	$&$	0.258	$&$	0.333	$&$	0.356	$&$	0.380	$\\
&$	200	$&$	0.547	$&$	0.421	$&$	0.569	$&$	0.680	$&$	0.688	$&$	0.703	$\\
&$	400	$&$	0.757	$&$	0.651	$&$	0.797	$&$	0.892	$&$	0.919	$&$	0.927	$\\
&$	800	$&$	0.987	$&$	0.963	$&$	0.992	$&$	1.000	$&$	1.000	$&$	1.000	$\\
\hline
\multirow{4}{*}{$n^{1/3}$}
&$	100	$&$	0.263	$&$	0.176	$&$	0.264	$&$	0.331	$&$	0.364	$&$	0.370	$\\
&$	200	$&$	0.486	$&$	0.381	$&$	0.507	$&$	0.632	$&$	0.672	$&$	0.688	$\\
&$	400	$&$	0.749	$&$	0.645	$&$	0.775	$&$	0.883	$&$	0.916	$&$	0.922	$\\
&$	800	$&$	0.978	$&$	0.950	$&$	0.988	$&$	1.000	$&$	1.000	$&$	1.000	$\\
\hline\hline
\end{tabular*}
\end{table}

\begin{table}[H]
\small\centering\setstretch{1.3}
\caption{Power for DGP (2) of Case 1 with dependent data ($\alpha=0.05$)}
\label{tab:case1dep DGP (2)}
\begin{tabular*}{15cm}{@{\extracolsep{\fill}}cccccccc}
\hline\hline
\multirow{2}{*}{$b(n)$} & \multirow{2}{*}{$n$} & \multicolumn{6}{c}{$\tau_n$} \\
\cline{3-8}
& & $\sqrt{\ln(n)/n}$ & $n^{-2/5}$ & $ n^{-1/3}$ & $ n^{-1/4}$ & $ n^{-1/5}$ & $ n^{-1/6}$ \\
\hline
\multirow{4}{*}{$n^{1/6}$}
&$	100	$&$	0.976	$&$	0.923	$&$	0.976	$&$	0.989	$&$	0.992	$&$	0.993	$\\
&$	200	$&$	1.000	$&$	0.998	$&$	1.000	$&$	1.000	$&$	1.000	$&$	1.000	$\\
&$	400	$&$	1.000	$&$	1.000	$&$	1.000	$&$	1.000	$&$	1.000	$&$	1.000	$\\
&$	800	$&$	1.000	$&$	1.000	$&$	1.000	$&$	1.000	$&$	1.000	$&$	1.000	$\\
\hline
\multirow{4}{*}{$n^{1/5}$}
&$	100	$&$	0.966	$&$	0.920	$&$	0.966	$&$	0.989	$&$	0.992	$&$	0.993	$\\
&$	200	$&$	1.000	$&$	0.999	$&$	1.000	$&$	1.000	$&$	1.000	$&$	1.000	$\\
&$	400	$&$	1.000	$&$	1.000	$&$	1.000	$&$	1.000	$&$	1.000	$&$	1.000	$\\
&$	800	$&$	1.000	$&$	1.000	$&$	1.000	$&$	1.000	$&$	1.000	$&$	1.000	$\\
\hline
\multirow{4}{*}{$n^{1/4}$}
&$	100	$&$	0.966	$&$	0.920	$&$	0.966	$&$	0.989	$&$	0.992	$&$	0.993	$\\
&$	200	$&$	1.000	$&$	0.999	$&$	1.000	$&$	1.000	$&$	1.000	$&$	1.000	$\\
&$	400	$&$	1.000	$&$	1.000	$&$	1.000	$&$	1.000	$&$	1.000	$&$	1.000	$\\
&$	800	$&$	1.000	$&$	1.000	$&$	1.000	$&$	1.000	$&$	1.000	$&$	1.000	$\\
\hline
\multirow{4}{*}{$n^{1/3}$}
&$	100	$&$	0.961	$&$	0.913	$&$	0.961	$&$	0.986	$&$	0.991	$&$	0.992	$\\
&$	200	$&$	1.000	$&$	0.999	$&$	1.000	$&$	1.000	$&$	1.000	$&$	1.000	$\\
&$	400	$&$	1.000	$&$	1.000	$&$	1.000	$&$	1.000	$&$	1.000	$&$	1.000	$\\
&$	800	$&$	1.000	$&$	1.000	$&$	1.000	$&$	1.000	$&$	1.000	$&$	1.000	$\\
\hline\hline
\end{tabular*}
\end{table}

\begin{table}[H]
\small\centering\setstretch{1.3}
\caption{Power for DGP (3) of Case 1 with dependent data ($\alpha=0.05$)}
\label{tab:case1dep DGP (3)}
\begin{tabular*}{15cm}{@{\extracolsep{\fill}}cccccccc}
\hline\hline
\multirow{2}{*}{$b(n)$} & \multirow{2}{*}{$n$} & \multicolumn{6}{c}{$\tau_n$} \\
\cline{3-8}
& & $\sqrt{\ln(n)/n}$ & $n^{-2/5}$ & $ n^{-1/3}$ & $ n^{-1/4}$ & $ n^{-1/5}$ & $ n^{-1/6}$ \\
\hline
\multirow{4}{*}{$n^{1/6}$}
&$	100	$&$	0.999	$&$	0.995	$&$	0.999	$&$	1.000	$&$	1.000	$&$	1.000	$\\
&$	200	$&$	1.000	$&$	1.000	$&$	1.000	$&$	1.000	$&$	1.000	$&$	1.000	$\\
&$	400	$&$	1.000	$&$	1.000	$&$	1.000	$&$	1.000	$&$	1.000	$&$	1.000	$\\
&$	800	$&$	1.000	$&$	1.000	$&$	1.000	$&$	1.000	$&$	1.000	$&$	1.000	$\\
\hline
\multirow{4}{*}{$n^{1/5}$}
&$	100	$&$	0.999	$&$	0.994	$&$	0.999	$&$	1.000	$&$	1.000	$&$	1.000	$\\
&$	200	$&$	1.000	$&$	1.000	$&$	1.000	$&$	1.000	$&$	1.000	$&$	1.000	$\\
&$	400	$&$	1.000	$&$	1.000	$&$	1.000	$&$	1.000	$&$	1.000	$&$	1.000	$\\
&$	800	$&$	1.000	$&$	1.000	$&$	1.000	$&$	1.000	$&$	1.000	$&$	1.000	$\\
\hline
\multirow{4}{*}{$n^{1/4}$}
&$	100	$&$	0.999	$&$	0.994	$&$	0.999	$&$	1.000	$&$	1.000	$&$	1.000	$\\
&$	200	$&$	1.000	$&$	1.000	$&$	1.000	$&$	1.000	$&$	1.000	$&$	1.000	$\\
&$	400	$&$	1.000	$&$	1.000	$&$	1.000	$&$	1.000	$&$	1.000	$&$	1.000	$\\
&$	800	$&$	1.000	$&$	1.000	$&$	1.000	$&$	1.000	$&$	1.000	$&$	1.000	$\\
\hline
\multirow{4}{*}{$n^{1/3}$}
&$	100	$&$	0.999	$&$	0.995	$&$	0.999	$&$	1.000	$&$	1.000	$&$	1.000	$\\
&$	200	$&$	1.000	$&$	1.000	$&$	1.000	$&$	1.000	$&$	1.000	$&$	1.000	$\\
&$	400	$&$	1.000	$&$	1.000	$&$	1.000	$&$	1.000	$&$	1.000	$&$	1.000	$\\
&$	800	$&$	1.000	$&$	1.000	$&$	1.000	$&$	1.000	$&$	1.000	$&$	1.000	$\\
\hline\hline
\end{tabular*}
\end{table}

\subsection{Performance Improvement in Conditional Moment Restriction Models with Weak Instruments}\label{sec: performance improvement}
Note that Cases 1 and 2 ($n=200$) with $\delta=0$ (under the null hypothesis) are identical to the DGPs in Tables 1 and 2 of \citet{jun2009semiparametric}, respectively. Thus, the results in Table \ref{tab:case1 null} with $n=100$ and Table \ref{tab:case2 null} with $n=200$ can be compared with those in Tables 1 and 2 of \citet{jun2009semiparametric}, respectively. We present the comparisons in Tables \ref{tab:comp tab1} and \ref{tab:comp tab2} below, where $\That_2(\thetahat_{*})$ and $\That_\mathrm{k}(\thetahat_{*})$ with $\thetahat_{*}\in \{\thetahat_{\mathrm{2SLS}}, \thetahat_{\mathrm{SP}}\}$ are two-step plug-in test statistics computed by using either a 2SLS or a semi-parametric estimator of $\theta_0$ as described in \citet{jun2009semiparametric}, and $\That_1(\thetahat_{\mathrm{CUE1}})$, $\That_2(\thetahat_{\mathrm{CUE2}})$, and $\That_\mathrm{k}(\thetahat_{\mathrm{CUEk}})$ are the test statistics proposed by \citet{jun2009semiparametric}.\footnote{The function $\That_\mathrm{k}(\cdot)$ is proposed by \citet{zheng1996consistent}, and the test statistic $\That_\mathrm{k}(\thetahat_{\mathrm{CUEk}})$ based on minimization of $\That_\mathrm{k}(\cdot)$ follows the idea of \citet{jun2009semiparametric}.} Our test uses $\tau_n=n^{-1/4}$ for illustration. The plug-in method suffers from substantial size distortion. The tests of \citet{jun2009semiparametric} improve upon the plug-in approach, but could be conservative as shown in their theoretical results. The proposed method achieves rejection rates closer to the nominal significance levels compared to the results of \citet{jun2009semiparametric}. These numerical observations provide supporting evidence for the theoretical results in the paper.

\begin{table}[H]
\small\centering\setstretch{1.3}
\caption{Comparison with Table 1 of \citet{jun2009semiparametric}}
\label{tab:comp tab1}
\begin{tabular*}{15cm}{@{\extracolsep{\fill}}ccccccc}
\hline\hline
\multirow{2}{*}{$\alpha$} & \multicolumn{2}{c}{Plug-in} & \multicolumn{3}{c}{\citet{jun2009semiparametric}} & Proposed Test \\
\cmidrule(lr){2-3} \cmidrule(lr){4-6} \cmidrule(lr){7-7}
& $\That_2(\thetahat_{\mathrm{2SLS}})$ & $\That_\mathrm{k}(\thetahat_{\mathrm{2SLS}})$ & $\That_1(\thetahat_{\mathrm{CUE1}})$ & $\That_2(\thetahat_{\mathrm{CUE2}})$ & $\That_\mathrm{k}(\thetahat_{\mathrm{CUEk}})$ & $\tau_n= n^{-1/4}$ \\
\hline
$0.01$  & $0.511$ & $0.480$ & $0.004$ & $0.012$ & $0.012$ & $0.011$ \\
$0.025$ & $0.533$ & $0.509$ & $0.007$ & $0.016$ & $0.022$ & $0.027$ \\
$0.05$  & $0.551$ & $0.531$ & $0.015$ & $0.025$ & $0.030$ & $0.051$ \\
$0.1$   & $0.584$ & $0.559$ & $0.027$ & $0.049$ & $0.045$ & $0.111$ \\
$0.2$   & $0.626$ & $0.602$ & $0.053$ & $0.078$ & $0.075$ & $0.241$ \\
\hline\hline
\end{tabular*}
\end{table}

\begin{table}[H]
\small\centering\setstretch{1.3}
\caption{Comparison with Table 2 of \citet{jun2009semiparametric}}
\label{tab:comp tab2}
\begin{tabular*}{15cm}{@{\extracolsep{\fill}}ccccccc}
\hline\hline
\multirow{2}{*}{$\alpha$} & \multicolumn{2}{c}{Plug-in} & \multicolumn{3}{c}{\citet{jun2009semiparametric}} &  Proposed Test \\
\cmidrule(lr){2-3} \cmidrule(lr){4-6} \cmidrule(lr){7-7}
& $\That_2(\thetahat_{\mathrm{SP}})$ & $\That_\mathrm{k}(\thetahat_{\mathrm{SP}})$ & $\That_1(\thetahat_{\mathrm{CUE1}})$ & $\That_2(\thetahat_{\mathrm{CUE2}})$ & $\That_\mathrm{k}(\thetahat_{\mathrm{CUEk}})$ & $\tau_n = n^{-1/4}$ \\
\hline
$0.01$  & $0.341$ & $0.362$ & $0.018$ & $0.018$ & $0.024$ & $0.009$ \\
$0.025$ & $0.360$ & $0.395$ & $0.027$ & $0.027$ & $0.030$ & $0.023$ \\
$0.05$  & $0.382$ & $0.419$ & $0.036$ & $0.036$ & $0.046$ & $0.049$ \\
$0.1$   & $0.422$ & $0.455$ & $0.052$ & $0.056$ & $0.067$ & $0.109$ \\
$0.2$   & $0.487$ & $0.519$ & $0.077$ & $0.093$ & $0.106$ & $0.233$ \\
\hline\hline
\end{tabular*}
\end{table}

\section{Conclusion}\label{sec:conclusion}
This paper provides a unified framework for inference on moment restriction models with nuisance parameters. We employ a new characterization that does not require the estimation of nuisance parameters, along with a numerical delta method to construct the test. The test is asymptotically size controlled and consistent. We conduct extensive Monte Carlo simulations to illustrate the finite sample properties of the proposed test. The numerical results show that the proposed method may achieve improvement in testing conditional moment restriction models with weak instruments. 
While the present paper focuses primarily on models with finite-dimensional nuisance parameters, the proposed approach might be extended to moment restrictions involving infinite-dimensional nuisance parameters that have been widely investigated in the literature. This framework may also be applied to other inference problems in the presence of nuisance parameters. We leave such extensions for future research.

\bibliographystyle{apalike}
\bibliography{references}	

\clearpage

\setcounter{page}{1}
\setcounter{equation}{0}
\setcounter{footnote}{0}
\renewcommand{\theequation}{\thesection.\arabic{equation}}
\begin{center}

\Large{\bf Unified Inference on Moment Restrictions with Nuisance Parameters}\\

\large{\bf Online Supplementary Appendix}\\
[0.75cm]

\large{	
Xingyu Li\qquad Xiaojun Song \qquad Zhenting Sun
}\\

\bigskip

\end{center}
\bigskip

The online supplementary appendix consists of five sections. Section \ref{appendix.auxiliary} provides auxiliary lemmas. 
Section \ref{sec:Analyses of Examples} verifies the assumptions for the examples in the main text. 
Section \ref{sec:multiple CDFs} extends the results for location-scale transformation to general parametric transformations on multiple CDFs. 
Section \ref{appendix.proofs of main results} contains the proofs of all main results. Section \ref{sec: additional simulations} provides additional simulation results. 

\begin{appendices}

\section{Auxiliary Results}\label{appendix.auxiliary}

\begin{lemma}\label{lemma:weak convergence with transformed index}
Let $\mathcal{H}=\lbr h_\xi :\xi\in \Xi \rbr$ be a class of real valued
functions indexed by $\Xi$. Assume that $\varphi, \varphi_1, \varphi_2,
\ldots$ are random elements taking values in $\ell^\infty(\mathcal{H})$.
For every $\xi\in\Xi$ and every $n\in\mathbb{Z}_+$, define
$\varrho(\xi)=\varphi\lp h_\xi \rp$ and $\varrho_n(\xi)=\varphi_n
\lp h_\xi \rp$.
If $\varphi_n \convd \varphi$ in $\ell^\infty(\mathcal{H})$ as $n\to\infty$,
then $\varrho_n\convd \varrho$ in $\ell^\infty(\Xi)$ as $n\to\infty$. Furthermore, if $\varphi$ is tight, then $\varrho$ is also tight.
\end{lemma}

\begin{pf}{ of Lemma \ref{lemma:weak convergence with transformed index}}
Define a map $\mathcal{I}: \ell^\infty(\mathcal{H})\to \ell^\infty(\Xi)$ such that $\mathcal{I}(\vartheta)(\xi)=\vartheta\lp h_\xi \rp$
for every $\vartheta\in \ell^\infty(\mathcal{H})$ and every $\xi\in\Xi$.
Then $\mathcal{I}$ is continuous on its domain. Indeed, for all
$\vartheta_1, \vartheta_2\in \ell^\infty(\mathcal{H})$, \begin{align*}
\lvv \mathcal{I}\lp \vartheta_1 \rp-\mathcal{I}\lp \vartheta_2 \rp
\rvv_\infty&= \sup_{\xi\in\Xi} \lv \mathcal{I}\lp \vartheta_1 \rp(\xi)
-\mathcal{I}\lp \vartheta_2 \rp(\xi) \rv 
= \sup_{\xi\in\Xi} \lv \vartheta_1\lp h_\xi \rp-
\vartheta_2 \lp h_\xi \rp \rv \\
&\le\sup_{h\in \mathcal{H}} \lv \vartheta_1(h)-\vartheta_2(h) \rv 
= \lvv \vartheta_1-\vartheta_2 \rvv_\infty.
\end{align*}
By Theorem 1.3.6 (continuous mapping) of \citet{van1996weak},
we have \begin{align*}
\varrho_n=\mathcal{I}\lp \varphi_n \rp \convd \mathcal{I} (\varphi)
=\varrho \text{ in } \ell^\infty(\Xi)
\end{align*}
as $n\to\infty$.

Since $\varphi$ is tight, for every $\varepsilon>0$, there exists a compact set $A\subset\ell^\infty(\mathcal{H})$ such that $\p(\varphi\in A)\ge 1-\varepsilon$. Define $\mathcal{I}(A)=\{\mathcal{I}(\varphi'):\varphi'\in A \}$. By the continuity of $\mathcal{I}$ and Theorem 2.34 of \citet{aliprantis2006infinite}, $\mathcal{I}(A)$ is compact in $\ell^\infty(\Xi)$. Moreover, \begin{align*}
\p(\varrho\in\mathcal{I}(A))=\p(\mathcal{I}(\varphi)\in\mathcal{I}(A)) \ge \p(\varphi\in A) \ge 1-\varepsilon,
\end{align*}
which implies the tightness of $\varrho$.
\end{pf}

The following lemma is an analog of Lemma \ref{lemma:weak convergence with transformed index}
for weak convergence conditional on the sample.

\begin{lemma}\label{lemma:conditional weak convergence of transformed index}
Let $\mathcal{H}=\lbr h_\xi :\xi\in \Xi \rbr$ be a class of real valued
functions indexed by $\Xi$. Assume that $\varphi$ is a tight random
element taking values in $\ell^\infty(\mathcal{H})$, and that for every $n\in\mathbb{Z}_+$,
$\mathcal{Z}_n$ is a random sample of size $n$ and $\varphi_n$ is
a random element taking values in $\ell^\infty(\mathcal{H})$.
For every $\xi\in\Xi$ and every $n\in\mathbb{Z}_+$, define
$\varrho(\xi)=\varphi\lp h_\xi \rp$ and $\varrho_n(\xi)=\varphi_n
\lp h_\xi \rp$. 
\begin{enumerate}[label=(\roman*),nosep]
\item If $\varphi_n \overset{\p}{\leadsto} \varphi $ as $n\to\infty$, then $\varrho_n \overset{\p}{\leadsto} \varrho $ as $n\to\infty$.
\item If $\varphi_n \overset{\text{a.s.}}{\leadsto} \varphi $ as $n\to\infty$, then $\varrho_n \overset{\text{a.s.}}{\leadsto} \varrho $ as $n\to\infty$.
\item If $\{\varphi_n\}$ is asymptotically measurable, then $\{\varrho_n\}$ is also asymptotically measurable. 
\end{enumerate}
\end{lemma}

\begin{pf}{ of Lemma \ref{lemma:conditional weak convergence of transformed index}}
Define a map $\mathcal{I}: \ell^\infty(\mathcal{H})\to \ell^\infty(\Xi)$ such that $\mathcal{I}(\vartheta)(\xi)=\vartheta\lp h_\xi \rp$
for every $\vartheta\in \ell^\infty(\mathcal{H})$ and every $\xi\in\Xi$.
As shown in the proof of Lemma \ref{lemma:weak convergence with transformed index},
for all $\vartheta_1, \vartheta_2\in \ell^\infty(\mathcal{H})$, \begin{align*}
\lvv \mathcal{I}\lp \vartheta_1 \rp-\mathcal{I}\lp \vartheta_2 \rp
\rvv_\infty	\le \lvv \vartheta_1-\vartheta_2 \rvv_\infty,
\end{align*}
which implies the Lipschitz continuity of $\mathcal{I}$. 
Results (i) and (ii) follow from Proposition 10.7(i) and 10.7(ii) of \citet{kosorok2008introduction}, respectively. The asymptotic measurability follows from the continuity of $\mathcal{I}$.
\end{pf}

\section{Analyses of Examples}
\label{sec:Analyses of Examples}

In this section, we study the sufficient conditions under which the examples discussed in the main text satisfy the assumptions for the test.

\begin{lemma}\label{lemma:examples satisfy model assumptions}
Examples \ref{exam:CMR.prob}--\ref{exam:LST.prob} satisfy Assumptions \ref{ass:para properties of g} and \ref{ass:para g mapsto f(g) is continuous} if the following conditions hold. \begin{enumerate}[label=(\roman*),nosep]
\item Example \ref{exam:CMR.prob}: The (marginal) distribution of $X$, denoted by $P_X$, has a Lebesgue probability density function $f$, and for every $\theta_0\in\Theta$ and every $\varepsilon>0$, there exists $\delta>0$ such that \begin{align*}
\E_P \Big[ \big(g(Y,\theta)-g(Y,\theta_0)\big)^2 \Big] < \varepsilon
\end{align*}
for all $\theta\in\Theta$ with $\lvv \theta-\theta_0 \rvv_2<\delta$.

\item Example \ref{exam:symmetry.prob}: The distribution function $G$ is continuous on $\mr$.
\item Example \ref{exam:fit.prob}: The distribution function $G$ is continuous on $\mr$, and the function $G_0$ is continuous in both arguments on $\mr\times\Theta$.
\item Example \ref{exam:LST.prob}: The distribution functions $F$ and $G$ are continuous on $\mr$.

\end{enumerate}
\end{lemma}

\begin{lemma}\label{lemma:examples are Donsker}

The function class $\Psi$ defined in Example \ref{exam:CMR.prob} is $P$-Donsker if the following conditions hold: (1) The parameter space $\Theta$ is compact in $\mr^{d_\theta}$ (Assumption \ref{ass:para Theta is compact}). (2) There exists a measurable function $m:\mr^{d_y}\to\mr_+$ with $\E_P[m(Y)^2]<\infty$ such that for all $y\in\mr^{d_y}$ and all $\theta_1, \theta_2\in \Theta$, \begin{align*}
\lv g(y,\theta_1)-g(y,\theta_2) \rv \le m(y) \lvv \theta_1-\theta_2 \rvv_2.
\end{align*}
(3) $\E_P [ \gbar(Y)^2 ]<\infty$, where $\gbar(y)=\sup_{\theta\in\Theta} |g(y,\theta)|$ for every $y\in\mr^{d_y}$.

Without further assumptions, the function classes $\Psi$ defined in Examples \ref{exam:symmetry.prob}--\ref{exam:LST.prob} are $P$-Donsker.
\end{lemma}

\begin{lemma}\label{lemma:examples satisfy second derivatives}
The functions $\phi_P$ in Examples \ref{exam:CMR.prob}--\ref{exam:LST.prob} satisfy Assumption \ref{ass:para G is 2nd differentiable} if the following conditions hold.
\begin{enumerate}[label=(\roman*),nosep]

\item Example \ref{exam:CMR.prob}: (1) Assumption \ref{ass:para Theta is compact} holds. (2) For all $\theta\in\Theta$, \begin{align*}
\E_P [|g(Y,\theta)|]<\infty \text{ and } \E_P\lbk \lvv \frac{\partial g(Y,\theta)}{\partial \theta} \rvv_2 \rbk <\infty.
\end{align*}
(3) The function $g$ is twice continuously differentiable with respect to its second argument $\theta$ at all $(y,\theta)\in \mr^{d_y}\times\Theta$. (4) The function  
\begin{align*}
\theta \mapsto \E_P \lbk \lvv \frac{\partial^2 g(Y,\theta)}{\partial \theta \partial \theta^\T} \rvv_2 \rbk
\end{align*}
is continuous on $\Theta$.
(5) The following two functions  
\begin{align*}
(x,\theta)\mapsto \E_P \lbk \frac{\partial g(Y,\theta)}{\partial \theta} \indicator\{X\le x\} \rbk \text{ and } (x,\theta)\mapsto \E_P \lbk \frac{\partial^2 g(Y,\theta)}{\partial \theta \partial \theta^\T} \indicator\{X\le x\} \rbk
\end{align*}
are continuous in $\theta$ at all $(x,\theta)\in\mr\times\Theta$.
(6) For every $\theta\in\Theta$ and every $x$, there is some $\delta>0$ such that 
\begin{align*}
&\E_P\lbk \int_{-\delta}^{\delta} \left\vert\frac{\partial g(Y,(\theta_{-j},\theta_j+\sigma))}{\partial \theta_j}\indicator\{X\le x\} \right\vert\mathrm{d}\sigma \rbk <\infty \text{ for all }j,\\
&\E_P\lbk \int_{-\delta}^{\delta} \left\vert\frac{\partial^2 g(Y,(\theta_{-k},\theta_k+\sigma))}{\partial \theta_j\partial \theta_k}\indicator\{X\le x\} \right\vert\mathrm{d}\sigma \rbk <\infty \text{ for all }j,k,
\end{align*}
where $\theta_{-j}=(\theta_1,\ldots,\theta_{j-1},\theta_{j+1},\ldots,\theta_{d_{\theta}})$ for all $j$.

\item Example \ref{exam:symmetry.prob}: The function $G$ has a bounded second order derivative, i.e., $\sup_{x\in\mr} | G''(x) |< \infty$. 
\item Example \ref{exam:fit.prob}: The function $G_0(x,\theta)$ is twice differentiable with respect to $\theta$, and \begin{align*}
\int_{\mr}	\sup_{\theta\in\Theta} \lvv  \left.	\frac{\partial^2 G_0 (z,\vartheta)}{\partial \vartheta \partial \vartheta^\T} \rv_{(z,\vartheta)=(x,\theta)} \rvv_{2}^2 \,\mathrm{d}\nu(x)	<\infty.
\end{align*}
\item Example \ref{exam:LST.prob}: (1) Assumption \ref{ass:para Theta is compact} holds and $\theta_2\ge\underline{\theta}_2$ for some $\underline{\theta}_2>0$. (2) The probability measure $\nu$ satisfies $\int_{\mr} x^4 \ddd \nu(x)<\infty$. (3) The function $G$ is twice differentiable with $\sup_{x\in\mr} | G'(x) |< \infty$ and  $\sup_{x\in\mr} | G''(x) |< \infty$.

\end{enumerate}
\end{lemma}

\begin{lemma}\label{lemma:examples satisfy identifiability}
Suppose Assumptions \ref{ass:para properties of g}--\ref{ass:para g mapsto f(g) is continuous} hold. If $\Theta_0=\varnothing$ (or equivalently, $\phi_P\notin\bD_0$), then Assumption \ref{ass:para strong identifiability} holds. For Examples \ref{exam:symmetry.prob}--\ref{exam:LST.prob}, if $\Theta_{0}\ne\varnothing$ (or equivalently, $\phi_P\in\bD_0$), then $\Theta_{0}=\Theta_{0}(\phi_P)$ is singleton, denoted by $\Theta_0=\{\theta_0\}$. In this case, \eqref{eq.identification para} holds for Examples \ref{exam:symmetry.prob}--\ref{exam:LST.prob} if there exist some $\kappa\in(0,1]$, some small $\varepsilonbar>0$, and some $C>0$ such that for all $\varepsilon\in(0,\varepsilonbar)$, \begin{align*}
\inf_{\theta\in\Theta: \lvv \theta-\theta_0 \rvv_2>\varepsilon} \sqrt{\int_\mr [G_0(x,\theta)-G_0(x,\theta_0)]^2 \ddd \nu(x)}\ge C \varepsilon^\kappa,
\end{align*}
where $G_0(x,\theta)=1-G(2\theta-x)$ for all $(x,\theta)\in\mr\times\Theta$ in Example \ref{exam:symmetry.prob},
and $G_0(x,\theta)=G((x-\theta_1)/\theta_2)$ for all $(x,\theta)\in \mr\times\Theta$ with $\theta=(\theta_1,\theta_2)$ in Example \ref{exam:LST.prob}.
\end{lemma}

\section{Transformations on Multiple CDFs}
\label{sec:multiple CDFs} 

Note that Example \ref{exam:LST.prob} (location-scale transformation) in the main text can be viewed as a special case of parametric transformation on two cumulative distribution functions (CDFs), for which the null hypothesis is \begin{align*}
\HH_{0}:\text{ For some } \theta\in\Theta, F(x)=G(g(x,\theta))\text{ for
all }x\in\mr,
\end{align*}
where $g:\mr\times\Theta\to\mr$ is a prespecified function. The problem of comparing two or multiple distributions has attracted considerable attention since the 1950s and remains a significant research topic. For example, \citet{chung2021permutation} consider testing within-group treatment effect heterogeneity.

In this section, we consider testing general parametric transformations on multiple cumulative distribution functions. These results may be generalized to other examples in Section \ref{sec:intro} with vector-valued $\psi_{x,\theta}$ under different conditions. Towards this end, let $F,G_{1},\ldots,G_{K}$ for some $K\geq 2$ be
unknown continuous CDFs on $\mr$. Let $\Theta_{k}\subset\mr^{d_{\theta_{k}}}$
for every $k\in\{1,\dots,K\}$ with $d_{\theta_{k}}\in\mathbb{Z}_{+}$. Let
$\Theta=\Theta_{1}\times\cdots\times\Theta_{K}$ equipped with a norm
$\Vert\cdot\Vert_{K2}$ such that for every $(\theta_{1},\ldots,\theta_{K}%
)\in\Theta$,
\[
\Vert(\theta_{1},\ldots,\theta_{K})\Vert_{K2}=\left(  \sum_{k=1}^{K}%
\Vert\theta_{k}\Vert_{2}^{2}\right)  ^{1/2}.
\]
For every $k\in\left\{  1,\ldots,K\right\}  $, let $g_{k}:\mathbb{R}\times
\Theta_{k}\rightarrow\mathbb{R}$ be some prespecified function. The null hypothesis of interest is
\begin{align}\label{eq:null multiple 1}
\HH_{0}:\text{ For some }(\theta_{1},\ldots,\theta_{K})\in\Theta
,F(x)=G_{1}(g_{1}(x,\theta_{1}))=\cdots=G_{K}(g_{K}(x,\theta_{K}))\text{ for
all }x\in\mr.
\end{align}
The parameter $(\theta_{1},\ldots,\theta_{K})$ in \eqref{eq:null multiple 1} is the nuisance parameter we need to take into account in the test. 

\begin{exam}[label=LSTK](Location-scale Transformations on Multiple CDFs)\label{exam:location-scale-multiple}
For every $k\in\{1,\ldots,K\}$, suppose that $Y_k$ is equivalent to $(X-\theta_{k1})/\theta_{k2}$ in distribution for some $\theta_{k1}\in\mathbb{R}$ and $\theta_{k2}\in\mathbb{R}_+$. Then the CDFs of $X$ and $Y_k$ satisfy
\begin{align*}
F(x)=\mathbb{P}(X\le x)=\mathbb{P}\left(\frac{X-\theta_{k1}}{\theta_{k2}}\le \frac{x-\theta_{k1}}{\theta_{k2}}\right) =\mathbb{P}\left(Y_k\le \frac{x-\theta_{k1}}{\theta_{k2}}\right)=G_k\left(\frac{x-\theta_{k1}}{\theta_{k2}}\right).
\end{align*}
For every $k\in\{1,\ldots,K\}$, let $\Theta_k=\lbk a_{k1}, b_{k1} \rbk\times \lbk a_{k2}, b_{k2} \rbk$, where $-\infty<a_{k1} < b_{k1} <\infty$ and $0<a_{k2}<b_{k2}<\infty$. Let $\Theta=\Theta_{1}\times\cdots\times\Theta_{K}$.
In this case, for every $k\in\{1,\ldots,K\}$, the parameter $\theta_k=\lp \theta_{k1},\theta_{k2} \rp\in\Theta_k$, and the function \begin{align*}
g_k(x,\theta_k)=\frac{x-\theta_{k1}}{\theta_{k2}} \text{ for all } x\in \mr \text{ and all } \theta_k\in\Theta_k.
\end{align*}
\end{exam}

Let $\nu$ be a probability measure on $\lp\mr,\borel_{\mr}\rp$. We now introduce the following assumptions for the transformations on multiple CDFs.

\begin{assum}\label{ass:properties of g multi}
For every $k\in\{1,\ldots,K\}$ and every $\theta_k\in\Theta_k$,
the function $x\mapsto g_k(x,\theta_k)$ is continuous and increasing.
\end{assum}

\begin{assum}\label{ass:nu and mu multi}
The probability measure $\nu$ on $\lp \mr, \borel_\mr \rp$ satisfies $\mu \ll \nu$, that is, if $\nu(B)=0$ for some $B\in\mathscr{B}_{\mathbb{R}}$, then $\mu(B)=0$.
\end{assum}

\begin{assum}\label{ass:Theta is compact multi}
The set $\Theta_k$ is compact in $\mathbb{R}^{d_{\theta_k}}$ for every $k\in\{1,\ldots,K\}$.
\end{assum}

\begin{assum}\label{ass:g mapsto f(g) is continuous multi}
For every $f\in  \mCb (\mr)$ and every $k$, the map $\theta_k\mapsto f(g_k(\cdot,\theta_k))$, from
$\Theta_k$ to $\lsv$, is continuous. That is, for an arbitrary fixed $\theta_{k0}\in\Theta_k$ and every $\varepsilon>0$,
there exists $\delta>0$ such that \begin{align*}
\int_\mr \lbk f\lp g_k(x,\theta_k) \rp-f\lp g_{k} (x,\theta_{k0}) \rp \rbk^2 \ddd \nu(x)<\varepsilon
\end{align*}
for all $\theta_k\in\Theta_k$ with $\lvv \theta_k-\theta_{k0} \rvv_{2} <\delta$.
\end{assum}

Assumptions \ref{ass:properties of g multi}%
--\ref{ass:g mapsto f(g) is continuous multi} are generalizations of Assumptions \ref{ass:para properties of g}--\ref{ass:para g mapsto f(g) is continuous}
in Section \ref{sec:Parametric Transformations} for transformations on multiple CDFs. For every $k\in\{1,\ldots,K\}$, define a function
space
\[
\bD_{\mL k}=\lbr\varphi_k\in\ell^{\infty}(\mr\times\Theta_{k}):\theta_{k}%
\mapsto\varphi_k(\cdot,\theta_{k}),\text{ as a map from }\Theta_{k}\text{ to
}\lsv,\text{ is continuous}\rbr.
\]
Then we define $\bD_{\mL0}=\prod_{k=1}^{K}\bD_{\mL k}$. For every
$k\in\{1,\ldots,K\}$ and every $f:\mr\rightarrow\mr$, we define a map $f\circ
g_{k}:\mr\times\Theta_{k}\rightarrow\mr$ such that $f\circ g_{k}(x,\theta
_{k})=f(g_{k}(x,\theta_{k}))$ for every $(x,\theta_{k})\in\mr\times\Theta_{k}%
$. Define a map $\phi_{k}:\mr\times\Theta_{k}\rightarrow\mr$ for every $k$
such that $\phi_{k}(x,\theta_{k})=F(x)-G_{k}\lp g_{k}(x,\theta_{k})\rp$ for
every $(x,\theta_{k})\in\mr\times\Theta_{k}$. Define $\phi:\mathbb{R}\times\Theta\to \mathbb{R}^K$ such that $\phi(x,\theta)=(\phi_{1}(x,\theta_1),\ldots,\phi_{K}(x,\theta_K))$ for every $(x,\theta)\in\mathbb{R}\times\Theta$, where $\theta=(\theta_1,\ldots,\theta_K)$ and $\theta_k\in\Theta_k$ for every $k$. The proposition below provides an
equivalent characterization of the null hypothesis in
\eqref{eq:null multiple 1}.

\begin{prop}\label{prop:equivalent null multi}
If Assumptions \ref{ass:properties of g multi}--\ref{ass:g mapsto f(g) is continuous multi} hold, then the null hypothesis in \eqref{eq:null multiple 1} is equivalent to
\begin{align}
\HH_0: \inf_{(\theta_1,\ldots,\theta_K)\in\Theta}\int_{\mr} \sum_{k=1}^K\lbk F(x)-G_k\lp g_k (x,\theta_k) \rp \rbk^2
\ddd \nu(x)=0.
\label{eq:working null multi}
\end{align}
\end{prop}

\subsection{Test Statistic}

\label{sec:Test Statistic multi}

Suppose that $\lbr X_{i} \rbr_{i=1}^{n_{x}}$ is a random sample drawn from $F$, and
$\lbr Y_{ki} \rbr_{i=1}^{n_{k}}$ is a random sample drawn from $G_{k}$ for every $k\in\{1,\ldots,K\}$.

\begin{assum}\label{ass:samples multi}
Each of the samples $\lbr X_i \rbr_{i=1}^{n_x}, \lbr Y_{1i} \rbr_{i=1}^{n_1}, \ldots, \lbr Y_{Ki} \rbr_{i=1}^{n_K}$ is independent
and identically distributed, and the samples $\lbr X_i \rbr_{i=1}^{n_x}, \lbr Y_{1i} \rbr_{i=1}^{n_1}, \ldots, \lbr Y_{Ki} \rbr_{i=1}^{n_K}$ are jointly independent.
\end{assum}

\begin{assum}\label{ass:ratio lambda multi}
The ratios ${n_x}/{n}\to \lambda_x\in (0,1)$ and ${n_k}/{n}\to \lambda_k\in (0,1)$ as $n \to\infty$ for every $k$, where $n=n_x+n_1+\cdots+n_K$.
\end{assum}
Assumption \ref{ass:samples multi} requires the multiple samples to be jointly
independent. In Assumption \ref{ass:ratio lambda multi}, $n_{x}$ and $n_{k}$
are viewed as functions of $n$. As $n\to\infty$, $n_{x}\to\infty$ and
$n_{k}\to\infty$ for every $k$.

Define a function space
\[
\bD_{\mL}=\bigg\{(\varphi_{1},\ldots,\varphi_{K})\in\prod_{k=1}^{K}%
\ell^{\infty}(\mathbb{R}\times\Theta_{k}):\int_{\mr}\sum_{k=1}^{K}%
\lbk\varphi_{k}(x,\theta_{k})\rbk^{2}\ddd\nu(x)<\infty\text{ for all }%
(\theta_{1},\ldots,\theta_{K})\in\Theta\bigg\}.
\]
Define a map $\mathcal{L}$ on $\bD_{\mL}$ such that $\mL(\varphi)=\inf
_{\theta\in\Theta}\int_{\mr}\sum_{k=1}^{K}\lbk\varphi_{k}(x,\theta
_{k})\rbk^{2}\ddd\nu(x)$ for every $\varphi\in\bD_{\mL}$ with $\varphi
=(\varphi_{1},\ldots,\varphi_{K})$ and $\theta=(\theta_{1},\ldots,\theta_{K}%
)$. Then under Assumptions \ref{ass:properties of g multi}%
--\ref{ass:g mapsto f(g) is continuous multi}, the null and the alternative
hypotheses can be expressed as
\[
\HH_{0}:\mL(\phi)=0\text{ and }\HH_{1}:\mL(\phi)>0.
\]
The CDFs $F$ and $G_{k}$ can be estimated by the empirical distribution
functions such that for every $x\in\mr$ and every $k$,
\[
\widehat{F}_{n_{x}}(x)=\frac{1}{n_{x}}\sum_{i=1}^{n_{x}}\indicator_{(-\infty
,x]}\lp X_{i}\rp\text{ and }\widehat{G}_{n_{k}}(x)=\frac{1}{n_{k}}\sum
_{i=1}^{n_{k}}\indicator_{(-\infty,x]}\lp Y_{ki}\rp.
\]
For every $x\in\mr$ and every $\theta\in\Theta$ with $\theta=(\theta
_{1},\ldots,\theta_{K})$, let
\[
\widehat{\phi}_{nk}(x,\theta_{k})=\widehat{F}_{n_{x}}(x)-\widehat{G}_{n_{k}%
}\lp g_{k}(x,\theta_{k})\rp\text{ and }\widehat{\phi}_{n}(x,\theta)=(\widehat{\phi}_{n1}(x,\theta_1),\ldots,\widehat{\phi}_{nK}(x,\theta_K)),
\]
and set the test statistic to be $T_{n}\mL(\phihat)$, where $T_{n}=n_{x}%
\cdot\prod_{k=1}^{K}(n_{k}/n)$.

\begin{lemma}
\label{prop:Donsker of phihat multi} Under Assumptions \ref{ass:samples multi}
and \ref{ass:ratio lambda multi}, we have
\[
\sqrt{T_{n}}(\phihat-\phi)\convd\bG_{0}\text{ in }\prod_{k=1}^{K}\ell^{\infty
}(\mr\times\Theta_{k})
\]
as $n\rightarrow\infty$, where $\mathbb{G}_{0}$ is a tight random element. If,
in addition, Assumption \ref{ass:g mapsto f(g) is continuous multi} holds,
then $\p\lp\bG_{0}\in\bD_{\mL0}\rp=1$.
\end{lemma}

Next, we show that the map $\mL$ is Hadamard directionally differentiable, but
its Hadamard directional derivative is also degenerate under $\HH_{0}$. Define
$\bD_{0}=\{ \varphi\in\bD_{\mL}:\mL(\varphi)=0 \}$.

\begin{lemma}
\label{lemma:multi-HDDL} If Assumptions
\ref{ass:Theta is compact multi} and
\ref{ass:g mapsto f(g) is continuous multi} hold, then $\mL$ is Hadamard
directionally differentiable at $\phi\in\bD_{\mL}$ tangentially to
$\bD_{\mL0}$ with the Hadamard directional derivative
\[
\mL_{\phi}^{\prime}(h)=2\inf_{\theta\in\Theta_{0}(\phi)}\int_{\mr}\sum
_{k=1}^{K}\phi_{k}(x,\theta_{k})h_{k}(x,\theta_{k})\ddd\nu(x)\text{ for all
}h\in\bD_{\mL0}\text{ with }h=(h_{1},\ldots,h_{K}),
\]
where $\Theta_{0}(\phi)=\argmin_{\theta\in\Theta}\int_{\mr}\sum_{k=1}^{K}\lbk\phi
_{k}(x,\theta_{k})\rbk^{2}\ddd\nu(x)$. Moreover, if $\phi\in\bD_{0}$, then the
derivative $\mL_{\phi}^{\prime}$ is well defined on the whole of $\prod
_{k=1}^{K}\ell^{\infty}(\mr\times\Theta_{k})$ with $\mL_{\phi}^{\prime}(h)=0$
for every $h\in\prod_{k=1}^{K}\ell^{\infty}(\mr\times\Theta_{k})$.
\end{lemma}

We now provide high level conditions for the existence of the second order Hadamard directional derivative of $\mL$. 

\begin{assum}\label{ass:G is 2nd differentiable multi}
For every $k\in\{1,\ldots,K\}$, the function $G_k\circ g_k$ is twice differentiable with respect to $\theta_k$, and the second partial derivative satisfies
\begin{align}\label{eq:second order derivative bound multi}
\int_{\mathbb{R}}	\sup_{\theta_k\in\Theta_k} \lvv \left.
\frac{\partial^2 ( G_k\circ g_k ) (z,\vartheta_k)}{\partial \vartheta_k \partial \vartheta_k^\T} 
\rv_{(z,\vartheta_k)=(x,\theta_k)} \rvv_{2}^2 \,\mathrm{d}\nu(x)
<\infty.
\end{align}	
\end{assum}

\begin{assum}\label{ass:strong identifiability multi}
The set $\Theta_0\equiv\{\theta\in\Theta:\int_\mr \sum_{k=1}^K [\phi_k(x,\theta_k)]^2\,\mathrm{d}\nu(x)=0\}\subset\mathrm{int}(\Theta)$,
and there exist some $\kappa\in(0,1]$ and some $C>0$ such 
that for all small $\varepsilon>0$, \begin{align*}
\inf_{\theta\in\Theta\setminus\Theta_0^\varepsilon}
\left\{\int_\mr \sum_{k=1}^K\lbk \phi_k(x,\theta_k) \rbk^2 \ddd \nu(x)\right\}^{1/2} \ge C \varepsilon^\kappa.
\end{align*}
\end{assum}
Assumptions \ref{ass:G is 2nd differentiable multi} and \ref{ass:strong identifiability multi} are generalized versions of Assumptions \ref{ass:para G is 2nd differentiable} and \ref{ass:para strong identifiability} for the transformations on multiple samples. 
We denote $\prod_{k=1}^K L^2(\nu)$ by $L^2_K(\nu)$. Define a norm $\Vert\cdot\Vert_{L^2_K(\nu)}$ on $L^2_K(\nu)$ such that for every $\psi\in L^2_K(\nu)$ with $\psi=(\psi_1,\ldots,\psi_K)$, 
\begin{align*}
\Vert\psi\Vert_{L^2_K(\nu)}=\left\{\sum_{k=1}^K\lvv \psi_k \rvv_\lsv^2 \right\}^{1/2}=\Vert (\Vert \psi_1\Vert_\lsv,\ldots, \Vert \psi_K\Vert_\lsv ) \Vert_2.
\end{align*}
For every $\theta\in\Theta$ with $\theta=(\theta_1,\ldots,\theta_K)$, define $\Phi_k'(\theta_k):\mr\to\mr^{d_{\theta_k}}$ such that 
\begin{align*}
\Phi_k'(\theta_k)(x)=-\left.
\frac{\partial ( G_k\circ g_k ) (z,\vartheta_k)}{\partial \vartheta_k} 
\rv_{(z,\vartheta_k)=(x,\theta_k)}
\quad \text{for every } x\in \mr.
\end{align*}
Let $\Phi^{\prime}(\theta,v)=(\Phi^{\prime}_1(\theta_1)^\T v_1,\ldots,\Phi^{\prime}_K(\theta_K)^\T v_K)$ for every $\theta=(\theta_1,\ldots,\theta_K)\in\Theta$ and every $v=(v_1,\ldots,v_K)\in\prod_{k=1}^K \mathbb{R}^{d_{\theta_k}}$.

\begin{lemma}\label{prop:second order Hadamard of L multi}
If Assumptions \ref{ass:Theta is compact multi}, \ref{ass:g mapsto f(g) is continuous multi},
\ref{ass:G is 2nd differentiable multi}, and \ref{ass:strong identifiability multi}
hold and $\phi\in\bD_0$, then the function $\mL$ is second order Hadamard directionally differentiable
at $\phi$ tangentially to $\bD_{\mL0}$ with the second order Hadamard
directional derivative \begin{align*}
\mL''_\phi(h)=\inf_{\theta\in\Theta_0(\phi)} \inf_{v\in \prod_{k=1}^K \mathbb{R}^{d_{\theta_k}}}
\lvv \Phi^{\prime}(\theta, v)+\sh (\theta) \rvv_{L_K^2(\nu)}^2 \text{ for all } h\in \bD_{\mL0} \text{ with }h=(h_1,\ldots,h_K),
\end{align*}	
where $\sh(\theta)(x)=(h_1(x,\theta_1),\ldots,h_K(x,\theta_K))$ for every
$(x,\theta)\in\mr\times\Theta$ with $\theta=(\theta_1,\ldots,\theta_K)$.
\end{lemma}

With Lemma \ref{prop:second order Hadamard of L multi}, the asymptotic distribution of the test statistic $T_n\mL( \phihat )$ under the null hypothesis is obtained by applying the second order delta method.

\begin{prop}\label{thry:asymptotic distribution of test stat multi}
If Assumptions \ref{ass:properties of g multi}--\ref{ass:strong identifiability multi}
hold and $\HH_0$ is true ($\phi\in\bD_0$), then 
\begin{align*}
T_n \mL(\phihat )	\convd \mL''_\phi\lp \bG_0 \rp \text{ as }n\to \infty.
\end{align*}
\end{prop}

\subsection{The Bootstrap}

We use the numerical second order Hadamard directional derivative $\mLhat$ proposed by \citet{hong2018numerical}
and \citet{chen2019inference} to approximate $\mL''_\phi$, which is defined as
\begin{align*}
\mLhat(h)=\frac{\mL( \phihat+\tau_n h )-\mL( \phihat )}{\tau_n^2}
\end{align*}
for all $h\in \prod_{k=1}^K\ell^\infty(\mr\times\Theta_k)$, where $\lbr \tau_n \rbr$ is a sequence of tuning parameters
satisfying the assumption below.

\begin{assum}\label{ass:rate of tau multi}
$\lbr \tau_n \rbr\subset\mr_+$ is a sequence of scalars 
such that $\tau_n\downarrow 0$
and $\tau_n \sqrt{T_n}\to\infty$ as $n\to\infty$.
\end{assum}

The next lemma establishes the consistency of $\mLhat$.

\begin{lemma}\label{prop:consistency of second derivative estimator multi}
If Assumptions \ref{ass:properties of g multi}--\ref{ass:rate of tau multi} hold
and $\HH_0$ is true ($\phi\in\bD_0$),
then for every sequence $\lbr h_n \rbr\subset \prod_{k=1}^K\ell^\infty
(\mr\times\Theta_k)$ and every $h\in\bD_{\mL0}$ such that $h_n\to h$
in $\prod_{k=1}^K\ell^\infty	(\mr\times\Theta_k)$ as $n\to\infty$, we have 
\begin{align*}
\mLhat\lp h_n \rp \convp \mL''_\phi (h) \text{ as }
n\to\infty.
\end{align*}
\end{lemma}

We approximate the distribution of $\bG_0$ via bootstrap.
Given the raw samples\linebreak $ \{ \lbr X_i \rbr_{i=1}^{n_x},
\lbr Y_{1i}\rbr_{i=1}^{n_1},\ldots,\lbr Y_{Ki}\rbr_{i=1}^{n_K} \} $, let the
bootstrap samples $ \{\lbr X_i^* \rbr_{i=1}^{n_x},\lbr Y_{1i}^*\rbr_{i=1}^{n_1},\ldots,\lbr Y_{Ki}^*\rbr_{i=1}^{n_K} \}$ be jointly independent, and drawn independently and identically from the empirical distributions
$\widehat{F}_{n_x}, \widehat{G}_{n_1},\ldots, \widehat{G}_{n_K}$, respectively.
Define for every $x\in\mr$ and every $k$,
\begin{align*}
\widehat{F}_{n_x}^*(x)=\frac{1}{n_x}\sum_{i=1}^{n_x} \indicator_{(-\infty,x]}\lp X^*_i \rp
\text{ and }
\widehat{G}_{n_k}^*(x)=\frac{1}{n_k}\sum_{i=1}^{n_k} \indicator_{(-\infty,x]}\lp Y^*_{ki} \rp.
\end{align*} 
For every $k$, let $\widehat{\phi}_{nk}^*(x,\theta_k)=\widehat{F}_{n_x}^* (x)-\widehat{G}_{n_k}^*\lp g_k (x,\theta_k) \rp$ for every $x\in \mr$ and every $\theta_k\in \Theta_k$. Let $\widehat{\phi}_{n}^{\ast}=(\widehat{\phi}_{n1}^{\ast},\ldots, \widehat{\phi}_{nK}^{\ast})$.

\begin{lemma}\label{prop:weak convergence of bootstrap multi}
If Assumptions \ref{ass:samples multi} and \ref{ass:ratio lambda multi} hold,
then \begin{align*}
&\sup_{\Gamma\in\mathrm{BL}_1\lp \prod_{k=1}^K\ell^\infty(\mr\times\Theta_k) \rp}
\lv \E \lbk \left. \Gamma \lp \sqrt{T_n} \lp \phihat^*-\phihat \rp \rp
\rv \lbr X_i \rbr_{i=1}^{n_x}, \lbr Y_{1i} \rbr_{i=1}^{n_1},\ldots,\lbr Y_{Ki} \rbr_{i=1}^{n_K} \rbk-
\E \lbk \Gamma \lp \bG_0 \rp \rbk \rv \\
&\convp 0, \text{ and } \sqrt{T_n}( \phihat^*-\phihat ) \text{ is asymptotically
measurable as } n\to \infty.
\end{align*}

\end{lemma}

The distribution
of $\mL''_\phi\lp \bG_0 \rp$ can be approximated by the conditional distribution of the bootstrap test statistic $\mLhat \{ 
\sqrt{T_n}( \phihat^*-\phihat ) \}$ given the raw samples. 

\begin{prop}\label{thry:consistency of approximation of test statistic distribution multi}
If Assumptions \ref{ass:properties of g multi}--\ref{ass:rate of tau multi} hold and
$\HH_0$ is true ($\phi\in\bD_0$),
then \begin{align*}
&\sup_{\Gamma\in\mathrm{BL}_1\lp \mr \rp}
\lv \E \lbk \left. \Gamma \lp \mLhat \lbk \sqrt{T_n} \lp \phihat^*
-\phihat \rp \rbk \rp \rv \lbr X_i \rbr_{i=1}^{n_x},
\lbr Y_{1i} \rbr_{i=1}^{n_1},\ldots,\lbr Y_{Ki} \rbr_{i=1}^{n_K} \rbk-\E \lbk \Gamma \lp \mL''_\phi \lp \bG_0 \rp
\rp \rbk \rv \\
&\convp 0 \text{ as }n\to\infty.
\end{align*}

\end{prop}

\subsection{Asymptotic Properties}

For a given level of significance $\alpha\in(0,1)$, define the bootstrap critical value 
\begin{align*}
\chat_{1-\alpha,n}=\inf\lbr c\in \mr: \p\lp\left. 
\mLhat \lbk \sqrt{T_n} \lp \phihat^*-\phihat \rp \rbk
\le c \rv \lbr X_i \rbr_{i=1}^{n_x},  \lbr Y_{1i} \rbr_{i=1}^{n_1},\ldots,\lbr Y_{Ki} \rbr_{i=1}^{n_K} \rp \ge 1-\alpha \rbr.
\end{align*}
We reject $\HH_0$ if and only if $T_n\mL ( \phihat )>\chat_{1-\alpha,n}$.
The next theorem shows that the proposed test is asymptotically size controlled and consistent.

\begin{thry}\label{thry:size and power multi}
Suppose that Assumptions \ref{ass:properties of g multi}--\ref{ass:rate of tau multi} hold.
\begin{enumerate}[label=(\roman*)]
\item If $\HH_0$ is true  and the CDF of $\mL''_\phi\lp \bG_0 \rp$
is strictly increasing and continuous at its $1-\alpha$ quantile, 
then \begin{align*}
\lim_{n\to\infty}
\p\lp T_n \mL( \phihat )>\chat_{1-\alpha,n} \rp = \alpha.
\end{align*}
\item If $\HH_0$ is false, then \begin{align*}
\lim_{n\to\infty}
\p\lp T_n \mL( \phihat )>\chat_{1-\alpha,n} \rp=1.
\end{align*}
\end{enumerate}
\end{thry}

The local power results for comparisons of multiple CDFs can be obtained analogously under settings similar to those in Section \ref{sec.local power}. 

\section{Proofs}\label{appendix.proofs of main results}

\subsection{Proofs for Section \ref{sec:Parametric Transformations}}

\begin{lemma}\label{lemma:para properties of DL}
If $\varphi_1,\varphi_2\in \bD_{\mL0}$, then $a_1\varphi_1+a_2\varphi_2\in
\bD_{\mL0}$ for all $a_1, a_2\in \mr$, and
the functions \begin{align*}
\theta \mapsto \int_{\mr} \lbk \varphi_1(x,\theta) \rbk^2 \ddd \nu(x) 
\text{ and } 
\theta \mapsto \int_{\mr} \varphi_1(x,\theta)\varphi_2(x,\theta) \ddd \nu(x)
\end{align*}
are continuous on $\Theta$.
\end{lemma}

\begin{pf}{ of Lemma \ref{lemma:para properties of DL}}
For all $\varphi_1,\varphi_2\in \bD_{\mL0}$ and
all $a_1,a_2\in \mr$, let $M=\lvv \varphi_1 \rvv_{\infty}
\vee \lvv \varphi_2 \rvv_{\infty} \vee 2a_1^2 \vee 2a_2^2$.
By the definition of $\bD_{\mL0}$, for every $\theta_0\in\Theta$ and
every $\varepsilon>0$, there exists $\delta \lp \theta_0,\varepsilon \rp>0$
such that 
\begin{align*}
\int_\mr \lbk \varphi_1\lp x,\theta \rp-\varphi_1\lp x,\theta_0 \rp \rbk^2 \ddd \nu(x) 
\vee
\int_\mr \lbk \varphi_2\lp x,\theta \rp-\varphi_2\lp x,\theta_0 \rp \rbk^2 \ddd \nu(x)
< \frac{\varepsilon}{2M} \wedge \lbk \frac{\varepsilon}{2M} \rbk^2
\end{align*}
whenever $\lvv \theta-\theta_0 \rvv_2 <\delta\lp \theta_0 ,\varepsilon \rp$.

To show the first claim, note that  \begin{align*}
&\phantom{=\:\:} \int_\mr \lbk a_1 \varphi_1 (x,\theta)+a_2 \varphi_2(x,\theta)
-a_1\varphi_1\lp x,\theta_0 \rp-a_2 \varphi_2 \lp x,\theta_0 \rp \rbk^2 \ddd \nu(x) \\
& \le 2 a_1^2 \int_\mr \lbk \varphi_1(x,\theta)-\varphi_1\lp x,\theta_0 \rp \rbk^2
\ddd \nu(x)+ 2 a_2^2  \int_\mr \lbk \varphi_2(x,\theta)-\varphi_2
\lp x,\theta_0 \rp \rbk^2 \ddd \nu(x) < \frac{\varepsilon}{2}+\frac{\varepsilon}{2}=\varepsilon
\end{align*}
whenever $\lvv \theta-\theta_0 \rvv_2<\delta\lp \theta_0, \varepsilon \rp$.
For the second claim, we have
\begin{align*}
&\phantom{=\:\:} \lv \int_\mr \lbk \varphi_1 (x,\theta) \rbk^2 \ddd \nu(x)
-\int_\mr \lbk \varphi_1 \lp x,\theta_0 \rp \rbk^2 \ddd \nu(x) \rv \\
& \le \int_\mr \lv \lbk \varphi_1(x,\theta)+\varphi_1 \lp x,\theta_0 \rp \rbk
\lbk \varphi_1(x,\theta)-\varphi_1\lp x,\theta_0 \rp \rbk \rv \ddd \nu(x) \\
& \le 2M \int_\mr \lv \varphi_1(x,\theta)-\varphi_1 \lp x,\theta_0 \rp \rv \ddd\nu(x) \le 2M \sqrt{\int_\mr \lbk \varphi_1(x,\theta)-\varphi_1\lp x,\theta_0 \rp \rbk^2
\ddd\nu(x)} <\varepsilon
\end{align*}		
whenever $\lvv \theta-\theta_0 \rvv_2<\delta\lp \theta_0, \varepsilon \rp$,
where the third inequality follows from the convexity of square functions
and Jensen's inequality.
The third claim can be proved analogously, since 
\begin{align*}
&\phantom{=\:\:} \lv \int_\mr \varphi_1(x,\theta)\varphi_2(x,\theta) \ddd \nu(x)
-\int_\mr \varphi_1\lp x,\theta_0 \rp \varphi_2 \lp x,\theta_0 \rp  \ddd \nu(x) \rv \\
&\le \int_\mr \lv \varphi_1(x,\theta) \lbk \varphi_2(x,\theta)-\varphi_2 \lp x,\theta_0 \rp \rbk
+\varphi_2\lp x,\theta_0 \rp \lbk \varphi_1(x,\theta)-\varphi_1\lp x,\theta_0 \rp \rbk \rv
\ddd \nu(x)\\
& \le M \int_\mr \lv \varphi_1 (x,\theta)-\varphi_1\lp x,\theta_0 \rp \rv \ddd \nu(x)
+M \int_\mr \lv \varphi_2 (x,\theta)-\varphi_2\lp x,\theta_0 \rp \rv \ddd \nu(x) \\
& \le M \sqrt{\int_\mr \lbk \varphi_1 (x,\theta)-\varphi_1\lp x,\theta_0 \rp \rbk^2 \ddd \nu(x)}
+M \sqrt{\int_\mr \lbk \varphi_2 (x,\theta)-\varphi_2\lp x,\theta_0 \rp \rbk^2 \ddd \nu(x)} <\varepsilon	
\end{align*}
whenever $\lvv \theta-\theta_0 \rvv_2<\delta\lp \theta_0, \varepsilon \rp$,
where the third inequality follows from the convexity of square functions
and Jensen's inequality.	
\end{pf}

\begin{pf}{ of Proposition \ref{prop:para equivalent null}}
If $\phi_P(x,\theta)=0$ for all $x\in\mr$ with some
$\theta\in \Theta$, then \eqref{eq:para working null} holds trivially.

Next, we show that \eqref{eq:para working null} implies \eqref{eq:para original null}.
Recall that $\mu$ is the Lebesgue measure on $\lp \mr, \borel(\mr) \rp$.
As shown in the main text above Proposition \ref{prop:para equivalent null}, Assumption \ref{ass:para g mapsto f(g) is continuous} implies that $\phi_P\in \bD_{\mL0}$.
Also, by Lemma \ref{lemma:para properties of DL}, the function $\theta
\mapsto \int_\mr [\phi_P(x,\theta)]^2 \ddd \nu(x)$
is continuous on $\Theta$. By Assumption \ref{ass:para Theta is compact},
there exists $\theta_0\in\Theta$ such that 
\begin{align}\label{eq.para integral nu equals 0}
\int_{\mr} [\phi_P(x,\theta_0)]^2 \ddd \nu(x)=\inf_{\theta\in \Theta}\int_{\mr} [\phi_P(x,\theta)]^2
\ddd \nu(x)=0.
\end{align}
Define $A=\lbr x\in \mr: \phi_P(x,\theta_{0}) \ne 0\rbr$. Then \eqref{eq.para integral nu equals 0} implies that $\nu(A)=0$ by Proposition 2.16 of \citet{folland2013real}. By the assumption that $\mu \ll \nu$, $\mu (A)=0$.
We now claim that $A=\varnothing$. Otherwise, there is an $x_0\in \mr$ such that
$\phi_P(x_0,\theta_{0})\ne 0$. Since $\phi_P(x,\theta_0)$ is continuous in $x$ by Assumption \ref{ass:para properties of g}, there exists $\delta>0$ such that 
$\phi_P(x,\theta_0)\ne 0$ for all $x\in \lbk x_0,	x_0+\delta \rbk$. This contradicts $\mu(A)=0$.
Thus, we have $\phi_P(x,\theta_0)=0$ for all $x\in \mr$.
\end{pf}

\begin{pf}{ of Lemma \ref{prop:para Donsker of phihat}}
Note that $\phihat(x,\theta)=\Phat(\psi_{x,\theta})$ and $\phi_P(x,\theta)=P(\psi_{x,\theta})$ for every $n\in\mathbb{Z}_+$ and every $(x,\theta)\in\mr\times\Theta$. For every $n\in\mathbb{Z}_+$, Assumption \ref{ass:para Donsker class} implies that $(\phihat-\phi_P)\in\ell^\infty(\mr\times\Theta)$. As a $P$-Donsker is necessarily $P$-Glivenko--Cantelli almost surely \citep[p.~82]{van1996weak}, we have \begin{align*}
\sup_{(x,\theta)\in\mr\times \Theta} \lv \phihat(x,\theta)-\phi_P(x,\theta) \rv= \sup_{(x,\theta)\in\mr\times \Theta} \lv \Phat(\psi_{x,\theta})-P(\psi_{x,\theta}) \rv =\sup_{f\in\Psi} \lv \Phat(f)-P(f) \rv \convas 0
\end{align*}
as $n\to\infty$. By Theorem 1.9.2(i) of \citet{van1996weak}, the above result implies convergence in probability. By Assumption \ref{ass:para Donsker class}, the tightness of $P$-Brownian bridges, and Lemma \ref{lemma:weak convergence with transformed index}, we have $\sqrt{n}(\phihat-\phi_P)\convd \bG_0$ in $\ell^\infty(\mr\times\Theta)$, where $\bG_0$ is tight and $\bG_0(x,\theta)=\bW(\psi_{x,\theta})$ for every $(x,\theta)\in\mr\times\Theta$.

Now we show $\p(\bG_0\in\bD_{\mL 0})=1$. Since the $P$-Brownian bridge $\bW$ is a Gaussian process indexed by $\Psi$, for all $(x_1,\theta_1),\ldots, (x_k,\theta_k)\in\mr\times\Theta$, we have
\begin{align*}
\Big( \bG_0(x_1,\theta_1), \ldots, \bG_0(x_k,\theta_k) \Big)=\Big(\bW(\psi_{x_1,\theta_1}), \ldots, \bW(\psi_{x_k,\theta_k}) \Big),
\end{align*}
which follows a $k$-variate Gaussian distribution. Hence $\bG_0$ is a Gaussian process indexed by $\mr\times\Theta$. Define an intrinsic semi-metric $\rho_2$ on $\mr\times\Theta$ such that for all $(x_1,\theta_1), (x_2,\theta_2)\in\mr\times\Theta$, \begin{align*}
\lbk \rho_2\big((x_1,\theta_1),(x_2,\theta_2)\big) \rbk^2&=\E_P\lbk \lv \bG_0(x_1,\theta_1)-\bG_0(x_2,\theta_2) \rv^2 \rbk=\E_P\lbk \lv \bW(\psi_{x_1,\theta_1})-\bW(\psi_{x_2,\theta_2}) \rv^2 \rbk\\
=&\,\E_P\lbk \bW^2(\psi_{x_1,\theta_1}) \rbk+\E_P\lbk \bW^2(\psi_{x_2,\theta_2}) \rbk-2\E_P[\bW(\psi_{x_1,\theta_1})\bW(\psi_{x_2,\theta_2})] \\
=&\,P\lp \psi_{x_1,\theta_1}^2 \rp-[P(\psi_{x_1,\theta_1})]^2+P\lp \psi_{x_2,\theta_2}^2 \rp-[P(\psi_{x_2,\theta_2})]^2-2P(\psi_{x_1,\theta_1}\psi_{x_2,\theta_2})\\
&+2P(\psi_{x_1,\theta_1})P(\psi_{x_2,\theta_2}) \\
=&\,P\lbk \lp \psi_{x_1,\theta_1}-\psi_{x_2,\theta_2} \rp^2 \rbk-\lbk P(\psi_{x_1,\theta_1})-P(\psi_{x_2,\theta_2}) \rbk^2.
\end{align*}
Since $\bG_{0}$ is a tight Gaussian process in $\ell^\infty(\mr\times\Theta)$, the discussion of \citet[p.~41]{van1996weak} implies that there exists $\Omega_0\subset \Omega$ with $\p(\Omega_0)=1$ such that for all $\omega\in\Omega_0$, the path $(x,\theta)\mapsto \bG_{0}(\omega)(x,\theta)$ is uniformly $\rho_2$-continuous. That is, for every $\varepsilon>0$, there exists $\delta_1>0$ such that for all $(x_1,\theta_1), (x_2,\theta_2)\in\mr\times\Theta$ with $\rho_2((x_1,\theta_1),(x_2,\theta_2))<\delta_1$, we have $\lv \bG_{0}(\omega)(x_1,\theta_1)-\bG_{0}(\omega)(x_2,\theta_2) \rv <\varepsilon$. By Assumption \ref{ass:para g mapsto f(g) is continuous}, for every $\theta_{0}\in\Theta$, there exists $\delta_2>0$ such that for all $\theta\in\Theta$ with $\lvv \theta-\theta_0 \rvv_2<\delta_2$, we have for all $x\in\mr$, \begin{align*}
&\rho_2\big((x,\theta),(x,\theta_0)\big)=\sqrt{P\lbk \lp \psi_{x,\theta}-\psi_{x,\theta_0} \rp^2 \rbk-\lbk P(\psi_{x,\theta})-P(\psi_{x,\theta_0}) \rbk^2} \\
\le&\, \sqrt{ P\lbk \lp \psi_{x,\theta}-\psi_{x,\theta_0} \rp^2 \rbk} \le \sqrt{ \sup_{x'\in\mr} P\lbk \lp \psi_{x',\theta}-\psi_{x',\theta_0} \rp^2 \rbk} < \delta_1,
\end{align*}
and thus \begin{align*}
\int_{\mr} \lbk \bG_{0}(\omega)(x,\theta)-\bG_{0}(\omega)(x,\theta_0)\rbk^2 \ddd \nu(x)<\varepsilon^2.
\end{align*}
This implies $\bG_{0}(\omega)\in\bD_{\mL 0}$ and $\p(\bG_{0}\in\bD_{\mL 0})=1$.
\end{pf}

We introduce the Hadamard directional differentiability following Definition A.1(ii) of \citet{chen2019inference}, which is equivalent to Condition (2.10) of \citet{shapiro2000statistical}.

\begin{definition}\label{def:first order Hadamard}
Let $\bH$ and $\bK$ be normed spaces equipped with norms
$\lvv \cdot\rvv_\bH$ and $\lvv \cdot \rvv_\bK$, respectively,
and $\mathcal{F}: \bH_\mathcal{F} \subset \bH \to \bK$.
The map $\mathcal{F}$ is said to be Hadamard directionally differentiable at $\phi\in\bH_\mathcal{F}$ tangentially to a set $\bH_0\subset \bH$, if there is a continuous and positively homogeneous map of degree one $\mathcal{F}'_\phi:\bH_0\to\bK$ such that
\begin{align*}
\lim_{n\to\infty} \lvv \frac{\mathcal{F}\lp \phi+t_n h_n \rp
-\mathcal{F}\lp \phi \rp}{t_n}-\mathcal{F}'_\phi(h) \rvv_\bK=0
\end{align*}
holds for all sequences $\lbr h_n \rbr\subset\bH$ and
$\lbr t_n \rbr\subset \mr_+$ such that $t_n \downarrow 0$,
$h_n\to h\in\bH_0$ as $n\to\infty$, and $\phi+t_n h_n \in\bH_\mathcal{F}$
for all $n$.
\end{definition}

\begin{pf}{ of Lemma \ref{lemma:para-HDDL}}
Define a map $\mathcal{S}:\bD_{\mL}\to \ell^{\infty}(\Theta)$ such that for every $\varphi\in \bD_{\mL}$
and every $\theta\in\Theta$, \begin{align*}
\mathcal{S}(\varphi)(\theta)=\int_\mr \lbk \varphi(x,\theta) \rbk^2 \ddd \nu(x).
\end{align*}
We show that the Hadamard directional derivative of $\mathcal{S}$ at
$\phi_P\in \bD_{\mL}$ is 
\begin{align*}
\mathcal{S}'_{\phi_P} (h)(\theta)=\int_\mr 2\phi_P(x,\theta)h(x,\theta)\ddd \nu(x)\text{ for all } h\in \bD_{\mL0}.
\end{align*}
By Assumption \ref{ass:para g mapsto f(g) is continuous} and Lemma \ref{lemma:para properties of DL}, $\mathcal{S}(\phi_P)\in \mathcal{C}(\Theta)$. Indeed, for all sequences $\lbr h_n \rbr_{n=1}^\infty \subset \ell^\infty(\mr\times\Theta)$
and $\lbr t_n \rbr_{n=1}^\infty \subset \mr_+$ such that $t_n \downarrow 0$,
$h_n\to h\in \bD_{\mL0}$ as $n\to \infty$, and $\phi_P+t_n h_n \in \bD_{\mL}$ for all $n$,
we have that $M=\sup_{n\in\mathbb{Z}_+} \lvv h_n \rvv_\infty<\infty$, and
\begin{align*}
&\phantom{=\:\,}	\sup_{\theta\in\Theta} \lv \frac{\mS\lp \phi_P+t_n h_n \rp(\theta)
-\mS (\phi_P)(\theta)}{t_n}-\mS'_{\phi_P} (h)(\theta) \rv \\
&= \sup_{\theta\in\Theta} \lv \int_\mr t_n h_n^2(x,\theta)
+2\phi_P(x,\theta)\lbk h_n(x,\theta)-h(x,\theta) \rbk
\ddd \nu(x) \rv \\
& \le \int_\mr  t_n M^2 + 2 \lvv \phi_P \rvv_{\infty} \lvv h_n- h \rvv_{\infty} \ddd \nu(x) = t_n M^2 +2 \lvv \phi_P \rvv_{\infty} \lvv h_n- h \rvv_{\infty} \to 0,
\end{align*}
since $t_n\downarrow 0$
and $h_n\to h$ in $\ell^\infty(\mr\times\Theta)$ as $n\to \infty$.

Define a function $\mR$ such that for every $\psi\in \ell^{\infty}(\Theta)$,
$\mR(\psi)=\inf_{\theta\in \Theta} \psi(\theta)$. By Lemma S.4.9
of \citet{fang2019inference}, $\mR$ is Hadamard directionally differentiable
at every $\psi\in \mC(\Theta)$ tangentially to $\mC(\Theta)$ with the Hadamard
directional derivative \begin{align*}
\mR'_\psi (f)=\inf_{\theta\in \Theta^*_0(\psi)} f(\theta) \text{ for all } f\in \mC(\Theta),
\end{align*}
where $ \Theta^*_0(\psi)=\argmin_{\theta\in\Theta} \psi(\theta)$.

Note that $\mL(\varphi)=\mR \lbk \mS (\varphi) \rbk=\mR \circ \mS (\varphi)$ for
every $\varphi \in \bD_{\mL}$.
By Proposition 3.6(i) of \citet{shapiro1990concepts}, $\mL$ is Hadamard
directionally differentiable at $\phi_P$ tangentially to
$\bD_{\mL0}$ with the Hadamard directional derivative \begin{align*}
\mL'_{\phi_P} (h)= \mR'_{\mS(\phi_P)} \lbk \mS'_{\phi_P}(h) \rbk
=\inf_{\theta\in\Theta^*_0 (\mS(\phi_P))} \int_{\mr} 2 \phi_P(x,\theta) h(x,\theta)
\ddd \nu(x) \text{ for all } h\in \bD_{\mL0}.
\end{align*}
Since $ \Theta^*_0 (\mS(\phi_P))=\argmin_{\theta\in\Theta}\int_{\mr} \lbk \phi_P (x,\theta) \rbk^2
\ddd \nu(x)$, the desired result follows.

Now we turn to the degeneracy of $\mL'_{\phi_P}$ under the condition that $\phi_P\in\bD_0$.
If $\phi_P \in \bD_0$, for every $\theta\in \Theta_0(\phi_P)$,
we have \begin{align*}
\int_\mr \lbk \phi_P(x,\theta) \rbk^2 \ddd \nu(x)=0,
\end{align*}
and consequently $\phi_P(x,\theta)=0$ holds for $\nu$-almost every $x$.
Therefore, $\mL'_{\phi_P}(h)=0$ for every $h\in \ell^\infty(\mr\times\Theta)$
whenever $\phi_P\in \bD_0$.
\end{pf}

For the second order Hadamard directional differentiability,
we introduce Definition A.2(ii) of \citet{chen2019inference},
which is equivalent to Condition (2.14)
of \citet{shapiro2000statistical} (with a difference by a factor of $1/2$ in the derivative).

\begin{definition}\label{def:second order Hadamard CF}
Let $\bH$ and $\bK$ be normed spaces equipped with norms
$\lvv \cdot\rvv_\bH$ and $\lvv \cdot \rvv_\bK$, respectively,
and $\mathcal{F}: \bH_\mathcal{F} \subset \bH \to \bK$.
Suppose that $\mathcal{F}:\bH_\mathcal{F}\to\bK$ is Hadamard directionally differentiable
tangentially to $\bH_0\subset\bH$ such that the derivative
$\mathcal{F}'_\phi:\bH_0\to\bK$ is well defined on $\bH$.
We say that $\mathcal{F}$ is second order Hadamard directionally differentiable at
$\phi\in\bH_\mathcal{F}$ tangentially to $\bH_0$ if there is a continuous and positively homogeneous map of degree two
$\mathcal{F}^{\prime\prime}_\phi:\bH_0\to\bK$ such that \begin{align*}
\lim_{n\to\infty}\lvv \frac{\mathcal{F}\lp \phi+t_n h_n \rp-\mathcal{F}
\lp\phi \rp-t_n\mathcal{F}'_\phi\lp h_n \rp}{t_n^2}
-\mathcal{F}''_\phi(h) \rvv_\bK=0
\end{align*}
holds for all sequences $\lbr h_n \rbr \subset \bH$
and $\lbr t_n \rbr \subset \mr_+$ such that $t_n \downarrow 0$,
$h_n\to h\in\bH_0$ as $n\to \infty$, and $\phi+t_n h_n\in\bH_\mathcal{F}$
for all $n$.
\end{definition}

\begin{pf}{ of Lemma \ref{prop:para second order Hadamard of L}}
The proof closely follows that of Lemma E.3 in \citet{chen2019inference}.
Define $\Phi:\Theta\to \lsv$ such that
$\Phi(\theta)(x)=\phi_P(x,\theta)$ for every $(x,\theta)\in\mr\times\Theta$.
Then it is easy to show that under the assumptions,
\begin{align*}
\mL(\phi_P)=\inf_{\theta\in\Theta}\int_\mr \lbk \phi_P(x,\theta) \rbk^2 \ddd \nu(x)
=\inf_{\theta\in\Theta} \lvv \Phi(\theta) \rvv^2_\lsv=0,
\end{align*}
and $ \Theta_0(\phi_P)=\{ \theta\in\Theta: \lvv \Phi(\theta) \rvv_\lsv=0 \}=\Theta_0$.
Consider all sequences $\lbr t_n \rbr_{n=1}^\infty\subset\mr_+$
and $\lbr h_n \rbr_{n=1}^\infty\subset \ell^\infty(\mr\times \Theta)$
such that $t_n\downarrow 0$, $h_n \to h\in \bD_{\mL0}$ in $\ell^\infty(\mr\times \Theta)$
as $n\to\infty$ with $h\neq 0$ (the case where $h= 0$ is trivial), and $\phi_P+t_n h_n \in \bD_{\mL}$ for all $n$.
For notational simplicity, define $\sh_n:\Theta\to\lsv$ for every $n\in\mathbb{Z}_+$
such that $\sh_n(\theta)(x)=h_n (x,\theta)$ for every $(x,\theta)\in\mr\times\Theta$,
and define $\sh:\Theta\to\lsv$ such that $\sh(\theta)(x)=h(x,\theta)$ for every
$(x,\theta)\in\mr\times\Theta$.

Since $h_n\to h\in \bD_{\mL0}$ in $\ell^\infty(\mr\times\Theta)$, it follows that $\lvv h \rvv_{\infty}\vee \sup_{n\in\mathbb{Z}_+} \lvv h_n \rvv_{\infty}
=M_1$ for some $M_1<\infty$. Then we have that
\begin{align*}
&  \lv \mL\lp \phi_P+t_n h_n \rp-\mL \lp\phi_P+t_n h \rp \rv =\lv \inf_{\theta\in\Theta} \lvv \Phi(\theta)+t_n \sh_n (\theta) \rvv_\lsv^2
-\inf_{\theta\in\Theta} \lvv \Phi(\theta)+t_n \sh (\theta) \rvv_\lsv^2 \rv \\
=&\,\lv \inf_{\theta\in\Theta} \lvv \Phi(\theta)+t_n \sh_n (\theta) \rvv_\lsv
+\inf_{\theta\in\Theta} \lvv \Phi(\theta)+t_n \sh (\theta) \rvv_\lsv \rv \\
& \cdot \lv \inf_{\theta\in\Theta} \lvv \Phi(\theta)+t_n \sh_n (\theta) \rvv_\lsv
-\inf_{\theta\in\Theta} \lvv \Phi(\theta)+t_n \sh (\theta) \rvv_\lsv \rv \\
\le&\, \lv \inf_{\theta\in\Theta_0(\phi_P)} \lvv \Phi(\theta)+t_n \sh_n (\theta) \rvv_\lsv
+\inf_{\theta\in\Theta_0(\phi_P)} \lvv \Phi(\theta)+t_n \sh (\theta) \rvv_\lsv \rv \\
& \cdot \lp t_n \sup_{\theta\in\Theta} \lvv \sh_n(\theta)-\sh(\theta) \rvv_\lsv \rp \\
\le&\, 2M_1 t_n^2 \lvv h_n-h \rvv_\infty = o\lp t_n^2 \rp,
\end{align*}
where the first inequality follows from the Lipschitz continuity of the supremum
map and the triangle inequality, and the second inequality follows from
the fact that $\Phi\lp \theta \rp=0$ $\nu$-almost everywhere for every $\theta\in\Theta_0(\phi_P)$.

Then for the $h$, let $a(h)>0$ be such that $C a(h)^\kappa = 3\lvv h \rvv_{\infty}$, where
$C$ and $\kappa$ are defined as in Assumption \ref{ass:para strong identifiability}. 
For sufficiently large $n\in\mathbb{Z}_+$ such that $t_n^\kappa \ge t_n$ and $a(h)t_n<\varepsilonbar$, we have that 
\begin{align}\label{eq.para second order 1}
& \inf_{\theta\in\Theta\setminus\Theta_0(\phi_P)^{a(h)t_n}}
\lvv \Phi(\theta)+t_n \sh(\theta) \rvv_\lsv \notag\\
\ge&\, \inf_{\theta\in\Theta\setminus\Theta_0(\phi_P)^{a(h)t_n}} \lvv \Phi(\theta)\rvv_\lsv
+\inf_{\theta\in\Theta\setminus\Theta_0(\phi_P)^{a(h)t_n}} \lbk -t_n \lvv \sh(\theta) 
\rvv_\lsv \rbk \notag\\
=&\, \inf_{\theta\in\Theta\setminus\Theta_0(\phi_P)^{a(h)t_n}} \lvv \Phi(\theta)\rvv_\lsv
-\sup_{\theta\in\Theta\setminus\Theta_0(\phi_P)^{a(h)t_n}} t_n \lvv \sh(\theta) 
\rvv_\lsv \notag \\
\ge&\, C \lp a(h) t_n \rp^\kappa -t_n \sup_{\theta\in\Theta\setminus\Theta_0(\phi_P)^{a(h)t_n}}
\lvv \sh(\theta) \rvv_\lsv \ge 3\lvv h \rvv_{\infty} t_n^\kappa- t_n \lvv h \rvv_{\infty} \notag\\
>&\, t_n \inf_{\theta\in\Theta_0(\phi_P)} \lvv \sh(\theta) \rvv_\lsv  = \inf_{\theta\in\Theta_0(\phi_P)} \lvv \Phi(\theta)+t_n \sh(\theta) \rvv_\lsv \ge \sqrt{\mL\lp \phi_P+t_n h \rp},
\end{align}
where the second inequality follows from Assumption \ref{ass:para strong identifiability}.

By Lemma \ref{lemma:para properties of DL}
and the fact that $\phi_P\in\bD_{\mL0}$ and $h\in\bD_{\mL0}$, the map
$\theta\mapsto \lvv \Phi(\theta)+t_n \sh(\theta) \rvv_\lsv^2$ is continuous
at every $\theta\in\Theta$ for every $n\in\mathbb{Z}_+$.
Since $\Theta$ and $\Theta_0(\phi_P)^{a(h)t_n}$
are compact sets in $\mr^{d_{\theta}}$, it follows that \begin{align*}
&\mL( \phi_P+t_n h ) =\min_{\theta\in\Theta}
\lvv \Phi(\theta)+t_n \sh(\theta) \rvv_\lsv^2 \\
&=\min \lbr \inf_{\theta\in \Theta\setminus\Theta_0(\phi_P)^{a(h)t_n}}
\lvv \Phi(\theta)+t_n \sh(\theta) \rvv_\lsv^2, \;
\min_{\theta\in \Theta\cap\Theta_0(\phi_P)^{a(h)t_n}}\lvv \Phi(\theta)+t_n \sh(\theta) \rvv_\lsv^2 \rbr.
\end{align*}
This, together with \eqref{eq.para second order 1}, implies that for large $n$, \begin{align*}
\mL\lp \phi_P+t_n h \rp =\min_{\theta\in \Theta\cap\Theta_0(\phi_P)^{a(h)t_n}}
\lvv \Phi(\theta)+t_n\sh(\theta)\rvv_\lsv^2.
\end{align*}

For every $a>0$, let $V(a)=\{ v\in\mr^{d_{\theta}}: \lvv v \rvv_2\le a \}$.
For every $\theta\in\Theta_0(\phi_P)$ and every $a>0$, define \begin{align*}
V_n(a,\theta)=\lbr v\in V(a): \theta+t_n v \in \Theta \rbr.
\end{align*}
It is easy to show that (with the compactness of $\Theta_0(\phi_P)$) 
\begin{align*}
\bigcup_{\theta\in\Theta_0(\phi_P)} \bigcup_{v\in V_n (a(h),\theta)} \lbr \theta+t_n v \rbr
=\Theta \cap \Theta_0(\phi_P)^{a(h)t_n} .
\end{align*}
Therefore, \begin{align*}
\mL\lp \phi_P+t_n h \rp =\inf_{\theta\in\Theta_0(\phi_P)} \inf_{v\in V_n(a(h),\theta)}
\lvv \Phi \lp \theta+t_n v \rp +t_n \sh \lp \theta+t_n v \rp \rvv_\lsv^2.
\end{align*}
Note that $0\in V_n(a(h),\theta)$. Then for every $\theta_0\in\Theta_0(\phi_P)$,
\begin{align*}
& \lv \mL\lp \phi_P+t_n h \rp-
\inf_{\theta\in\Theta_0(\phi_P)} \inf_{v\in V_n(a(h),\theta)}
\lvv \Phi \lp \theta+t_n v \rp +t_n \sh \lp \theta\rp \rvv_\lsv^2 \rv \\
=&\, \lv \inf_{\theta\in\Theta_0(\phi_P)} \inf_{v\in V_n(a(h),\theta)}
\lvv \Phi \lp \theta+t_n v \rp +t_n \sh \lp \theta+t_n v \rp \rvv_\lsv \right. \\
& \left.	+\inf_{\theta\in\Theta_0(\phi_P)} \inf_{v\in V_n(a(h),\theta)}
\lvv \Phi \lp \theta+t_n v \rp +t_n \sh \lp \theta \rp \rvv_\lsv \rv \\
& \phantom{==} \cdot \lv \inf_{\theta\in\Theta_0(\phi_P)} \inf_{v\in V_n(a(h),\theta)}
\lvv \Phi \lp \theta+t_n v \rp +t_n \sh \lp \theta+t_n v \rp \rvv_\lsv \right. \\
& \phantom{===} \left. -\inf_{\theta\in\Theta_0(\phi_P)} \inf_{v\in V_n(a(h),\theta)}
\lvv \Phi \lp \theta+t_n v \rp +t_n \sh \lp \theta \rp \rvv_\lsv \rv \\
\le&\, 2 \lvv \Phi\lp \theta_0 \rp+t_n \sh \lp \theta_0 \rp \rvv_\lsv
\sup_{\theta\in\Theta_0(\phi_P)} \sup_{v\in V_n(a(h),\theta)} t_n
\lvv \sh \lp \theta+t_n v \rp-\sh(\theta) \rvv_\lsv \\
\le&\, 2t_n^2 \lvv h \rvv_{\infty} 
\sup_{\theta_1,\theta_2\in\Theta: \lvv \theta_1-\theta_2 \rvv_2 \le a(h) t_n}
\lvv \sh \lp \theta_1 \rp-\sh \lp \theta_2 \rp \rvv_\lsv = o( t_n^2 ),
\end{align*}
where the last equality follows from the definition of $\bD_{\mL0}$ and
the compactness of $\Theta$.

For every $\theta\in\Theta$, define $\Phi'(\theta):\mr\to\mr^{d_{\theta}}$ such that 
\begin{align*}
\Phi'(\theta)(x)=\left.	\frac{\partial \phi_P (z,\vartheta)}{\partial \vartheta} 			\rv_{(z,\vartheta)=(x,\theta)}	\quad \text{for every } x\in \mr.
\end{align*}
Using an argument similar
to the previous result, we have \begin{align*}
&\phantom{=\:\:} \lv \inf_{\theta\in\Theta_0(\phi_P)} \inf_{v\in V_n(a(h),\theta)}
\lvv \Phi\lp \theta+t_n v \rp +t_n \sh(\theta) \rvv_\lsv^2 \right. \\
&\phantom{==} \left. - \inf_{\theta\in\Theta_0(\phi_P)} \inf_{v\in V_n(a(h),\theta)}
\lvv \Phi(\theta)+t_n \lbk \Phi'(\theta) \rbk^\T v+t_n \sh(\theta) \rvv_\lsv^2 \rv \\
& \le 2 t_n^2 \lvv h \rvv_{\infty}
\sup_{\theta\in\Theta_0(\phi_P)} \sup_{v\in V_n(a(h),\theta)}
\lvv \frac{\Phi\lp \theta+t_n v \rp-\Phi(\theta)}{t_n}-\lbk \Phi'(\theta) \rbk^\T v \rvv_\lsv.
\end{align*}
Since $\Theta_0(\phi_P)\subset \mathrm{int}(\Theta)$ and $\Theta_0(\phi_P)$ is compact,
for sufficiently large $n$, we have $V_n(a(h),\theta)=V(a(h))$ for all $\theta\in\Theta_0(\phi_P)$.
Then Assumption \ref{ass:para G is 2nd differentiable} implies that when $n$ is large,
for all $\theta\in\Theta_0(\phi_P)$ and all $v\in V_n(a(h),\theta)$, 
\begin{align*}
&\phantom{=\:\:} \lvv \frac{\Phi\lp \theta+t_n v \rp-\Phi(\theta)}{t_n}
-\lbk \Phi'(\theta) \rbk^\T v \rvv_\lsv^2 \\
& =\int_\mr \lbk \frac{\phi_P(x,\theta+t_n v)-\phi_P(x,\theta)}{t_n}
-\lp \left.
\frac{\partial \phi_P (z,\vartheta)}{\partial \vartheta} 
\rv_{(z,\vartheta)=(x,\theta)} \rp^\T v \rbk^2 \ddd \nu(x) \\
&=\int_\mr \lbk \frac{t_n}{2} v^\T \lp 	\left.
\frac{\partial^2 \phi_P (z,\vartheta)}{\partial \vartheta \partial \vartheta^\T} 
\rv_{(z,\vartheta)=(x,\theta+t_n^*(x) v)}	\rp v
\rbk^2 \ddd \nu(x) \\
& \le \frac{a(h)^4 t_n^2}{4}\int_\mr \sup_{\theta^{\ast}\in\Theta}
\lvv \left.
\frac{\partial^2 \phi_P (z,\vartheta)}{\partial \vartheta \partial \vartheta^\T} 
\rv_{(z,\vartheta)=(x,\theta^{\ast})} \rvv_2^2 \,\mathrm{d}\nu(x)=O(t_n^2),
\end{align*}
where $0\le t_n^{\ast}(x) \le t_n $ for all $x$, and the last inequality
follows from 
the property of the $\ell^2$ operator norm.	Then it follows that 
\begin{align*}
\sup_{\theta\in\Theta_0(\phi_P)} \sup_{v\in V_n(a(h),\theta)}
\lvv \frac{\Phi\lp \theta+t_n v \rp-\Phi(\theta)}{t_n}-
\lbk \Phi'(\theta) \rbk^\T v \rvv_\lsv
=o(1).
\end{align*} 
Combining the above results yields \begin{align}\label{eq:second limit}
\lv \mL\lp \phi_P+t_n h_n \rp - t_n^2 \inf_{\theta\in\Theta_0(\phi_P)} \inf_{v\in V(a(h))}
\lvv \lbk \Phi'(\theta) \rbk^\T v+\sh (\theta) \rvv_\lsv^2 \rv=o\lp t_n^2 \rp.
\end{align}	

Because the limit in \eqref{eq:second limit} as $n\to\infty$ is unique, by similar arguments, we can show  that for all $a\geq a\left(  h\right)  $,
\begin{align*}
\inf_{\theta\in\Theta_0(\phi_P)}\inf_{v\in V\left(  a\right)  }\left\Vert \Phi^{\prime}\left(  \theta\right)^{\T}v+\mathscr{H}\left(  \theta\right)  \right\Vert _{L^{2}\left( \nu\right)  }^{2}=\inf_{\theta\in\Theta_0(\phi_P)}\inf_{v\in V\left(  a\left( h \right)  \right)  }\left\Vert \Phi^{\prime}\left(  \theta\right)^{\T}v+\mathscr{H}\left(  \theta\right) \right\Vert _{L^{2}\left(  \nu\right)  }^{2}.
\end{align*}
For every $v^{\prime}\in\mathbb{R}^{d_{\theta}}$, if $\left\Vert v^{\prime}\right\Vert_{2}\geq a\left(  h\right)  $, then \begin{align*}
\inf_{\theta\in\Theta_0(\phi_P)}\left\Vert \Phi^{\prime}\left(  \theta\right)^{\T}v^{\prime}+\mathscr{H}\left(  \theta\right)  \right\Vert _{L^{2}\left(  \nu\right)}^{2}  & \geq\inf_{\theta\in\Theta_0(\phi_P)}\inf_{v\in V\left(  \left\Vert v'\right\Vert_{2}\right)}\left\Vert \Phi^{\prime}\left(  \theta\right)  ^{\T}v+\mathscr{H}\left(\theta\right)  \right\Vert _{L^{2}\left(  \nu\right)  }^{2}\\
& =\inf_{\theta\in\Theta_0(\phi_P)}\inf_{v\in V\left(  a\left(  h\right)  \right)  }\left\Vert \Phi^{\prime}\left(  \theta\right)  ^{\T}v+\mathscr{H}\left(  \theta\right)  \right\Vert_{L^{2}\left(  \nu\right)  }^{2};
\end{align*}
if $\left\Vert v^{\prime}\right\Vert _{2}<a\left(  h\right)  $, then \begin{align*}
\inf_{\theta\in\Theta_0(\phi_P)}\left\Vert \Phi^{\prime}\left(  \theta\right)^{\T}v^{\prime} +\mathscr{H}\left(  \theta\right)  \right\Vert _{L^{2}\left(  \nu\right)}^{2}\geq&\inf_{\theta\in\Theta_0(\phi_P)}\inf_{v\in V\left(  a\left(  h\right)  \right)  }\left\Vert \Phi^{\prime}\left(  \theta\right)  ^{\T}v+\mathscr{H}\left(  \theta\right) \right\Vert _{L^{2}\left(  \nu\right)  }^{2}\\
=&\inf_{v\in V\left(  a\left(  h\right)  \right)  }\inf_{\theta\in\Theta_0(\phi_P)}\left\Vert \Phi^{\prime}\left(  \theta\right)  ^{\T}v+\mathscr{H}\left(  \theta\right) \right\Vert _{L^{2}\left(  \nu\right)  }^{2}.
\end{align*}
This implies that 
\begin{align*}
\inf_{v\in \mr^{d_\theta}}\inf_{\theta\in\Theta_0(\phi_P)}\left\Vert \Phi^{\prime}\left(  \theta\right)^{\T}v+\mathscr{H}\left(  \theta\right)  \right\Vert_{L^{2}\left(  \nu\right)  }^{2}
\ge\inf_{v\in V\left(  a\left(  h\right)  \right)  }\inf_{\theta\in\Theta_0(\phi_P)}\left\Vert \Phi^{\prime}\left(  \theta\right)  ^{\T}v+\mathscr{H}\left(  \theta\right) \right\Vert _{L^{2}\left(  \nu\right)  }^{2}. 
\end{align*}

On the other hand, $V(a(h))\subset\mr^{d_\theta}$ by definition. Thus, \begin{align*}
\inf_{\theta\in\Theta_0(\phi_P)}\inf_{v\in \mr^{d_\theta}}\left\Vert \Phi^{\prime}\left(  \theta\right)^{\T}v+\mathscr{H}\left(  \theta\right)  \right\Vert_{L^{2}\left(  \nu\right)  }^{2}=&\inf_{v\in \mr^{d_\theta}}\inf_{\theta\in\Theta_0(\phi_P)}\left\Vert \Phi^{\prime}\left(  \theta\right)^{\T}v+\mathscr{H}\left(  \theta\right)  \right\Vert_{L^{2}\left(  \nu\right)  }^{2}\\
\le&\inf_{v\in V\left(  a\left(  h\right)  \right)  }\inf_{\theta\in\Theta_0(\phi_P)}\left\Vert \Phi^{\prime}\left(  \theta\right)  ^{\T}v+\mathscr{H}\left(  \theta\right) \right\Vert _{L^{2}\left(  \nu\right)  }^{2}\\
=&\inf_{\theta\in\Theta_0(\phi_P)}\inf_{v\in V\left(  a\left(  h\right)  \right)  }\left\Vert \Phi^{\prime}\left(  \theta\right)  ^{\T}v+\mathscr{H}\left(  \theta\right) \right\Vert _{L^{2}\left(  \nu\right)  }^{2}
. 
\end{align*}
\end{pf}

\begin{pf}{ of Proposition \ref{thry:para asymptotic distribution of test stat}}
Note that both $\ell^\infty(\mr\times\Theta)$ and $\mr$ are normed spaces.
By Lemma \ref{prop:para second order Hadamard of L}, the map
$\mL$ is second order Hadamard directionally differentiable at $\phi_P$
tangentially to $\bD_{\mL0}$. Lemma \ref{prop:para Donsker of phihat}
shows that $\sqrt{n}( \phihat-\phi_P )\convd \bG_0$ in $\ell^\infty(\mr\times\Theta)$
as $n\to\infty$ and $\bG_0$ is tight with $\bG_0\in\bD_{\mL0}$ almost surely.
Hence, Assumptions 2.1(i), 2.1(ii), 2.2(i), and 2.2(ii) of \citet{chen2019inference}
are satisfied. The desired result follows from
Theorem 2.1 of \citet{chen2019inference}, the facts that $\mL(\phi_P)=0$ and
$\mL'_{\phi_P}(h) =0$ for all $h\in\ell^\infty(\mr\times\Theta)$ whenever $\phi_P\in\bD_0$,
and that $( \phihat-\phi_P)\in \ell^\infty(\mr\times\Theta)$ for every $n\in\mathbb{Z}_+$.
\end{pf}

\begin{pf}{ of Lemma \ref{prop:para consistency of second derivative estimator}}
Note that both $\ell^\infty(\mr\times\Theta)$ and $\mr$ are normed spaces,
and by Lemma \ref{prop:para second order Hadamard of L}, the map
$\mL$ is second order Hadamard directionally differentiable at $\phi_P\in\bD_0$
tangentially to $\bD_{\mL0}$. By Lemma \ref{lemma:para-HDDL},
$\mL'_{\phi_P}(h)=0$ for all $h\in\ell^\infty(\mr\times\Theta)$ whenever $\phi_P
\in\bD_0$.
Lemma \ref{prop:para Donsker of phihat}
shows that $\sqrt{n}( \phihat-\phi_P )\convd \bG_0$ 
in $\ell^\infty(\mr\times\Theta)$
as $n\to\infty$, where $\bG_0$ is tight with $\bG_0\in\bD_{\mL0}$ almost surely.
Hence, Assumptions 2.1, 2.2(i), 2.2(ii), and 3.5 of \citet{chen2019inference}
hold, and the desired result follows from Proposition 3.1 of
\citet{chen2019inference}.
\end{pf}	

\begin{pf}{ of Lemma \ref{prop:para weak convergence of bootstrap}}
By Theorem 2.6 of \citet{kosorok2008introduction}, as $n\to\infty$, \begin{align*}
\sup_{\Gamma\in\mathrm{BL}_1(\ell^\infty(\Psi))} \lv \E \lbk \left. \Gamma \lp \sqrt{n}(\Phat^*-\Phat) \rp \rv \bZn \rbk -\E[ \Gamma(\bW) ] \rv \convp 0,
\end{align*}
and the sequence $\sqrt{n}(\Phat^*-\Phat)$ is asymptotically measurable. By construction, $\phihat^*(x,\theta)=\Phat^*(\psi_{x,\theta})$ and $\phihat(x,\theta)=\Phat(\psi_{x,\theta})$ for every $(x,\theta)\in\mr\times\Theta$ and every $n\in\mathbb{Z}_+$. From the proof of Lemma \ref{prop:para Donsker of phihat}, $\bG_0(x,\theta)=\bW(\psi_{x,\theta})$ for every $(x,\theta)\in\mr\times\Theta$. The desired result follows from Lemma \ref{lemma:conditional weak convergence of transformed index}.
\end{pf}

\begin{pf}{ of Proposition \ref{prop:para consistency of approximation of test statistic distribution}}
Note that both $\ell^\infty(\mr\times\Theta)$ and $\mr$ are normed spaces, and by Lemma \ref{prop:para second order Hadamard of L}, the map $\mL$ is second order Hadamard directionally differentiable at $\phi_P\in\bD_0$ tangentially to $\bD_{\mL0}$. Lemma \ref{prop:para Donsker of phihat} shows that $\sqrt{n}( \phihat-\phi_P )\convd \bG_0$ in $\ell^\infty(\mr\times\Theta)$ as $n\to\infty$ and $\bG_0$ is tight with $\bG_0\in\bD_{\mL0}$ almost surely. By Lemma \ref{lemma:para properties of DL}, $\bD_{\mL0}$ is closed under vector addition, that is, $\varphi_1+\varphi_2\in \bD_{\mL0}$ whenever $\varphi_1,\varphi_2\in \bD_{\mL0}$. By construction, the random weights used to construct the bootstrap samples are independent of the data set, and $f(\sqrt{n}( \phihat^*-\phihat ))$ is a measurable function of the random weights for every continuous and bounded $f:\ell^\infty(\mr\times\Theta)\to\mathbb{R}$ given every sample. By Lemma \ref{prop:para weak convergence of bootstrap}, \begin{align*}
\sup_{\Gamma\in\mathrm{BL}_1\lp \ell^\infty(\mr\times\Theta) \rp} \lv \E \lbk \left. \Gamma \lp \sqrt{n} \lp \phihat^*-\phihat \rp \rp \rv \bZn\rbk- \E \lbk \Gamma \lp \bG_0 \rp \rbk \rv \convp 0,
\end{align*}
and $\sqrt{n}( \phihat^*-\phihat )$ is asymptotically measurable as $n\to \infty$. Lemma \ref{prop:para consistency of second derivative estimator} establishes the consistency of $\mLhat$ for $\mathcal{L}^{\prime\prime}_{\phi_P}$. Hence, Assumptions 2.1(i), 2.1(ii), 2.2, 3.1, 3.2, and 3.4 of \citet{chen2019inference} are satisfied, and the result follows from Theorem 3.3 of \citet{chen2019inference}.
\end{pf}

\begin{pf}{ of Theorem \ref{thry:para size and power}}
We first prove Claim (i). 
The proof closely follows that of Theorem S.1.1 in \citet{fang2019inference}.
Let $\Pi_0$ be the cumulative distribution function of $\mL''_{\phi_P}\lp \bG_0 \rp$
and $c_{1-\alpha}$ be the $1-\alpha$ quantile for $\mL''_{\phi_P}\lp \bG_0 \rp$.
Define \begin{align*}
\Pihat(c)=\p \lp\left. \mLhat\lbk \sqrt{n} \lp \phihat^*-\phihat \rp \rbk
\le c \rv \bZn \rp
\end{align*}
for every $n\in\mathbb{Z}_+$ and every $c\in\mr$.
Let $C_{\Pi_0}\subset \mr$ be the set of continuity points of $\Pi_0$, and $\mathbb{L}(\mr)$
be the set of all Lipschitz continuous functions $\Gamma:\mr\to[0,1]$.
For every $\Gamma\in\mathbb{L}(\mr)$, let $M=1\vee L_\Gamma$, where $L_\Gamma$ is the Lipschitz constant of $\Gamma$. Then $\Gamma/M\in \mathrm{BL}_1(\mr)$, and by
Proposition \ref{prop:para consistency of approximation of test statistic distribution},
\begin{align}
\E \lbk \left. \Gamma \lp \mLhat \lbk \sqrt{n} \lp \phihat^*
-\phihat \rp \rbk \rp \rv \bZn \rbk \convp \E \lbk \Gamma \lp \mL''_{\phi_P} \lp \bG_0 \rp
\rp \rbk  \label{eq:para pointwise convergence in probability}
\end{align}
as $n\to\infty$ if $\HH_0$ is true. By Lemma 10.11(i) of
\citet{kosorok2008introduction}, we have $\Pihat(c)\convp\Pi_0(c)$
for every $c\in C_{\Pi_0}$.
Because $\Pi_0$ is strictly increasing and continuous at $c_{1-\alpha}$
and a cumulative distribution function has at most countably many discontinuity points, for every $\varepsilon>0$, there exist $a_1, a_2 \in C_{\Pi_0}$ such that
$a_1<c_{1-\alpha}<a_2$, $\lv a_1- c_{1-\alpha} \rv<\varepsilon$,
$\lv a_2- c_{1-\alpha} \rv<\varepsilon$, and \begin{align*}
\delta=\frac{1}{2}\lbk \lv \Pi_0 \lp a_1 \rp-(1-\alpha) \rv
\wedge \lv \Pi_0 \lp a_2 \rp-(1-\alpha) \rv \rbk>0.
\end{align*}
From the definition of $\chat_{1-\alpha,n}$, it follows that \begin{align*}
\p \lp \lv \chat_{1-\alpha,n}-c_{1-\alpha} \rv>\varepsilon \rp
& \le \p \lp \chat_{1-\alpha,n}< a_1 \rp +
\p \lp \chat_{1-\alpha,n}> a_2 \rp \\
& \le \p \lp \Pihat \lp a_1 \rp \ge 1-\alpha \rp +
\p \lp \Pihat \lp a_2 \rp < 1-\alpha \rp \\
& \le \p \lp \lv \Pihat\lp a_1 \rp -\Pi_0 \lp a_1 \rp \rv >\delta \rp
+\p \lp \lv \Pihat\lp a_2 \rp -\Pi_0 \lp a_2 \rp \rv >\delta \rp,
\end{align*}
and the last line converges to $0$ since $\Pihat\lp a_1 \rp \convp\Pi_0\lp a_1 \rp$
and $\Pihat\lp a_2 \rp \convp\Pi_0\lp a_2 \rp$ as $n\to\infty$.
This implies that $\chat_{1-\alpha,n}\convp c_{1-\alpha}$ as $n\to\infty$.

By Proposition \ref{thry:para asymptotic distribution of test stat} of this paper, if $\HH_0$ is true ($\phi_P\in\bD_0$), then $n \mL ( \phihat ) \convd \mL''_{\phi_P} \lp \bG_0 \rp$ as $n\to\infty$. By Lemma 2.8(i) of \citet{van1998asymptotic}, $n \mL ( \phihat )-\chat_{1-\alpha,n}\convd  \mL''_{\phi_P} \lp \bG_0 \rp-c_{1-\alpha}$ as $n\to\infty$. Since the cumulative distribution function of $\mL''_{\phi_P} \lp \bG_0 \rp$ is continuous and strictly increasing at $c_{1-\alpha}$,
the cumulative distribution function of $\mL''_{\phi_P} \lp \bG_0 \rp-c_{1-\alpha}$ is
continuous at 0 and $\p ( \mL''_{\phi_P} \lp \bG_0 \rp-c_{1-\alpha}>0 )=\alpha$.
By Lemma 2.2(i) (portmanteau) of \citet{van1998asymptotic},
we have \begin{align*}
&\lim_{n\to\infty}
\p\lp n \mL \lp \phihat \rp>\chat_{1-\alpha,n} \rp
=  \lim_{n\to\infty}
\p\lp n \mL \lp \phihat \rp-\chat_{1-\alpha,n}>0 \rp =\p \lp \mL''_{\phi_P} \lp \bG_0 \rp-c_{1-\alpha}>0 \rp=\alpha.
\end{align*}

Now we prove Claim (ii).
For all $\theta\in\Theta$ and all $\phi_1,\phi_2\in \bD_{\mL}$, 
\begin{align*}
&\phantom{=\:\:}\lv \int_\mr \lbk \phi_1 (x,\theta) \rbk^2 \ddd \nu(x)-
\int_\mr \lbk \phi_2 (x,\theta) \rbk^2 \ddd \nu(x) \rv \\
& \le \int_\mr \lv \lbk \phi_1 (x,\theta)+\phi_2(x,\theta) \rbk
\lbk \phi_1(x,\theta)-\phi_2(x,\theta) \rbk \rv \ddd \nu(x) \le (\Vert \phi_1\Vert_{\infty}+\Vert \phi_2\Vert_{\infty}) \lvv \phi_1-\phi_2 \rvv_\infty.
\end{align*}
This implies that 
\begin{align*}
\lv \mL \lp \phi_1 \rp -\mL \lp \phi_2 \rp \rv \le (\Vert \phi_1\Vert_{\infty}+\Vert \phi_2\Vert_{\infty}) \lvv \phi_1-\phi_2 \rvv_\infty.
\end{align*}
Thus the function $\varphi\mapsto \mL(\varphi)$ is continuous.		
If $\HH_0$ is false, then by Lemma \ref{prop:para Donsker of phihat} of this paper and Theorem 1.9.5 (continuous mapping) of \citet{van1996weak}, we have $\mL( \phihat ) \convp	\mL(\phi_P)>0$ as $n\to\infty$. Combining this with Assumption \ref{ass:para rate of tau} yields $ 1/[n\tau_n^2\mL( \phihat )] \convp0$ as $n\to\infty$. 

By definition, for every $h\in\ell^{\infty}(\mathbb{R}\times\Theta)$, \begin{align*}
\tau_{n}^2\mathcal{\widehat{L}}_{n}^{\prime\prime}(h) &= \mathcal{L}\left(  \widehat{\phi}_{n}+\tau_{n}h\right)-\mathcal{L}\left(  \widehat{\phi}_{n}\right) \leq\sup_{\theta\in\Theta}\left\vert \int_{\mathbb{R}}\lbk 2\tau_{n}h\left(  x,\theta\right)  \widehat{\phi}_{n}\left(  x,\theta\right)+\tau_{n}^{2}h^{2}\left(  x,\theta\right) \rbk \ddd \nu\left(  x\right)\right\vert \\
& \leq 2\tau_n \Vert  h \Vert_{\infty} \Vert \phihat \Vert_{\infty}+\tau_{n}^2 \Vert h \Vert_{\infty}^2  \le 2\tau_n \Vert \phihat \Vert_{\infty}+\lp 2\tau_{n} \Vert \phihat \Vert_{\infty}+\tau_{n}^2 \rp \Vert h \Vert_{\infty}^2,
\end{align*} 
where the last inequality follows from the fact that $\Vert h \Vert_{\infty}\le 1\vee \lvv h \rvv_{\infty}^2 \le 1+ \lvv h \rvv_{\infty}^2$. Define $\widehat{\mathcal{L}}_{b,n}\left(  h\right)  =\bhat_{0,n}+\bhat_{1,n} \Vert h \Vert_{\infty}^2$ for every $h\in\ell^{\infty}(\mathbb{R}\times\Theta)$, where $\bhat_{0,n}= 2\tau_n \Vert \phihat \Vert_{\infty}$ and $\bhat_{1,n}=2\tau_{n} \Vert \phihat \Vert_{\infty}+\tau_{n}^2$. Recall that $\Vert \phihat-\phi_P \Vert_{\infty}\convp 0$ as $n\to\infty$ and $\phi_P\in\bD_{\mL 0}\subset\ell^\infty(\mr\times\Theta)$. Since $\Vert \phihat \Vert_{\infty}\le \Vert \phi_P \Vert_{\infty}+\Vert \phihat-\phi_P \Vert_{\infty}$, we have $ \Vert \phihat \Vert_{\infty}=O_\p(1)$ as $n\to\infty$. This implies that $\bhat_{0,n}\convp 0 $ and $\bhat_{1,n}\convp 0$ as $n\to\infty$.

The functional $h\mapsto \Vert h \Vert_{\infty}^2$ is continuous at every $h\in\ell^\infty(\mr\times\Theta)$. Indeed, for any $h_0\in\ell^\infty(\mr\times\Theta)$ and any $\varepsilon>0$, we can pick $\delta>0$ such that $2\Vert h_0 \Vert_{\infty}\delta+\delta^2 <\varepsilon$. Then for all $h\in \ell^\infty(\mr\times\Theta)$ with $\Vert h-h_0 \Vert_{\infty}<\delta$, we have \begin{align*}
\lv \lvv h \rvv_{\infty}^2 -\lvv h_0 \rvv_{\infty}^2 \rv =&\,\lp \lvv h \rvv_{\infty} +\lvv h_0 \rvv_{\infty} \rp \lv \lvv h \rvv_{\infty} -\lvv h_0 \rvv_{\infty} \rv \\
\le&\, \lp 2\lvv h_0 \rvv_{\infty}+\lvv h-h_0 \rvv_{\infty} \rp \lvv h-h_0 \rvv_{\infty} \le \lp 2\lvv h_0 \rvv_{\infty}+\delta \rp \delta <\varepsilon.
\end{align*}
Furthermore, the set $\bD_{\mL 0}$ is closed. To see this, consider any sequence $\{\varphi_k\}_{k=1}^{\infty}\subset \bD_{\mL 0}$ satisfying $\varphi_k\to \varphi$ in $\ell^\infty(\mr\times\Theta)$. For every $\theta_0\in\Theta$ and any $\varepsilon>0$, there exist $k\in\mathbb{Z}_+$ and $\delta>0$, so that $\lvv \varphi_k-\varphi \rvv_{\infty}^2<\varepsilon$ and \begin{align*}
\int_{\mr} [\varphi_k(x,\theta)-\varphi_k(x,\theta_0)]^2 \ddd \nu(x) <\varepsilon
\end{align*}
for all $\theta\in\Theta$ with $\lvv \theta-\theta_{0} \rvv_2<\delta$. Thus \begin{align*}
&\int_{\mr} [\varphi(x,\theta)-\varphi(x,\theta_0)]^2 \ddd \nu(x) \\
=&\,\int_{\mr} [(\varphi_k+\varphi-\varphi_k)(x,\theta)-(\varphi_k+\varphi-\varphi_k)(x,\theta_0)]^2 \ddd \nu(x) \\
=&\,\int_{\mr} [\varphi_k(x,\theta)-\varphi_k(x,\theta_{0})+(\varphi-\varphi_k)(x,\theta)-(\varphi-\varphi_k)(x,\theta_0) ]^2 \ddd \nu(x)\\
\le&\, 2 \int_{\mr} [\varphi_k(x,\theta)-\varphi_k(x,\theta_{0})]^2 \ddd \nu(x)+8 \lvv \varphi-\varphi_k \rvv_{\infty}^2 <10 \varepsilon
\end{align*}
for all $\theta\in\Theta$ with $\lvv \theta-\theta_{0} \rvv_2<\delta$, which implies that $\varphi\in\bD_{\mL 0}$.

Note that both $\ell^\infty(\mr\times\Theta)$ and $\mr$ are Banach spaces. We have established that $\bD_{\mL 0}\subset\ell^\infty(\mr\times\Theta)$ is closed and that $h\mapsto \lvv h \rvv_{\infty}^2$ is continuous at all points in $\ell^\infty(\mr\times\Theta)$. By construction,  $f(\sqrt{n}(\phihat^*-\phihat))$ is a measurable function of the random weights for every continuous and bounded $f:\ell^\infty(\mr\times\Theta)\to\mathbb{R}$ given every sample. By Lemma \ref{prop:para weak convergence of bootstrap}, as $n\to\infty$, \begin{align*}
\sup_{\Gamma\in\mathrm{BL}_1\lp \ell^\infty(\mr\times\Theta) \rp}\lv \E \lbk \left. \Gamma \lp \sqrt{n} \lp \phihat^*-\phihat \rp \rp	\rv \bZn \rbk-\E \lbk \Gamma \lp \bG_0 \rp \rbk \rv \convp 0,
\end{align*}
where $\bG_{0}$ is tight and $\p(\bG_{0}\in\bD_{\mL 0})=1$. Applying Theorem 10.8 of \citet{kosorok2008introduction} yields that \begin{align}
\sup_{\Gamma\in\mathrm{BL}_1\lp \mr \rp}\lv \E \lbk \left. \Gamma \lp \lvv \sqrt{n} \lp \phihat^*-\phihat \rp \rvv_{\infty}^2 \rp	\rv \bZn \rbk-\E \lbk \Gamma \lp \lvv  \bG_0  \rvv_{\infty}^2 \rp \rbk \rv \convp 0 \label{eq:conditional weak convergence of squared norms}
\end{align}
as $n\to\infty$.

Since $\bG_0$ takes values in $\bD_{\mL 0}\subset\ell^\infty(\mr\times\Theta)$ almost surely, then $\p\lp\Vert \bG_0 \Vert_{\infty}^2\in \mr \rp=1$. Hence for $\alpha\in(0,1)$, the $(1-\alpha)$ quantile of $ \Vert \bG_0 \Vert_{\infty}^2$, denoted by $c'_{1-\alpha}$, is finite. Since a cumulative distribution function has at most countably many discontinuity points, there exists $c''_{1-\alpha}\in(c'_{1-\alpha},\infty)$ such that the cumulative distribution function of $\Vert\bG_0 \Vert_{\infty}^2 $ is continuous at $c''_{1-\alpha}$ and $\p \lp \lvv \bG_0 \rvv_{\infty}^2 \le c''_{1-\alpha} \rp > 1-\alpha$. Using an argument analogous to \eqref{eq:para pointwise convergence in probability}, we can use \eqref{eq:conditional weak convergence of squared norms} to conclude that \begin{align*}
\E \lbk \left. \Gamma \lp \lvv \sqrt{n} \lp \phihat^*-\phihat \rp \rvv_{\infty}^2 \rp	\rv \bZn \rbk  \convp \E \lbk \Gamma \lp \lvv  \bG_0  \rvv_{\infty}^2 \rp \rbk  
\end{align*}
as $n\to\infty$ for every $\Gamma\in\mathbb{L}(\mr)$. By Lemma 10.11(i) of \citet{kosorok2008introduction}, as $n\to\infty$, \begin{align*}
\p \lp \left. \lvv \sqrt{n}\lp \phihat^*-\phihat \rp \rvv_{\infty}^2 \le c''_{1-\alpha} \rv \bZn \rp \convp \p \lp \lvv \bG_0 \rvv_{\infty}^2 \le c''_{1-\alpha} \rp.
\end{align*}

Recall that $\tau_{n}^2\mathcal{\widehat{L}}_{n}^{\prime\prime}(h) \le \widehat{\mathcal{L}}_{b,n}(h)$ for all $h\in\ell^\infty(\mr\times\Theta)$. Above results imply that as $n\to\infty$, \begin{align*}
&\p \lp\left.  \tau_{n}^2\mathcal{\widehat{L}}_{n}^{\prime\prime}\lp \sqrt{n} \lp \phihat^*-\phihat \rp  \rp \le 1+c''_{1-\alpha} \rv \bZn \rp \ge \p \lp \left. \widehat{\mathcal{L}}_{b,n} \lp \sqrt{n} \lp \phihat^*-\phihat \rp  \rp \le 1+c''_{1-\alpha} \rv \bZn \rp \\
\ge&\, \p \lp \left.  \lbr \lvv \sqrt{n}\lp \phihat^*-\phihat \rp \rvv_{\infty}^2 \le c''_{1-\alpha} \rbr \cap \lbr \bhat_{0,n} \le 1 \rbr \cap \lbr \bhat_{1,n} \le 1 \rbr \rv \bZn \rp \\
=&\,1-\p\lp \left. \lbr \lvv \sqrt{n}\lp \phihat^*-\phihat \rp \rvv_{\infty}^2 > c''_{1-\alpha} \rbr \cup \lbr \bhat_{0,n} > 1 \rbr \cup \lbr \bhat_{1,n} > 1 \rbr \rv \bZn \rp   \\
\ge&\, 1- \p \lp \left. \lvv \sqrt{n}\lp \phihat^*-\phihat \rp \rvv_{\infty}^2 > c''_{1-\alpha} \rv \bZn \rp - \p\lp \left. \bhat_{0,n} > 1 \rv \bZn \rp -\p \lp \left. \bhat_{1,n} > 1 \rv \bZn \rp \\
\ge&\, \p \lp \left. \lvv \sqrt{n}\lp \phihat^*-\phihat \rp \rvv_{\infty}^2 \le c''_{1-\alpha} \rv \bZn \rp -\indicator \lbr \bhat_{0,n}>1 \rbr -\indicator \lbr \bhat_{1,n}>1 \rbr\convp \p \lp \lvv \bG_0 \rvv_{\infty}^2 \le c''_{1-\alpha} \rp\\
>&\,1-\alpha.
\end{align*}

Combining all these results, we have that as $n\to\infty$, \begin{align*}
& \p\lp n\mL\lp \phihat \rp>\chat_{1-\alpha,n} \rp \ge \p \lp \lbr \tau_{n}^2 \chat_{1-\alpha,n} \le 1+c''_{1-\alpha} \rbr\cap \lbr n\tau_{n}^2 \mL\lp \phihat \rp >1+c''_{1-\alpha} \rbr \rp \\
=\;& 1-\p \lp \lbr \tau_{n}^2 \chat_{1-\alpha,n} > 1+c''_{1-\alpha} \rbr\cup \lbr n\tau_{n}^2 \mL\lp \phihat \rp \le 1+c''_{1-\alpha} \rbr \rp \\
\ge\;& 1- \p \lp \tau_{n}^2 \chat_{1-\alpha,n} > 1+c''_{1-\alpha} \rp-\p \lp  n\tau_{n}^2 \mL\lp \phihat \rp \le 1+c''_{1-\alpha} \rp \\
=\;& \p \lp \tau_{n}^2 \chat_{1-\alpha,n} \le 1+c''_{1-\alpha}  \rp -\p \lp  n\tau_{n}^2 \mL\lp \phihat \rp \le 1+c''_{1-\alpha} \rp  \\
\ge\;& \p \lbk \p \lp\left.  \tau_{n}^2\mathcal{\widehat{L}}_{n}^{\prime\prime}\lp \sqrt{n} \lp \phihat^*-\phihat \rp  \rp \le 1+c''_{1-\alpha} \rv \bZn \rp >1-\alpha \rbk -\p \lp \frac{1}{n\tau_{n}^2 \mL\lp \phihat \rp} \ge \frac{1}{1+c''_{1-\alpha}} \rp \\
\convp&\, 1-0=1.
\end{align*} 
\end{pf}

\begin{pf}{ of Proposition \ref{prop:local power}}
Under Assumptions \ref{ass:para Donsker class} and \ref{ass:local sequence matched pairs}, we can use Theorem 3.10.12 of \citet{van1996weak} to conclude that $\sqrt{n}(\Phat-P)\convd \bW+V_P$ under $P_n$ in $\ell^\infty(\Psi)$ as $n\to\infty$, where $\bW$ is a tight Brownian bridge and $V_P(f)=P(fv_0)$ for every $f\in\Psi$. Note that $\bW+V_P$ is also tight. Since $\phihat(x,\theta)=\Phat(\psi_{x,\theta})$ and $\phi_P(x,\theta)=P(\psi_{x,\theta})$ for every $(x,\theta)\in\mr\times\Theta$ and $n\in\mathbb{Z}_+$, by Lemma \ref{lemma:weak convergence with transformed index}, we have $\sqrt{n}(\phihat-\phi_P)\convd \bG_0+\zeta_P$ under $P_n$ in $\ell^\infty(\mr\times\Theta)$ as $n\to\infty$, where both $\bG_0$ and $\bG_0+\zeta_P$ are tight, and $\bG_0(x,\theta)=\bW(\psi_{x,\theta})$ and $\zeta_P(x,\theta)=P(\psi_{x,\theta}v_0)$ for every $(x,\theta)\in\mr\times\Theta$.

Next, we show that $\p(\bG_0+\zeta_P\in\bD_{\mL 0})=1$. Observe that for every $\omega\in\Omega$ and every $\theta,\theta_{0}\in\Theta$, 
\begin{align*}
&\int_{\mr} \lbk (\bG_0+\zeta_P)(\omega)(x,\theta)- (\bG_0+\zeta_P)(\omega)(x,\theta_0) \rbk^2 \ddd \nu(x) \\
=&\,\int_{\mr} \lbk \bG_0(\omega)(x,\theta)-\bG_0(\omega)(x,\theta_0)+\zeta_P(x,\theta)- \zeta_P(x,\theta_0) \rbk^2 \ddd \nu(x) \\
\le&\, 2\int_{\mr} \lbk \bG_0(\omega)(x,\theta)-\bG_0(\omega)(x,\theta_0)\rbk^2\ddd \nu(x)+2\int_{\mr} \lbk \zeta_P(x,\theta)- \zeta_P(x,\theta_0) \rbk^2\ddd \nu(x).
\end{align*}
By Cauchy--Schwarz inequality, \begin{align*}
\lbk \zeta_P(x,\theta)- \zeta_P(x,\theta_0) \rbk^2=\lp P[(\psi_{x,\theta}-\psi_{x,\theta_0})v_0]\rp^2 \le P \lbk (\psi_{x,\theta}-\psi_{x,\theta_0})^2 \rbk P\lp v_0^2\rp.
\end{align*}
Assumption \ref{ass:local sequence matched pairs} implies that $P(v_0)=0$ and $P(v_0^2)<\infty$ by Lemma 3.10.11 of \citet{van1996weak}. By a similar proof of Lemma \ref{prop:para Donsker of phihat}, there exists $\Omega_0\subset \Omega$ with $\p(\Omega_0)=1$, such that for all $\omega\in\Omega_0$, for every $\theta_0\in\Theta$, and for any $\varepsilon>0$, there exists $\delta>0$, so that for all $\theta\in\Theta$ with $\lvv \theta-\theta_0\rvv_2<\delta$, we have \begin{align*}
&\int_{\mr} \lbk (\bG_0+\zeta_P)(\omega)(x,\theta)- (\bG_0+\zeta_P)(\omega)(x,\theta_0) \rbk^2 \ddd \nu(x) \le 2\lbk 1+P\lp v_0^2 \rp \rbk \varepsilon^2.
\end{align*}
This implies that $(\bG_0+\zeta_P)(\omega)\in\bD_{\mL 0}$ and thus $\p(\bG_0+\zeta_P\in\bD_{\mL 0})=1$.

The above results, together with Lemma \ref{prop:para second order Hadamard of L}, verify Assumptions 2.1(i), 2.1(ii) and 2.2(i), 2.2(ii) of \citet{chen2019inference} under $P_n$. Recall that 
$\mL'_{\phi_P}(h) =0$ for all $h\in\ell^\infty(\mr\times\Theta)$ whenever $\phi_P\in\bD_0$,
and that $\mathbb{D}_{\mathcal{L}0}$ is closed under vector addition by Lemma \ref{lemma:para properties of DL}. 
Then Assumptions 2.1(iii) and 2.2(iii) of \citet{chen2019inference} hold.
By assumption, $P$ satisfies $\mathrm{H}_0$, that is, $\mathcal{L}(\phi_P)=0$. We let $\phi_{P_n}(x,\theta)=P_n(\psi_{x,\theta})$ for all $(x,\theta)$.
By Theorem 3.10.12 of \citet{van1996weak}, 
\begin{align*}
\sqrt{n}(\Phat-P_n)&\convd \bW \text{ under }P_n,\\
\sup_{(x,\theta)\in\mathbb{R}\times\Theta}|\sqrt{n}(\phi_{P_n}(x,\theta)-\phi_P(x,\theta))-P(\psi_{x,\theta}v_0)|
&=\sup_{f\in\Psi}|\sqrt{n}(P_n(f)-P(f))-P(fv_0)|\to0.
\end{align*}
By Lemma \ref{lemma:weak convergence with transformed index}, $\sqrt{n}(\phihat-\phi_{P_n})\convd \bG_0$ under $P_n$ in $\ell^\infty(\mr\times\Theta)$ as $n\to\infty$.
By Lemma C.1 of \citet{chen2019inference}, $n\mL(\phihat)\convd\mL''_{\phi_P}(\bG_0+\zeta_P)$ under $P_n$ as $n\to\infty$. As shown in the proof of Theorem \ref{thry:para size and power}(i), $\chat_{1-\alpha,n}\convp c_{1-\alpha}$ under $P$ as $n\to\infty$. By the discussion after (3.10.10) of \citet[p.~406]{van1996weak}, the two sequences of distributions, $\{P_n^n\}$ and $\{P^n\}$, are contiguous. By Theorem 12.3.2(i) of \citet{lehmann2005testing}, $\chat_{1-\alpha,n}\convp c_{1-\alpha}$ under $P_n$ as $n\to\infty$. By Example 1.4.7 (Slutsky's lemma) of \citet{van1996weak}, we have $n\mL(\phihat)-\chat_{1-\alpha,n}\convd \mL''_{\phi_P}(\bG_0+\zeta_P)-c_{1-\alpha}$ under $P_n$ as $n\to\infty$. Since $(0,\infty)$ is an open set, Theorem 1.3.4 of \citet{van1996weak} (Portmanteau) implies that \begin{align*}
\liminf_{n\to\infty}\p\lp n \mL( \phihat )>\chat_{1-\alpha,n} \rp \ge \mathbb{P}(\mathcal{L}''_{\phi_P}(\mathbb{G}_0+\zeta_P)>c_{1-\alpha}).
\end{align*}
\end{pf}

\subsection{Proofs for Section \ref{sec:Extension to Dependent Data}}

\begin{pf}{ of Lemma \ref{lemma:dep weak convergence}}
Recall that $\phihat(x,\theta)=\Phat(\psi_{x,\theta})$ and $\phi_P(x,\theta)=P(\psi_{x,\theta})$ for every $(x,\theta)\in\mr\times\Theta$ and every $n\in\mathbb{Z}_+$. By Assumption \ref{ass:dep VC class},  $(\phihat-\phi_P)\in\ell^\infty(\mr\times\Theta)$ for every $n\in\mathbb{Z}_+$.

Note that $\beta_k=O(k^{-q})$ for some $q>p/(p-2)$ is sufficient for Condition (2.4) of \citet{arcones1994central}. Under Assumptions \ref{ass:dep VC class} and \ref{ass:dep beta mixing} of this paper, we apply Theorem 2.1 of \citet{arcones1994central} to conclude that \begin{align*}
\sqrt{n}\lp \Phat-P \rp \convd \bW \text{ in } \ell^\infty(\Psi)
\end{align*}
as $n\to\infty$, where $\bW$ is a Gaussian process with almost surely uniformly bounded and uniformly continuous paths with respect to the $\lvv \cdot \rvv_{L^2(P)}$ norm. By Lemma \ref{lemma:weak convergence with transformed index}, \begin{align*}
\sqrt{n}\lp \phihat -\phi_P \rp \convd \bG_0 \text{ in } \ell^\infty(\mr\times\Theta)
\end{align*}
as $n\to\infty$, where $\bG_0(x,\theta)=\bW(\psi_{x,\theta})$ for every $(x,\theta)\in\mr\times\Theta$. By Example 1.4.7 (Slutsky's lemma), Theorem 1.3.6, and Lemma 1.10.2(iii) of \citet{van1996weak}, the above result implies that \begin{align*}
\sup_{(x,\theta)\in\mr\times \Theta}\lv \phihat(x,\theta)-\phi_P(x,\theta) \rv\convp  0
\end{align*}
as $n\to\infty$.

By Assumption \ref{ass:dep VC class}, the set $\Psi$ is totally bounded under the metric induced by $\lvv \cdot \rvv_{L^2(P)}$. Then by 
Theorems 1.3.6, 1.3.4(iii), 1.5.7, and 1.5.4 
of \citet{van1996weak}, the Gaussian process $\bW$ is tight in $\ell^\infty(\Psi)$. By Lemma \ref{lemma:weak convergence with transformed index}, $\bG_0$ is tight.

Since $\bW$ almost surely has uniformly bounded and uniformly continuous paths with respect to the $\lvv \cdot \rvv_{L^2(P)}$ norm, there exists $\Omega_0\subset \Omega$ with $\p(\Omega_0)=1$ such that for every $\omega\in\Omega_0$ and every $\varepsilon>0$, $\bG_0(\omega)$ is uniformly bounded and there exists $\delta_1>0$ such that \begin{align*}
\lv \bG_0(\omega)(x_1,\theta_1)-\bG_0(\omega)(x_2,\theta_2) \rv = \lv \bW(\omega)(\psi_{x_1,\theta_1})-\bW(\omega)(\psi_{x_2,\theta_2})\rv<\varepsilon,
\end{align*}
whenever \begin{align*}
\lvv \psi_{x_1,\theta_1}-\psi_{x_2,\theta_2}\rvv_{L^2(P)}=\sqrt{P\lbk (\psi_{x_1,\theta_1}-\psi_{x_2,\theta_2})^2 \rbk}<\delta_1.
\end{align*}
By Assumption \ref{ass:para g mapsto f(g) is continuous}, for every $\theta_0\in\Theta$ and every $\varepsilon>0$, there is $\delta_2>0$ such that $\sup_{x\in\mr} P[(\psi_{x,\theta}-\psi_{x,\theta_0})^2]<\delta_1^2$ whenever $\lvv \theta-\theta_0 \rvv_2<\delta_2$,
and thus \begin{align*}
\int_\mr \lbk \bG_{0}(\omega)(x,\theta)-\bG_{0}(\omega)(x,\theta_0) \rbk^2 \ddd \nu(x)<\varepsilon^2
\end{align*}
for all $\theta\in\Theta$ with $\lvv \theta-\theta_0 \rvv_2<\delta_2$. This implies that $\bG_0(\omega)\in\bD_{\mL 0}$ and hence $\p(\bG_{0}\in\bD_{\mL 0})=1$.
\end{pf}

\begin{pf}{ of Lemma \ref{lemma:dep conditional weak convergence of bootstrap}}
Under Assumptions \ref{ass:dep VC class}--\ref{ass:dep block size b} of this paper, we apply Theorem 1 of \citet{radulovic1996bootstrap} to conclude that \begin{align*}
\sup_{\Gamma\in\mathrm{BL}_1\lp \ell^\infty(\Psi) \rp}\lv \E \lbk \left. \Gamma \lp \sqrt{n} \lp \Phat^*-\Phat \rp \rp	\rv \bZn \rbk-\E \lbk \Gamma \lp \bW \rp \rbk \rv \convp 0
\end{align*}
as $n\to\infty$, where $\bW$ is defined in the proof of Lemma \ref{lemma:dep weak convergence}. Recall that $\phihat^*(x,\theta)=\Phat^*(\psi_{x,\theta})$, $\phihat(x,\theta)=\Phat(\psi_{x,\theta})$, and $\bG_{0}(x,\theta)=\bW(\psi_{x,\theta})$ for every $n\in\mathbb{Z}_+$ and all $(x,\theta)\in\mr\times\Theta$. The conditional weak convergence of $\sqrt{n}(\phihat^*-\phihat)$ in probability follows from Lemma \ref{lemma:conditional weak convergence of transformed index}.
\end{pf}

\begin{pf}{ of Theorem \ref{thry:dep size and power}}
Note that Lemmas \ref{lemma:para-HDDL}--\ref{prop:para consistency of second derivative estimator}, Propositions \ref{prop:para equivalent null}--\ref{prop:para consistency of approximation of test statistic distribution}, and Theorem \ref{thry:para size and power} do not directly rely on the i.i.d.\ nature of the data observations, possibly given the consistency and weak convergence of $\phihat$ (Lemma \ref{prop:para Donsker of phihat}) and the conditional weak convergence of $\phihat^*$ in probability (Lemma \ref{prop:para weak convergence of bootstrap}). Thus, it suffices to establish the consistency and weak convergence of $\phihat$ and the conditional weak convergence of $\phihat^*$ in probability for dependent data, which has been accomplished in Lemmas \ref{lemma:dep weak convergence} and \ref{lemma:dep conditional weak convergence of bootstrap}. The remaining parts of the proof are analogous to the proof of Theorem \ref{thry:para size and power}.
\end{pf}

\subsection{Proofs for Appendix \ref{sec:Analyses of Examples}}

\begin{pf}{ of Lemma \ref{lemma:examples satisfy model assumptions}}
We first show that Assumption \ref{ass:para properties of g} holds. The continuity of $x\mapsto\phi_P(x,\theta)$ for every $\theta\in\Theta$ is obvious in Examples \ref{exam:symmetry.prob}--\ref{exam:LST.prob}. In Example \ref{exam:CMR.prob}, define $e(X,\theta)=\E_P[g(Y,\theta)|X]$. Since $P_X$ has Lebesgue probability density function $f$, applying the law of iterated expectations yields \begin{align*}
\phi_P(x,\theta)=\E_{P}\lbk e(X,\theta) \indicator\{X\le x\} \rbk=\int_{-\infty}^x e(z_1,\theta) f(z_1) \ddd z_1,
\end{align*}
which implies that for every $\theta\in\Theta$, $\phi_P(x,\theta)$ is differentiable with respect to $x$ and thus continuous in $x$.

To show that Assumption \ref{ass:para g mapsto f(g) is continuous} holds in Example \ref{exam:CMR.prob}, we note that \begin{align*}
\lbk \psi_{x,\theta}(Z)-\psi_{x,\theta_0}(Z) \rbk^2 &= \lbk g(Y,\theta)\indicator\{X\le x\}-g(Y,\theta_0)\indicator\{X\le x\} \rbk^2 \\
&=\lbk g(Y,\theta)-g(Y,\theta_0) \rbk^2 \indicator\{X\le x\},
\end{align*}
where $Z=(X,Y)$. Thus \begin{align*}
\sup_{x\in\mr} P\lbk (\psi_{x,\theta}-\psi_{x,\theta_0})^2 \rbk = \sup_{x\in\mr} \E_P \big( [g(Y,\theta)-g(Y,\theta_0)]^2 \indicator\{X\le x\} \big) \le \E_P \big( [g(Y,\theta)-g(Y,\theta_0)]^2 \big),
\end{align*}
and the desired result is implied by the condition in Lemma \ref{lemma:examples satisfy model assumptions}(i).

Now we show that Assumption \ref{ass:para g mapsto f(g) is continuous} holds in Examples \ref{exam:symmetry.prob}--\ref{exam:LST.prob}. It suffices to show that \begin{align}
\lim_{k\to\infty} \sup_{x\in\mr} P\lbk (\psi_{x,\theta_k}-\psi_{x,\theta_0})^2 \rbk=0 \label{eq:uniform convergence of psi}
\end{align}
for all sequences $\{\theta_k\}_{k=1}^\infty\subset \Theta$ with $\lim_{k\to\infty} \lvv \theta_k-\theta_0 \rvv_2=0$. \begin{enumerate}[label=(\roman*),nosep,start=2]
\item In Example \ref{exam:symmetry.prob}, \begin{align*}
\lbk \psi_{x,\theta_k}(Z)-\psi_{x,\theta_0}(Z)\rbk^2&=\lbk \indicator\{Z\le 2\theta_k-x\}-\indicator\{Z\le 2\theta_0-x\} \rbk^2\\
&=\indicator \{2(\theta_k\wedge\theta_{0})-x < Z\le 2(\theta_k\vee\theta_{0})-x\},
\end{align*}
and hence \begin{align*}
P\lbk (\psi_{x,\theta_k}-\psi_{x,\theta_0})^2 \rbk =\E_P\lp \lbk \psi_{x,\theta_k}(Z)-\psi_{x,\theta_0}(Z)\rbk^2 \rp=\lv G(2\theta_k-x)-G(2\theta_{0}-x) \rv.
\end{align*}
Define $G_*(x)=1-G(2\theta_0-x)$ and $G_k(x)=1-G(2\theta_k-x)$ for every $x\in\mr$ and $k\in\mathbb{Z}_+$.
By assumption, $G_*$ is continuous on $\mr$, and $\lim_{k\to \infty} | G_k(x)-G_*(x) |=0$ for every $x\in\mr$. By Lemma 2.11 of \citet{van1998asymptotic}, 
\begin{align*}
\lim_{k\to\infty} \sup_{x\in\mr} \lv G_k(x)-G_*(x) \rv=0,
\end{align*}
and the result in \eqref{eq:uniform convergence of psi} follows.

\item In Example \ref{exam:fit.prob}, \begin{align*}
P\lbk (\psi_{x,\theta_k}-\psi_{x,\theta_0})^2 \rbk &=\E_P\lp [\psi_{x,\theta_k}(Z)-\psi_{x,\theta_0}(Z)]^2 \rp \\
&=\E_P\lp [G_0(x,\theta_k)-G_0(x,\theta_{0})]^2 \rp=[G_0(x,\theta_k)-G_0(x,\theta_{0})]^2.
\end{align*}
Define $G_*(x)=G_0(x,\theta_{0})$ and $G_k(x)=G_0(x,\theta_k)$ for every $x\in\mr$ and $k\in\mathbb{Z}_+$.
By assumption, $G_*$ is continuous on $\mr$, and $\lim_{k\to \infty} | G_k(x)-G_*(x) |=0$ for every $x\in\mr$. The result in \eqref{eq:uniform convergence of psi} follows from Lemma 2.11 of \citet{van1998asymptotic}.

\item In Example \ref{exam:LST.prob} with $Z=(X,Y)$, \begin{align*}
\lbk \psi_{x,\theta_k}(Z)-\psi_{x,\theta_0}(Z) \rbk^2&=\lbk \indicator\lbr Y \le \frac{x-\theta_{0,1}}{\theta_{0,2}} \rbr-\indicator\lbr Y \le \frac{x-\theta_{k,1}}{\theta_{k,2}} \rbr \rbk^2 \\
&=\indicator \lbr \frac{x-\theta_{0,1}}{\theta_{0,2}} \wedge \frac{x-\theta_{k,1}}{\theta_{k,2}} < Y \le \frac{x-\theta_{0,1}}{\theta_{0,2}} \vee \frac{x-\theta_{k,1}}{\theta_{k,2}} \rbr,
\end{align*}
and hence \begin{align*}
P\lbk (\psi_{x,\theta_k}-\psi_{x,\theta_0})^2 \rbk =\E_P\lp \lbk \psi_{x,\theta_k}(Z)-\psi_{x,\theta_0}(Z)\rbk^2 \rp= \lv G \lp \frac{x-\theta_{k,1}}{\theta_{k,2}} \rp - G \lp \frac{x-\theta_{0,1}}{\theta_{0,2}} \rp \rv.
\end{align*}
Define $G_*(x)=G[(x-\theta_{0,1})/\theta_{0,2}]$ and $G_k(x)=G[(x-\theta_{k,1})/\theta_{k,2}]$ for every $x\in\mr$ and $k\in\mathbb{Z}_+$.
By assumption, $G_*$ is continuous on $\mr$, and $\lim_{k\to \infty} | G_k(x)-G_*(x) |=0$ for every $x\in\mr$. The result in \eqref{eq:uniform convergence of psi} follows from Lemma 2.11 of \citet{van1998asymptotic}.
\end{enumerate}
\end{pf}

\begin{pf}{ of Lemma \ref{lemma:examples are Donsker}}
Example \ref{exam:CMR.est}: Condition (2) implies that for every $y\in\mr^{d_y}$, the function $\theta\mapsto g(y,\theta)$ is continuous in $\theta$. Combining this with Condition (1) yields $\sup_{\theta\in\Theta}|g(y,\theta)|<\infty$ for all $y\in\mr^{d_y}$. By Condition (3), $\sup_{\theta\in\Theta} \E_P[|g(Y,\theta)|]<\infty$.  Define $\cF_1=\{g(\cdot,\theta):\theta\in\Theta\}$ and $\cF_2=\{\indicator_{(-\infty,x]}:x\in\mr\}$. For every $z=(z_1,z_2)\in \mr\times\mr^{d_y}$ and $(x_1,\theta_1), (x_2,\theta_2)\in\mr\times\Theta$, 
\begin{align*}
&\lv g(z_2,\theta_1)\indicator\{z_1\le x_1\}-g(z_2,\theta_2)\indicator\{z_1\le x_2\}\rv\\
=&\,\lv g(z_2,\theta_1)[\indicator\{z_1\le x_1\}-\indicator\{z_1\le x_2\}]+[g(z_2,\theta_1)-g(z_2,\theta_2)]\indicator\{z_1\le x_2\} \rv \\
\le&\, \lv g(z_2,\theta_1)\rv \lv \indicator\{z_1\le x_1\}-\indicator\{z_1\le x_2\} \rv + \lv g(z_2,\theta_1)-g(z_2,\theta_2) \rv \indicator\{z_1\le x_2\} \\
\le&\, \gbar(z_2) \lv \indicator\{z_1\le x_1\}-\indicator\{z_1\le x_2\} \rv+\lv g(z_2,\theta_1)-g(z_2,\theta_2) \rv,
\end{align*}
where the first inequality follows from the triangle inequality and the second inequality is implied by the definition of $\gbar$. 
Then it follows that 
\begin{align*}
&\lv g(z_2,\theta_1)\indicator\{z_1\le x_1\}-g(z_2,\theta_2)\indicator\{z_1\le x_2\}\rv^2\\
\le&\, 2\lv g(z_2,\theta_1)-g(z_2,\theta_2) \rv^2 + 2\gbar(z_2)^2 \lv \indicator\{z_1\le x_1\}-\indicator\{z_1\le x_2\} \rv^2.
\end{align*}
Thus, Condition (2.10.12) of \citet{van1996weak} is satisfied with $L_{\alpha,1}(z)=\sqrt{2}$ and $L_{\alpha,2}(z)=\sqrt{2} \gbar(z_2)$ for every $z=(z_1,z_2)\in \mr\times\mr^{d_y}$. By Conditions (1) and (2) in this lemma and Example 19.7 of \citet{van1998asymptotic}, the class $\cF_1$ is $P$-Donsker, and hence $L_{\alpha,1}\cF_1=\{\sqrt{2}g(\cdot,\theta):\theta\in\Theta \}$ is also Donsker. By Condition (2), the function $\theta\mapsto \E_P [\sqrt{2}g(Y,\theta)]$ is Lipschitz continuous on $\Theta$. By Condition (1), \begin{align*}
\sup_{f\in L_{\alpha,1}\cF_1} |P(f)|=\sup_{\theta\in\Theta} | \E_P [\sqrt{2}g(Y,\theta)] |<\infty.
\end{align*}
By Example 2.6.1 of \citet{van1996weak} and Lemma 9.8 of \citet{kosorok2008introduction}, the class $\cF_2$ is VC-subgraph, where $\cF_2$ can be seen as a class of indicator functions $\indicator_{(-\infty,x]\times\mathbb{R}^{d_y}}$. Since $L_{\alpha,2}$ is a fixed function, the class $L_{\alpha,2}\cF_2=\{\sqrt{2}\gbar\indicator_{(-\infty,x]}:x\in\mr\}$ is VC-subgraph by Lemma 2.6.18(vi) of \citet{van1996weak}. Clearly, $\sqrt{2}\gbar$ is an envelope function of $L_{\alpha,2}\cF_2$ and square integrable with respect to $P$ by Condition (3). By Theorem 2.5.2 of \citet{van1996weak}, the class $L_{\alpha,2}\cF_2$ is $P$-Donsker. Moreover, $\sup_{f\in L_{\alpha,2}\cF_2} |P(f)|<\infty$. Under the conditions in this lemma, every function in the class $\Psi=\{g(\cdot,\theta)\indicator_{(-\infty,x]}:(x,\theta)\in\mr\times\Theta\}$ is square integrable with respect to $P$. By Corollary 2.10.13 of \citet{van1996weak}, the class $\Psi$ is $P$-Donsker.

Example \ref{exam:symmetry.prob}: Clearly, $\{\indicator_{(-\infty,2\theta-x]}:(x,\theta)\in\mr\times\Theta\}\subset \{\indicator_{(-\infty,x]}:x\in\mr\}$. By Example 2.5.4 of \citet{van1996weak}, the class $\{\indicator_{(-\infty,x]}:x\in\mr\}$ is $P$-Donsker. Since $\sup_{x\in\mr} |P(\indicator_{(-\infty,x]})|\le 1$, the class $\Psi$ is $P$-Donsker by Theorem 2.10.1 and Example 2.10.7 of \citet{van1996weak}.

Example \ref{exam:fit.prob}: Note that the class $\{ G_0(x,\theta):(x,\theta)\in\mr\times\Theta \}$ consists of bounded constant functions, and thus it is trivially $P$-Donsker. Furthermore, $\sup_{(x,\theta)\in\mr\times\Theta}|P[G_0(x,\theta)]|\le 1$, $\sup_{x\in\mr} |P(\indicator_{(-\infty,x]})|\le 1$, and the class $\{\indicator_{(-\infty,x]}:x\in\mr\}$ is $P$-Donsker. By Example 2.10.7 and Theorem 2.10.1 of \citet{van1996weak}, the class $\Psi$ is $P$-Donsker.

Example \ref{exam:LST.prob}: Note that $\{\indicator_{(-\infty,(x-\theta_1)/\theta_2 ]}:(x,\theta)\in\mr\times\Theta\}\subset \{\indicator_{(-\infty,x]}:x\in\mr\}$, where $\theta=(\theta_1,\theta_2)$. Then the proof is analogous to that for Example \ref{exam:symmetry.prob}.
\end{pf}

\begin{pf}{ of Lemma \ref{lemma:examples satisfy second derivatives}}
Example \ref{exam:CMR.prob}: Recall that $\psi_{x,\theta}(Z)=g(Y,\theta)\indicator\{X\le x\}$. Under the conditions for Example \ref{exam:CMR.prob} in Lemma \ref{lemma:examples satisfy second derivatives}, both $\psi_{x,\theta}(Z)$ and $\partial \psi_{x,\theta}(Z)/ \partial \theta$ satisfy the conditions of Theorem A.5.1 of \citet{durrett2019probability}. Applying this theorem twice yields that \begin{align*}
\frac{\partial^2 \phi_P(x,\theta)}{\partial\theta\partial\theta^\T}=\frac{\partial^2 \E_P[g(Y,\theta)\indicator\{X\le x\}]}{\partial\theta\partial\theta^\T}=\E_P \lbk \frac{\partial^2 g(Y,\theta)}{\partial\theta\partial\theta^\T} \indicator\{X\le x\} \rbk.
\end{align*}
Furthermore, for all $(x,\theta)\in\mr\times\Theta$, \begin{align*}
\lvv \frac{\partial^2 \phi_P(x,\theta)}{\partial\theta\partial\theta^\T} \rvv_2 \le \E_P \lbk \lvv \frac{\partial^2 g(Y,\theta)}{\partial\theta\partial\theta^\T} \rvv_2 \rbk,
\end{align*}
and the result follows from Conditions (1) and (4).

Example \ref{exam:symmetry.prob}: Under the conditions, \begin{align*}
\frac{\partial^2 \phi_P(x,\theta)}{\partial\theta^2}=4G''(2\theta-x),
\end{align*}
and thus, \begin{align*}
\int_{\mr} \sup_{\theta\in\Theta} \lvv \frac{\partial^2 \phi_P(x,\theta)}{\partial\theta^2} \rvv_2^2 \ddd \nu(x)= \int_{\mr} \sup_{\theta\in\Theta} 16 \lv G''(2\theta-x) \rv^2 \ddd \nu(x) \le 16 \lp \sup_{x\in\mr} \lv G''(x) \rv \rp^2 < \infty.
\end{align*}

Example \ref{exam:fit.prob}: Under the conditions, for every $x$,
\begin{align*}
\frac{\partial^2 \phi_P(x,\theta)}{\partial\theta\partial \theta^\T}=-\frac{\partial^2 G_0 (x,\theta)}{\partial \theta\partial \theta^\T},
\end{align*}
and the desired result follows from the conditions in the lemma.

Example \ref{exam:LST.prob}: Under Condition (3), \begin{align*}
\frac{\partial^2 \phi_P(x,\theta)}{\partial\theta_1^2}&=-\frac{1}{\theta_2^2}G''\lp \frac{x-\theta_1}{\theta_2} \rp, \\
\frac{\partial^2 \phi_P(x,\theta)}{\partial\theta_1\partial\theta_2}&=-G''\lp \frac{x-\theta_1}{\theta_2} \rp \frac{x-\theta_1}{\theta_2^3}-G'\lp \frac{x-\theta_1}{\theta_2} \rp\frac{1}{\theta_2^2}, \text{ and }\\
\frac{\partial^2 \phi_P(x,\theta)}{\partial\theta_2^2}&=-G''\lp \frac{x-\theta_1}{\theta_2}\rp\frac{(x-\theta_1)^2}{\theta_2^4}-2G'\lp \frac{x-\theta_1}{\theta_2} \rp \frac{x-\theta_1}{\theta_2^3}.
\end{align*}
Conditions (1) and (3) imply that there exists an $M>0$, such that for all $(x,\theta)$, \begin{align*}
\lv \frac{\partial^2 \phi_P(x,\theta)}{\partial\theta_1^2} \rv \le  M, \lv \frac{\partial^2 \phi_P(x,\theta)}{\partial\theta_1\partial\theta_2} \rv \le M|x|+M, \text{ and } \lv \frac{\partial^2 \phi_P(x,\theta)}{\partial\theta_2^2} \rv \le Mx^2+M|x|+M.
\end{align*}
Since the Frobenius norm of a symmetric matrix dominates its spectral norm ($\ell_2$ operator norm), the above inequalities imply that there is some $C>0$ such that for all $x\in\mr$, \begin{align*}
\sup_{\theta\in\Theta} \lvv \frac{\partial^2 \phi_P(x,\theta)}{\partial\theta\partial\theta^\T} \rvv_2^2 \le C x^4+C|x|^3+C x^2+C|x|+C,
\end{align*}
and the desired results follow from Condition (2).
\end{pf}

\begin{pf}{ of Lemma \ref{lemma:examples satisfy identifiability}}
As shown in the proof of Proposition \ref{prop:para equivalent null}, under Assumptions \ref{ass:para properties of g}--\ref{ass:para g mapsto f(g) is continuous}, \begin{align*}
\inf_{\theta\in\Theta} \int_{\mr} [\phi_P(x,\theta)]^2 \ddd \nu(x)= \min_{\theta\in\Theta} \int_{\mr} [\phi_P(x,\theta)]^2 \ddd \nu(x).
\end{align*}

Consider the case where $\Theta_{0}=\varnothing$. It implies that $\varepsilonbar:=\inf_{\theta\in\Theta} \{\int_{\mr} [\phi_P(x,\theta)]^2 \ddd \nu(x)\}^{1/2}>0$. By definition, $\Theta_{0}^\varepsilon=\varnothing$ for all $\varepsilon>0$. Let $\kappa=1$ and $C=1$, and then Assumption \ref{ass:para strong identifiability} holds.

Now consider the case where $\Theta_{0}\ne\varnothing$ for Examples \ref{exam:symmetry.prob}--\ref{exam:LST.prob}. Let $G_0$ be defined as in this lemma. Under the conditions in Examples \ref{exam:symmetry.prob}--\ref{exam:LST.prob}, the parameter $\theta$ is identified by $G_0$ in the sense that for all $\theta,\theta'\in\Theta$ with $\theta\ne\theta'$, there exists $x_0\in\mr$ such that $G_0(x_0,\theta)\ne G_0(x_0,\theta')$. By Proposition \ref{prop:para equivalent null}, $\Theta_{0}\ne\varnothing$ is equivalent to the case where there exists some $\theta_0\in\Theta$ such that $\phi_P(x,\theta_0)=0$ for all $x\in\mr$. The identifiability of $\theta$ implies that such a $\theta_0$ is unique and thus $\Theta_{0}=\{\theta_{0}\}$. Note that for Examples \ref{exam:symmetry.prob}--\ref{exam:LST.prob}, \begin{align*}
\inf_{\theta\in\Theta\setminus\Theta_0^\varepsilon}	\int_\mr \lbk \phi_P(x,\theta) \rbk^2 \ddd \nu(x)&=\inf_{\theta\in\Theta: \lvv \theta-\theta_0 \rvv_2>\varepsilon} \int_\mr \lbk \phi_P(x,\theta) \rbk^2 \ddd \nu(x) \\
&=\inf_{\theta\in\Theta: \lvv \theta-\theta_0 \rvv_2>\varepsilon} \int_\mr \lbk \phi_P(x,\theta)-\phi_P(x,\theta_0) \rbk^2 \ddd \nu(x) \\
&=\inf_{\theta\in\Theta: \lvv \theta-\theta_0 \rvv_2>\varepsilon} \int_\mr \lbk G_0(x,\theta)-G_0(x,\theta_0) \rbk^2 \ddd \nu(x),
\end{align*}
and Assumption \ref{ass:para strong identifiability} holds under the conditions of the lemma.
\end{pf}

\subsection{Proofs for Appendix \ref{sec:multiple CDFs}}

\begin{lemma}\label{lemma:properties of DL multi}
For every $k\in\{1,\ldots,K\}$, if $\varphi_1,\varphi_2\in \bD_{\mL k}$, then $a_1\varphi_1+a_2\varphi_2\in
\bD_{\mL k}$ for all $a_1, a_2\in \mr$, and
the functions \begin{align*}
\theta_k \mapsto \int_{\mr} \lbk \varphi_1(x,\theta_k) \rbk^2 \ddd \nu(x) 
\text{ and } 
\theta_k \mapsto \int_{\mr} \varphi_1(x,\theta_k)\varphi_2(x,\theta_k) \ddd \nu(x)
\end{align*}
are continuous at every $\theta_k\in\Theta_k$.
\end{lemma}

\begin{pf}{ of Lemma \ref{lemma:properties of DL multi}}
The proof is similar to that of Lemma \ref{lemma:para properties of DL}.
\end{pf}

\begin{pf}{ of Proposition \ref{prop:equivalent null multi}}
If $F(x)=G_k\lp g_k (x,\theta_k) \rp$ for all $x\in\mr$ with some
$\theta_k\in \Theta_k$ for all $k\in\{1,\ldots,K\}$, then \eqref{eq:working null multi} holds trivially.

Next, we show that \eqref{eq:working null multi} implies \eqref{eq:null multiple 1}.
Recall that $\mu$ is the Lebesgue measure on $\lp \mr, \borel_\mr \rp$.
Since $G_k\in \mCb (\mr)$, Assumption \ref{ass:g mapsto f(g) is continuous multi}
implies that $G_k\circ g_k\in \bD_{\mL k}$ and hence $\phi_k \in \bD_{\mL k}$.
By Lemma \ref{lemma:properties of DL multi}, the function $\theta_k
\mapsto \int_\mr \lbk F(x)-G_k\lp g_k (x,\theta_k) \rp \rbk^2 \ddd \nu(x)$
is continuous on $\Theta_k$. Thus, the function $(\theta_1,\ldots,\theta_K)
\mapsto \int_\mr\sum_{k=1}^K \lbk F(x)-G_k\lp g_k (x,\theta_k) \rp \rbk^2 \ddd \nu(x)$
is continuous on $\Theta$. By Assumption \ref{ass:Theta is compact multi},
there exists $\theta_{0}\in\Theta$ with $\theta_0=(\theta_{01},\ldots,\theta_{0K})$ such that 
\begin{align}\label{eq.integral nu equals 0 multi}
\int_{\mr} \sum_{k=1}^K\lbk F(x)-G_k\lp g_k (x,\theta_{0k}) \rp \rbk^2 \ddd \nu(x)=\inf_{(\theta_1,\ldots,\theta_K)\in\Theta}\int_{\mr}\sum_{k=1}^K \lbk F(x)-G_k\lp g_k (x,\theta_k) \rp \rbk^2
\ddd \nu(x)=0.
\end{align}
Define $A_k=\lbr x\in \mr: F(x)\ne G_k \lp g_{k} (x,\theta_{0k}) \rp \rbr$ for every $k\in\{1,\ldots,K\}$. Then \eqref{eq.integral nu equals 0 multi} implies that $\nu(A_k)=0$ by Proposition 2.16 of \citet{folland2013real}. By the assumption that $\mu \ll \nu$, $\mu (A_k)=0$.
We now claim that $A_k=\varnothing$. Otherwise, there is an $x_0\in \mr$ such that
$F\lp x_0 \rp \ne G_k\lp g_{k} \lp x_0,\theta_{0k} \rp \rp$. Since both $F$ and $G_k$ are
continuous and $g_{k}(\cdot,\theta_{0k})$ is continuous, there exists $\delta>0$ such that $F\lp x \rp \ne G_k\lp g_{k} \lp x,\theta_{0k} \rp \rp$ for all $x\in \lbk x_0,
x_0+\delta \rbk$. This contradicts $\mu(A_k)=0$.
Therefore, we have $F(x)=G_k\lp g_{k}(x,\theta_{0k}) \rp$ for all $x\in \mr$ and all $k$.
\end{pf}

\begin{lemma}\label{prop:G-C of phihat multi}
Under Assumptions \ref{ass:samples multi} and \ref{ass:ratio lambda multi},
we have \begin{align*}
\lim_{n\to\infty} \sup_{(x,\theta)\in\mr\times \Theta}
\left\Vert \phihat ( x, \theta )-\phi (x, \theta) \right\Vert_2=0 \text{ almost surely}.
\end{align*}
\end{lemma}

\begin{pf}{ of Lemma \ref{prop:G-C of phihat multi}}
By Theorem 19.1 of \citet{van1998asymptotic} and Assumption \ref{ass:ratio lambda multi}, we have \begin{align*}
&\lim_{n\to\infty} \sup_{x\in \mr} | \widehat{F}_{n_x} (x)-F(x) |=0 \text{ almost surely},\\ 
&\text{and }
\lim_{n\to\infty} \sup_{x\in \mr} | \widehat{G}_{n_k} (x)-G_{k}(x) |=0
\text{ almost surely for every } k.	
\end{align*}
Note that for every $(x,\theta_k)\in \mr\times \Theta_k$, 
\begin{align*}
\lv \widehat{G}_{n_k} \lp g_k (x,\theta_k) \rp-G_{k}\lp g_k (x, \theta_k) \rp \rv \le
\sup_{z\in \mr} \lv \widehat{G}_{n_k}(z) -G_{k}(z) \rv,
\end{align*}
which implies \begin{align*}
\lim_{n\to\infty} \sup_{(x,\theta_k)\in\mr\times \Theta_k}
\lv \widehat{G}_{n_k} \lp g_k (x,\theta_k) \rp-G_{k}\lp g_k (x,\theta_k) \rp \rv=0 \text{ almost surely}.
\end{align*}
Then the desired result follows from the definitions of $\phihat$
and $\phi$.
\end{pf}

\begin{pf}{ of Lemma \ref{prop:Donsker of phihat multi}}
By Theorem 19.3 of	\citet{van1998asymptotic}, we have 
\begin{align*}
\sqrt{n_x}( \widehat{F}_{n_x}-F ) \convd \bW_F \text{ in } \ell^\infty (\mr), \text{ and for all $k\in\{1,\ldots,K\}$}, \sqrt{n_k}( \widehat{G}_{n_k}-G_k ) \convd \bW_{G_k} \text{ in } \ell^\infty (\mr)
\end{align*}
as $n\to \infty$, where $\mathbb{W}_F,\mathbb{W}_{G_1},\ldots,\mathbb{W}_{G_K}$ are jointly independent.
Define classes of indicator functions 
\begin{align*}
\mathcal{G}_0=\lbr \indicator_{(-\infty,x]}: x\in \mr\rbr \text{ and }			\mathcal{G}_k=\lbr \indicator_{\lp -\infty, g_k (x,\theta_k) \rbk }:	(x,\theta_k)\in \mr\times\Theta_k \rbr \text{ for all } k.
\end{align*}
Let $\widehat{\mathcal{Y}}_{n_k}$ be a stochastic process and
$\cy_k$ be a real-valued function such that 
\begin{align*}
\widehat{\mathcal{Y}}_{n_k} (f)=\frac{1}{n_k}\sum_{i=1}^{n_k} f \lp Y_{ki} \rp 
\text{ and }
\cy_k(f)=\E \lbk f\lp Y_{ki} \rp \rbk
\end{align*}
for all measurable $f$.
By Example 2.5.4 of \citet{van1996weak}, $\mathcal{G}_0$ is a
Donsker class. 
Therefore, $\sqrt{n_k}( \widehat{\mathcal{Y}}_{n_k}-\cy_k ) \convd \by_k$ in $\ell^\infty\lp \mathcal{G}_0 \rp$
as $n\to\infty$, where $\by_k$ is a tight measurable centered Gaussian process. Since $\mathcal{G}_k\subset\mathcal{G}_0$, it follows that for every $h\in\mathcal{C}_\mathrm{b}(\ell^\infty\lp \mathcal{G}_k \rp)$, $h\in\mathcal{C}_\mathrm{b}(\ell^\infty\lp \mathcal{G}_0 \rp)$ and 
\begin{align*}
\mathbb{E}[h(\sqrt{n_k}( \widehat{\mathcal{Y}}_{n_k}-\cy_k ))]\to \mathbb{E}[h(\by_k)],
\end{align*}
which implies that $\sqrt{n_k}( \widehat{\mathcal{Y}}_{n_k}-\cy_k ) \convd \by_k$ in $\ell^\infty\lp \mathcal{G}_k \rp$
as $n\to\infty$.
It is easy to show that
$ \widehat{G}_{n_k}\circ g_k  (x,\theta_k)=\widehat{\mathcal{Y}}_{n_k}( \indicator_{\lp -\infty, g_k (x,\theta_k) \rbk } )$
and $G_k\circ g_k (x,\theta_k)=\cy_k ( \indicator_{\lp -\infty, g_k (x,\theta_k) \rbk } )$
for every $(x,\theta_k)\in\mr\times\Theta_k$.
Define a random element $W_k\in\ell^{\infty}(\mathbb{R}\times\Theta_k)$ such that $W_k(x,\theta_k)=\mathbb{Y}_k(\indicator_{\lp -\infty, g_k(x,\theta_k) \rbk })$ for all $(x,\theta_k)\in\mathbb{R}\times\Theta_k$. 
By Lemma \ref{lemma:weak convergence with transformed index},
$\sqrt{n_k}( \widehat{G}_{n_k}\circ g_k -G_k\circ g_k ) \convd W_k$ in $\ell^\infty
(\mr\times\Theta_k)$ as $n\to\infty$. Let $\lambda_{-x}=\prod_{k=1}^K\lambda_k$ and $\lambda_{-k}=(\lambda_x\cdot\prod_{j=1}^K\lambda_j)/\lambda_k$.
By the joint independence of the samples, Assumption \ref{ass:ratio lambda multi}
of this paper, and Example 1.4.6 of \citet{van1996weak}, we have the joint weak convergence \begin{align*}
\begin{bmatrix}
\sqrt{T_n}\lp \widehat{F}_{n_x}-F \rp \\ \sqrt{T_n}\lp \widehat{G}_{n_1}\circ g_1 -G_1\circ g_1\rp\\
\vdots\\
\sqrt{T_n}\lp \widehat{G}_{n_K}\circ g_K -G_K\circ g_K\rp
\end{bmatrix} \convd 
\begin{bmatrix}
\sqrt{\lambda_{-x}} \bW_F \\ \sqrt{\lambda_{-1}} W_1 \\ \vdots \\ \sqrt{\lambda_{-K}} W_K
\end{bmatrix}  \text{ in } \ell^\infty(\mr)\times \ell^\infty(\mr\times\Theta_1) \times \cdots \times \ell^\infty(\mr\times\Theta_K)
\end{align*}
as $n \to \infty$, where $\bW_F, W_1,\ldots, W_K$ are jointly independent.
Define
\begin{align*}
\mathbb{A}= \ell^\infty(\mr)\times \ell^\infty(\mr\times\Theta_1) \times \cdots \times \ell^\infty(\mr\times\Theta_K) \text{ and }
\mathbb{B}=  \ell^\infty(\mr\times\Theta_1) \times \cdots \times \ell^\infty(\mr\times\Theta_K).
\end{align*} 
Define the norms $\Vert\cdot\Vert_{\mathbb{A}}$ and $\Vert\cdot\Vert_{\mathbb{B}}$ on $\mathbb{A}$ and $\mathbb{B}$, respectively, with $\lvv \lp f,h_1,\ldots,h_K \rp \rvv_\mathbb{A}=
\lvv f \rvv_{\infty}+\sum_{k=1}^K\lvv h_k \rvv_{\infty}$ for every $\lp f,h_1,\ldots,h_K \rp \in \mathbb{A}$ and $\lvv \lp h_1,\ldots,h_K \rp \rvv_\mathbb{B}=
\sum_{k=1}^K\lvv h_k \rvv_{\infty}$ for every \linebreak $\lp h_1,\ldots,h_K \rp \in \mathbb{B}$.
Let $\mathcal{I}: \mathbb{A}\to \mathbb{B}$ be such
that
\begin{align*}
\mathcal{I}\lp f, h_1,\ldots,h_K \rp = (f -h_1 ,\ldots,f-h_K)
\end{align*}
for every $\lp f, h_1,\ldots,h_K \rp\in \mathbb{A}$. Note that 
\begin{align*}
&\lvv \mathcal{I}\lp f^{\prime}, h_1^{\prime},\ldots,h_K^{\prime} \rp-\mathcal{I} \lp f, h_1,\ldots,h_K \rp \rvv_\mathbb{B}=\sum_{k=1}^K
\sup_{(x,\theta_k)\in\mr\times \Theta_k} \lv f^{\prime}(x)-h_k^{\prime}(x,\theta_k)-f(x)+h_k(x,\theta_k) \rv \\
&\le\, K\sup_{x\in\mr} \lv f^{\prime}(x)-f(x) \rv+
\sum_{k=1}^K\sup_{(x,\theta_k)\in\mr\times \Theta_k} \lv h_k^{\prime}(x,\theta_k)-h_k(x,\theta_k) \rv 
\end{align*}
for all $\lp f^{\prime}, h_1^{\prime},\ldots,h_K^{\prime} \rp, \lp f, h_1,\ldots,h_K \rp\in \mathbb{A}$,
and therefore $\mathcal{I}$ is continuous.
The weak convergence of $\sqrt{T_n} ( \phihat-\phi ) $ to a tight random element $\mathbb{G}_0 = \mathcal{I}(\sqrt{\lambda_{-x}}\bW_F, \sqrt{\lambda_{-1}}W_1,\ldots,\sqrt{\lambda_{-K}} W_K)$
follows from Theorem 1.3.6 (continuous mapping) of \citet{van1996weak}.	
Furthermore, by the proof similar to that of Lemma \ref{prop:para Donsker of phihat}, $\p (\bG_0\in \bD_{\mL0} )=1$.
\end{pf}	

\begin{pf}{ of Lemma \ref{lemma:multi-HDDL}}
Define a map $\mathcal{S}:\bD_{\mL}\to \ell^{\infty}(\Theta)$ such that for every $\varphi\in \bD_{\mL}$
and every $\theta\in\Theta$ with $\varphi=(\varphi_1,\ldots,\varphi_K)$ and $\theta=(\theta_1,\ldots,\theta_K)$, 
\begin{align*}
\mathcal{S}(\varphi)(\theta)=\int_\mr\sum_{k=1}^K \lbk \varphi_k(x,\theta_k) \rbk^2 \ddd \nu(x).
\end{align*}
We show that the Hadamard directional derivative of $\mathcal{S}$ at
$\phi\in \bD_{\mL}$ is 
\begin{align*}
\mathcal{S}'_\phi (h)(\theta)=\int_\mr 2\sum_{k=1}^K\phi_k(x,\theta_k)h_k(x,\theta_k)\ddd \nu(x)\text{ for all } h\in \bD_{\mL0} \text{ with } h=(h_1,\ldots,h_K).
\end{align*}
Because $F,G_k\in\mathcal{C}_{\mathrm{b}}(\mathbb{R})$, by Assumption \ref{ass:g mapsto f(g) is continuous multi} and Lemma \ref{lemma:properties of DL multi}, $\mathcal{S}(\phi)\in \mathcal{C}(\Theta)$. Indeed, for all sequences $\lbr h_n \rbr_{n=1}^\infty \subset \prod_{k=1}^K\ell^\infty(\mr\times\Theta_k)$ with $h_n=(h_{n1},\ldots,h_{nK})$
and $\lbr t_n \rbr_{n=1}^\infty \subset \mr_+$ such that $t_n \downarrow 0$,
$h_n\to h\in \bD_{\mL0}$ as $n\to \infty$ with $h=(h_1,\ldots,h_K)$, and $\phi+t_nh_n \in \bD_{\mL}$ for all $n$,
we have that $M=\max_{k\in\{1,\ldots,K\}}\sup_{n\in\mathbb{Z}_+} \lvv h_{nk} \rvv_\infty<\infty$, and
\begin{align*}
&\phantom{=\:\,}	\sup_{\theta\in\Theta} \lv \frac{\mS\lp \phi+t_n h_n \rp(\theta)
-\mS (\phi)(\theta)}{t_n}-\mS'_\phi (h)(\theta) \rv \\
&= \sup_{\theta\in\Theta} \lv \sum_{k=1}^K \int_\mr t_n h_{nk}^2(x,\theta_k)
+2\phi_k(x,\theta_k)\lbk h_{nk}(x,\theta_k)-h_k(x,\theta_k) \rbk
\ddd \nu(x) \rv \\
& \le \sum_{k=1}^K\int_\mr  t_n M^2 + 2 \lvv \phi_k \rvv_{\infty} \lvv h_{nk}- h_k \rvv_{\infty} \ddd \nu(x)  \to 0,
\end{align*}
since $t_n\downarrow 0$
and $h_n\to h$ in $\prod_{k=1}^K\ell^\infty(\mr\times\Theta_k)$ as $n\to \infty$.

Define a function $\mR$ such that for every $\psi\in \ell^{\infty}(\Theta)$,
$\mR(\psi)=\inf_{\theta\in \Theta} \psi(\theta)$. By Lemma S.4.9
of \citet{fang2019inference}, $\mR$ is Hadamard directionally differentiable
at every $\psi\in \mC(\Theta)$ tangentially to $\mC(\Theta)$ with the Hadamard
directional derivative \begin{align*}
\mR'_\psi (f)=\inf_{\theta\in \Theta^*_0(\psi)} f(\theta) \text{ for all } f\in \mC(\Theta),
\end{align*}
where $ \Theta^*_0(\psi)=\argmin_{\theta\in\Theta} \psi(\theta)$.

Note that $\mL(\varphi)=\mR \lbk \mS (\varphi) \rbk=\mR \circ \mS (\varphi)$ for
every $\varphi \in \bD_{\mL}$.
By Proposition 3.6(i) of \citet{shapiro1990concepts}, $\mL$ is Hadamard
directionally differentiable at $\phi$ tangentially to
$\bD_{\mL0}$ with the Hadamard directional derivative 
\begin{align*}
&\mL'_\phi (h)= \mR'_{\mS(\phi)} \lbk \mS'_\phi(h) \rbk
=\inf_{\theta\in\Theta^*_0 (\mS(\phi))} \int_\mr 2\sum_{k=1}^K\phi_k(x,\theta_k)h_k(x,\theta_k)\ddd \nu(x)\text{ for all } h\in \bD_{\mL0} \\ &\text{with } h=(h_1,\ldots,h_K).
\end{align*}
Since $ \Theta^*_0 (\mS(\phi))=\argmin_{\theta\in\Theta} \int_{\mr} \sum_{k=1}^K\lbk \phi_k (x,\theta_k) \rbk^2
\ddd \nu(x)$, the desired result follows.

Now we turn to the degeneracy of $\mL'_\phi$ under the condition that $\phi\in\bD_0$.
If $\phi \in \bD_0$, for every $\theta\in \Theta_0(\phi)$ with $\theta=(\theta_1,\ldots,\theta_K)$,
we have \begin{align*}
\int_\mr \sum_{k=1}^K \lbk \phi_k(x,\theta_k) \rbk^2 \ddd \nu(x)=0,
\end{align*}
and consequently $\phi_k(x,\theta_k)=0$ holds for $\nu$-almost every $x$ and every $k$.
Therefore, $\mL'_\phi(h)=0$ for every $h\in \prod_{k=1}^K\ell^\infty(\mr\times\Theta_k)$
whenever $\phi\in \bD_0$.
\end{pf}

\begin{pf}{ of Lemma \ref{prop:second order Hadamard of L multi}}
This proof extends that of Lemma \ref{prop:para second order Hadamard of L} with more complications.
For every $k$, define $\Phi_k:\Theta_k\to \lsv$ such that
$\Phi_k(\theta_k)(x)=\phi_k(x,\theta_k)$ for every $(x,\theta_k)\in\mr\times\Theta_k$. Define $\Phi:\Theta\to\prod_{k=1}^K L^2(\nu)$ such that for every $\theta\in\Theta$ with $\theta=(\theta_1,\ldots,\theta_K)$, $\Phi(\theta)=(\Phi_1(\theta_1),\ldots,\Phi_K(\theta_K))$.
Then it is easy to show that
\begin{align*}
\mL(\phi)=\inf_{\theta\in\Theta}\int_\mr\sum_{k=1}^K \lbk \phi_k(x,\theta_k) \rbk^2 \ddd \nu(x)
=\inf_{\theta\in\Theta} \sum_{k=1}^K\lvv \Phi_k(\theta_k) \rvv^2_\lsv=\inf_{\theta\in\Theta} \lvv \Phi(\theta) \rvv^2_{L^2_K(\nu)}=0,
\end{align*}
and $\Theta_0(\phi)=\{ \theta\in\Theta: \sum_{k=1}^K\lvv \Phi_k(\theta_k) \rvv_\lsv^2=0 \}=\Theta_0$.
Consider all sequences $\lbr t_n \rbr_{n=1}^\infty\subset\mr_+$
and $\lbr h_n \rbr_{n=1}^\infty\subset\prod_{k=1}^K \ell^\infty(\mr\times \Theta_k)$
such that $t_n\downarrow 0$, $h_n \to h\in \bD_{\mL0}$
as $n\to\infty$, and $\phi+t_n h_n \in \bD_{\mL}$ for all $n$, where $h_{n}=(h_{n1},\ldots,h_{nK})$ and $h=(h_1,\ldots,h_K)\neq0$ (the case where $h=0$ is trivial).
For notational simplicity, for every $k$ and every $n$, define $\sh_{nk}:\Theta_k\to\lsv$
such that $\sh_{nk}(\theta_k)(x)=h_{nk} (x,\theta_k)$ for every $(x,\theta_k)\in\mr\times\Theta_k$,
and define $\sh_k:\Theta_k\to\lsv$ such that $\sh_k(\theta_k)(x)=h_k(x,\theta_k)$ for every
$(x,\theta_k)\in\mr\times\Theta_k$.
For every $\theta\in\Theta$ with $\theta=(\theta_1,\ldots,\theta_K)$, let $\mathscr{H}_n(\theta)=(\mathscr{H}_{n1}(\theta_1),\ldots,\mathscr{H}_{nK}(\theta_K))$ and $\mathscr{H}(\theta)=(\mathscr{H}_{1}(\theta_1),\ldots,\mathscr{H}_{K}(\theta_K))$.
Since $h_n\to h\in \bD_{\mL0}\subset\prod_{k=1}^K\ell^\infty(\mr\times\Theta_k)$, it follows that $\max_{k\in\{1,\ldots,K\}}
(\lvv h_{k} \rvv_{\infty}\vee \sup_{n\in\mathbb{Z}_+} \lvv h_{nk} \rvv_{\infty})
=M_1$ for some $M_1<\infty$. Then we have that
\begin{align*}
&  \lv \mL\lp \phi+t_n h_n \rp-\mL \lp\phi+t_n h \rp \rv 
=	\,\lv \inf_{\theta\in\Theta} \lvv \Phi(\theta)+t_n \sh_{n} (\theta) \rvv^2_{L_K^2(\nu)}
-\inf_{\theta\in\Theta} \lvv \Phi(\theta)+t_n \sh (\theta) \rvv^2_{L_K^2(\nu)} \rv \\
=&\,\lv \inf_{\theta\in\Theta} \lvv \Phi(\theta)+t_n \sh_n (\theta) \rvv_{L_K^2(\nu)}
+\inf_{\theta\in\Theta} \lvv \Phi(\theta)+t_n \sh (\theta) \rvv_{L_K^2(\nu)} \rv \\
& \cdot \lv \inf_{\theta\in\Theta} \lvv \Phi(\theta)+t_n \sh_n (\theta) \rvv_{L_K^2(\nu)}
-\inf_{\theta\in\Theta} \lvv \Phi(\theta)+t_n \sh (\theta) \rvv_{L_K^2(\nu)} \rv \\
\le&\, \lv \inf_{\theta\in\Theta_0(\phi)} \lvv \Phi(\theta)+t_n \sh_n (\theta) \rvv_{L_K^2(\nu)}
+\inf_{\theta\in\Theta_0(\phi)} \lvv \Phi(\theta)+t_n \sh (\theta) \rvv_{L_K^2(\nu)} \rv \\
& \cdot \lp t_n \sup_{\theta\in\Theta} \lvv \sh_n(\theta)-\sh(\theta) \rvv_{L_K^2(\nu)} \rp \\
=&\, O\left( t_n^2 \left\{\sum_{k=1}^K\lvv h_{nk}-h_k \rvv_\infty^2\right\}^{1/2}\right) = o\lp t_n^2 \rp,
\end{align*}
where the inequality follows from the Lipschitz continuity of the supremum
map and the triangle inequality, and the third equality follows from
the fact that $\Phi\lp \theta \rp=0$ $\nu$-almost everywhere for every $\theta\in\Theta_0(\phi)$.

Then for the $h$, pick an $a(h)>0$ such that $C a(h)^\kappa = 3({ \sum_{k=1}^K\lvv h_k \rvv_{\infty}^2})^{1/2}$, where
$C$ and $\kappa$ are defined as in Assumption \ref{ass:strong identifiability multi}. 
For sufficiently large $n\in\mathbb{Z}_+$ such that $t_n^\kappa \ge t_n$, we have that 
\begin{align}\label{eq.second order 1 multi}
& \inf_{\theta\in\Theta\setminus\Theta_0(\phi)^{a(h)t_n}}
\lvv \Phi(\theta)+t_n \sh(\theta) \rvv_{L_K^2(\nu)} \notag\\
\ge&\, \inf_{\theta\in\Theta\setminus\Theta_0(\phi)^{a(h)t_n}} \lvv \Phi(\theta)\rvv_{L_K^2(\nu)}
+\inf_{\theta\in\Theta\setminus\Theta_0(\phi)^{a(h)t_n}} \lbk -t_n \lvv \sh(\theta) 
\rvv_{L_K^2(\nu)} \rbk \notag\\
=&\, \inf_{\theta\in\Theta\setminus\Theta_0(\phi)^{a(h)t_n}} \lvv \Phi(\theta)\rvv_{L_K^2(\nu)}
-\sup_{\theta\in\Theta\setminus\Theta_0(\phi)^{a(h)t_n}} t_n \lvv \sh(\theta) 
\rvv_{L_K^2(\nu)} \notag \\
\ge&\, C \lp a(h) t_n \rp^\kappa -t_n \sup_{\theta\in\Theta\setminus\Theta_0(\phi)^{a(h)t_n}}
\lvv \sh(\theta) \rvv_{L_K^2(\nu)} \ge 3\left({ \sum_{k=1}^K\lvv h_k \rvv_{\infty}^2}\right)^{1/2} t_n^\kappa- t_n\left({ \sum_{k=1}^K\lvv h_k \rvv_{\infty}^2}\right)^{1/2} \notag\\
>&\, t_n \inf_{\theta\in\Theta_0(\phi)} \lvv \sh(\theta) \rvv_{L_K^2(\nu)}  = \inf_{\theta\in\Theta_0(\phi)} \lvv \Phi(\theta)+t_n \sh(\theta) \rvv_{L_K^2(\nu)} \ge \sqrt{\mL\lp \phi+t_n h \rp},
\end{align}
where the second inequality follows from Assumption \ref{ass:strong identifiability multi}.

By Lemma \ref{lemma:properties of DL multi}
and the fact that $\phi\in\bD_{\mL0}$ and $h\in\bD_{\mL0}$, the map
$\theta\mapsto \lvv \Phi(\theta)+t_n \sh(\theta) \rvv_{L_K^2(\nu)}^2$ is continuous
at every $\theta\in\Theta$ for every $n\in\mathbb{Z}_+$.
Since $\Theta$ and $\Theta_0(\phi)^{a(h)t_n}$
are compact sets in $\prod_{k=1}^K\mathbb{R}^{d_{\theta_k}}$, it follows that \begin{align*}
&\mL\lp \phi+t_n h \rp =\min_{\theta\in\Theta}
\lvv \Phi(\theta)+t_n \sh(\theta) \rvv_{L_K^2(\nu)}^2 \\
=&\,\min \lbr \inf_{\theta\in \Theta\setminus\Theta_0(\phi)^{a(h)t_n}}
\lvv \Phi(\theta)+t_n \sh(\theta) \rvv_{L_K^2(\nu)}^2, \;
\min_{\theta\in \Theta\cap\Theta_0(\phi)^{a(h)t_n}}\lvv \Phi(\theta)+t_n \sh(\theta) \rvv_{L_K^2(\nu)}^2 \rbr.
\end{align*}
This, together with \eqref{eq.second order 1 multi}, implies that for large $n$, \begin{align*}
\mL\lp \phi+t_n h \rp =\min_{\theta\in \Theta\cap\Theta_0(\phi)^{a(h)t_n}}
\lvv \Phi(\theta)+t_n\sh(\theta)\rvv_{L_K^2(\nu)}^2.
\end{align*}

For every $a>0$, let $V(a)=\{ v\in\prod_{k=1}^K\mathbb{R}^{d_{\theta_k}}: \lvv v \rvv_{K2}\le a \}$.
For every $\theta\in\Theta_0(\phi)$ and every $a>0$, define \begin{align*}
V_n(a,\theta)=\lbr v\in V(a): \theta+t_n v \in \Theta \rbr.
\end{align*}
It is easy to show that (with the compactness of $\Theta_0(\phi)$)
\begin{align*}
\bigcup_{\theta\in\Theta_0(\phi)} \bigcup_{v\in V_n (a(h),\theta)} \lbr \theta+t_n v \rbr
=\Theta \cap \Theta_0(\phi)^{a(h)t_n} .
\end{align*}
Therefore, \begin{align*}
\mL\lp \phi+t_n h \rp =\inf_{\theta\in\Theta_0(\phi)} \inf_{v\in V_n(a(h),\theta)}
\lvv \Phi \lp \theta+t_n v \rp +t_n \sh \lp \theta+t_n v \rp \rvv_{L_K^2(\nu)}^2.
\end{align*}
Note that $0\in V_n(a(h),\theta)$. Then for every $\theta_0\in\Theta_0(\phi)$,
\begin{align*}
& \lv \mL\lp \phi+t_n h \rp-
\inf_{\theta\in\Theta_0(\phi)} \inf_{v\in V_n(a(h),\theta)}
\lvv \Phi \lp \theta+t_n v \rp +t_n \sh \lp \theta\rp \rvv_{L_K^2(\nu)}^2 \rv \\
=&\, \lv \inf_{\theta\in\Theta_0(\phi)} \inf_{v\in V_n(a(h),\theta)}
\lvv \Phi \lp \theta+t_n v \rp +t_n \sh \lp \theta+t_n v \rp \rvv_{L_K^2(\nu)} \right. \\
& \left.	+\inf_{\theta\in\Theta_0(\phi)} \inf_{v\in V_n(a(h),\theta)}
\lvv \Phi \lp \theta+t_n v \rp +t_n \sh \lp \theta \rp \rvv_{L_K^2(\nu)} \rv \\
& \phantom{==} \cdot \lv \inf_{\theta\in\Theta_0(\phi)} \inf_{v\in V_n(a(h),\theta)}
\lvv \Phi \lp \theta+t_n v \rp +t_n \sh \lp \theta+t_n v \rp \rvv_{L_K^2(\nu)} \right. \\
&  \phantom{===}\left. -\inf_{\theta\in\Theta_0(\phi)} \inf_{v\in V_n(a(h),\theta)}
\lvv \Phi \lp \theta+t_n v \rp +t_n \sh \lp \theta \rp \rvv_{L_K^2(\nu)} \rv \\
\le&\, 2 \lvv \Phi\lp \theta_0 \rp+t_n \sh \lp \theta_0 \rp \rvv_{L_K^2(\nu)}
\sup_{\theta\in\Theta_0(\phi)} \sup_{v\in V_n(a(h),\theta)} t_n
\lvv \sh \lp \theta+t_n v \rp-\sh(\theta) \rvv_{L_K^2(\nu)} \\
\le&\, 2t_n^2 \left\{\sum_{k=1}^K\lvv h_k \rvv_{\infty} ^2\right\}^{1/2}
\sup_{\theta_1,\theta_2\in\Theta: \lvv \theta_1-\theta_2 \rvv_{K2} \le a(h) t_n}
\lvv \sh \lp \theta_1 \rp-\sh \lp \theta_2 \rp \rvv_{L_K^2(\nu)} = o( t_n^2 ),
\end{align*}
where the last equality follows from the definition of $\bD_{\mL0}$ and
the compactness of $\Theta$.

For every $\theta\in\Theta$ with $\theta=(\theta_1,\ldots,\theta_K)$, define $\Phi_k'(\theta_k):\mr\to\mr^{d_{\theta_k}}$ such that 
\begin{align*}
\Phi_k'(\theta_k)(x)=-\left.
\frac{\partial ( G_k\circ g_k ) (z,\vartheta_k)}{\partial \vartheta_k} 
\rv_{(z,\vartheta_k)=(x,\theta_k)}
\quad \text{for every } x\in \mr.
\end{align*}
For every $\theta=(\theta_1,\ldots,\theta_K)$ and every $v=(v_1,\ldots,v_K)$,
let
\begin{align*}
\Phi^{\prime}(\theta,v)(x)=(\Phi^{\prime}_1(\theta_1)(x)^\T v_1,\ldots,\Phi^{\prime}_K(\theta_K)(x)^\T v_K)    
\end{align*}
for all $x$. Using an argument similar
to the previous result, we have \begin{align*}
&\phantom{=\:\:} \lv \inf_{\theta\in\Theta_0(\phi)} \inf_{v\in V_n(a(h),\theta)}
\lvv \Phi\lp \theta+t_n v \rp +t_n \sh(\theta) \rvv_{L_K^2(\nu)}^2 \right. \\
&\phantom{==} \left. - \inf_{\theta\in\Theta_0(\phi)} \inf_{v\in V_n(a(h),\theta)}
\lvv \Phi(\theta)+t_n \Phi'(\theta,v)+t_n \sh(\theta) \rvv_{L_K^2(\nu)}^2 \rv \\
& \le 2O(t_n^2)
\sup_{\theta\in\Theta_0(\phi)} \sup_{v\in V_n(a(h),\theta)}
\left\{\sum_{k=1}^K\lvv \frac{\Phi_k\lp \theta_k+t_n v_k \rp-\Phi_k(\theta_k)}{t_n}-[\Phi'_k(\theta_k)]^\T v_k \rvv^2_{L^2(\nu)}\right\}^{1/2}.
\end{align*}
Since $\Theta_0(\phi)\subset \mathrm{int}(\Theta)$,
for sufficiently large $n$, we have $V_n(a(h),\theta)=V(a(h))$ for all $\theta\in\Theta_0(\phi)$. 
For every $\theta\in\Theta_0(\phi)$ and every $v\in V_n(a(h),\theta)$, Assumption
\ref{ass:G is 2nd differentiable multi} implies that when $n$ is large, 
\begin{align*}
&\phantom{=\:\:} \lvv \frac{\Phi_k\lp \theta_k+t_n v_k \rp-\Phi_k(\theta_k)}{t_n}
-\lbk \Phi_k'(\theta_k) \rbk^\T v_k \rvv_\lsv^2 \\
& =\int_\mr \lbk \frac{G_k\lp g_k(x,\theta_k+t_n v_k)\rp-G_k\lp g_k(x,\theta_k)\rp}{t_n}
-\lp \left.
\frac{\partial ( G_k\circ g_k ) (z,\vartheta_k)}{\partial \vartheta_k} 
\rv_{(z,\vartheta_k)=(x,\theta_k)} \rp^\T v_k \rbk^2 \ddd \nu(x) \\
&=\int_\mr \lbk \frac{t_n}{2} v_k^\T \lp 	\left.
\frac{\partial^2 ( G_k\circ g_k ) (z,\vartheta_k)}{\partial \vartheta_k \partial \vartheta_k^\T} 
\rv_{(z,\vartheta_k)=(x,\theta_k+t_{kn}^*(x) v_k)}	\rp v_k
\rbk^2 \ddd \nu(x) \\
& \le \frac{a(h)^4 t_n^2}{4}\int_\mr \sup_{\theta_k^{\ast}\in\Theta_k}
\lvv \left.
\frac{\partial^2 ( G_k\circ g_k ) (z,\vartheta_k)}{\partial \vartheta_k \partial \vartheta_k^\T} 
\rv_{(z,\vartheta_k)=(x,\theta_k^{\ast})} \rvv_2^2 \,\mathrm{d}\nu(x)=O(t_n^2),
\end{align*}
where $0\le t_{kn}^{\ast}(x) \le t_n $ for all $x$ and all $k$, and the last inequality
follows from 
the property of the $\ell^2$ operator norm.
Then it follows that 
\begin{align*}
\sup_{\theta\in\Theta_0(\phi)} \sup_{v\in V_n(a(h),\theta)}
\left\{\sum_{k=1}^K\lvv \frac{\Phi_k\lp \theta_k+t_n v_k \rp-\Phi_k(\theta_k)}{t_n}-\Phi'_k(\theta_k)^\T v_k \rvv^2_{L^2(\nu)}\right\}^{1/2}
=o(1).
\end{align*}
Combining the above results yields \begin{align*}
\lv \mL\lp \phi+t_n h_n \rp - t_n^2 \inf_{\theta\in\Theta_0(\phi)} \inf_{v\in V(a(h))}
\lvv \Phi'(\theta,v)+\sh (\theta) \rvv_{L_K^{2}(\nu)}^2 \rv=o\lp t_n^2 \rp.
\end{align*}

By similar arguments, we can show that for all $a\geq a\left(  h\right)  $,
\begin{align*}
\inf_{\theta\in\Theta_0(\phi)}\inf_{v\in V\left(  a\right)  }\left\Vert \Phi'(\theta,v)+\mathscr{H}\left(  \theta\right)  \right\Vert _{L^{2}\left( \nu\right)  }^{2}=\inf_{\theta\in\Theta_0(\phi)}\inf_{v\in V\left(  a\left( h \right)  \right)  }\left\Vert \Phi'(\theta,v)+\mathscr{H}\left(  \theta\right) \right\Vert _{L^{2}\left(  \nu\right)  }^{2}.
\end{align*}
For every $v^{\prime}\in\prod_{k=1}^K\mr^{d_{\theta_{k}}}$, if $\left\Vert v^{\prime}\right\Vert_{2}\geq a\left(  h\right)  $, then \begin{align*}
\inf_{\theta\in\Theta_0(\phi)}\left\Vert \Phi'(\theta,v^{\prime})+\mathscr{H}\left(  \theta\right)  \right\Vert _{L^{2}\left(  \nu\right)}^{2}  & \geq\inf_{\theta\in\Theta_0(\phi)}\inf_{v\in V\left(  \left\Vert v'\right\Vert_{2}\right)}\left\Vert \Phi'(\theta,v)+\mathscr{H}\left(\theta\right)  \right\Vert _{L^{2}\left(  \nu\right)  }^{2}\\
& =\inf_{\theta\in\Theta_0(\phi)}\inf_{v\in V\left(  a\left(  h\right)  \right)  }\left\Vert \Phi'(\theta,v)+\mathscr{H}\left(  \theta\right)  \right\Vert_{L^{2}\left(  \nu\right)  }^{2};
\end{align*}
if $\left\Vert v^{\prime}\right\Vert _{2}<a\left(  h\right)  $, then \begin{align*}
\inf_{\theta\in\Theta_0(\phi)}\left\Vert \Phi'(\theta,v^{\prime}) +\mathscr{H}\left(  \theta\right)  \right\Vert _{L^{2}\left(  \nu\right)}^{2}\geq\inf_{\theta\in\Theta_0(\phi)}\inf_{v\in V\left(  a\left(  h\right)  \right)  }\left\Vert \Phi'(\theta,v)+\mathscr{H}\left(  \theta\right) \right\Vert _{L^{2}\left(  \nu\right)  }^{2}.
\end{align*}
On the other hand, $V(a(h))\subset\prod_{k=1}^K\mr^{d_{\theta_{k}}}$ by definition. Thus, \begin{align*}
\inf_{\theta\in\Theta_0(\phi)}\inf_{v\in \prod_{k=1}^K\mr^{d_{\theta_{k}}}}\left\Vert \Phi'(\theta,v)+\mathscr{H}\left(  \theta\right)  \right\Vert_{L^{2}\left(  \nu\right)  }^{2}&=\inf_{v\in \prod_{k=1}^K\mr^{d_{\theta_{k}}}}\inf_{\theta\in\Theta_0(\phi)}\left\Vert \Phi'(\theta,v)+\mathscr{H}\left(  \theta\right)  \right\Vert_{L^{2}\left(  \nu\right)  }^{2}\\
&= \inf_{v\in V\left(  a\left(  h\right)  \right)  }\inf_{\theta\in\Theta_0(\phi)}\left\Vert \Phi'(\theta,v)+\mathscr{H}\left(  \theta\right)  \right\Vert_{L^{2}\left(  \nu\right)  }^{2}\\
&= \inf_{\theta\in\Theta_0(\phi)}\inf_{v\in V\left(  a\left(  h\right)  \right)  }\left\Vert \Phi'(\theta,v)+\mathscr{H}\left(  \theta\right)  \right\Vert_{L^{2}\left(  \nu\right)  }^{2}.
\end{align*}
\end{pf}

\begin{pf}{ of Proposition \ref{thry:asymptotic distribution of test stat multi}}
Note that both $\prod_{k=1}^K\ell^\infty(\mr\times\Theta_k)$ and $\mr$ are normed spaces.
By Lemma \ref{prop:second order Hadamard of L multi}, the map
$\mL$ is second order Hadamard directionally differentiable at $\phi$
tangentially to $\bD_{\mL0}$. Lemma \ref{prop:Donsker of phihat multi}
shows that $\sqrt{T_n}( \phihat-\phi )\convd \bG_0$ in $\prod_{k=1}^K\ell^\infty(\mr\times\Theta_k)$
as $n\to\infty$ and $\bG_0$ is tight with $\bG_0\in\bD_{\mL0}$ almost surely.
Therefore, Assumptions 2.1(i), 2.1(ii), 2.2(i), and 2.2(ii) of \citet{chen2019inference}
are satisfied. The desired result follows from
Theorem 2.1 of \citet{chen2019inference}, the fact that $\mL(\phi)=0$ and
$\mL'_\phi(h) =0$ for all $h\in\prod_{k=1}^K\ell^\infty(\mr\times\Theta_k)$ whenever $\phi\in\bD_0$,
and that $( \phihat-\phi)\in \prod_{k=1}^K\ell^\infty(\mr\times\Theta_k)$ for every $n\in\mathbb{Z}_+$.
\end{pf}

\begin{pf}{ of Lemma \ref{prop:consistency of second derivative estimator multi}}
Note that both $\prod_{k=1}^K\ell^\infty(\mr\times\Theta_k)$ and $\mr$ are normed spaces,
and by Lemma \ref{prop:second order Hadamard of L multi}, the map
$\mL$ is second order Hadamard directionally differentiable at $\phi\in\bD_0$
tangentially to $\bD_{\mL0}$. By Lemma \ref{lemma:multi-HDDL},
$\mL'_\phi(h)=0$ for all $h\in\prod_{k=1}^K\ell^\infty(\mr\times\Theta_k)$ whenever $\phi
\in\bD_0$.
Lemma \ref{prop:Donsker of phihat multi}
shows that $\sqrt{T_n}( \phihat-\phi )\convd \bG_0$ 
in $\prod_{k=1}^K\ell^\infty(\mr\times\Theta_k)$
as $n\to\infty$ and $\bG_0$ is tight with $\bG_0\in\bD_{\mL0}$ almost surely.
Therefore, Assumptions 2.1, 2.2(i), 2.2(ii), and 3.5 of \citet{chen2019inference}
hold, and the desired result follows from Proposition 3.1 of
\citet{chen2019inference}.
\end{pf}

\begin{pf}{ of Lemma \ref{prop:weak convergence of bootstrap multi}}
Define \begin{align*}
\mathcal{F}=\lbr \indicator_{(-\infty,x]} : x\in \mr\rbr \text{ and }
\mathcal{G}_k=\lbr \indicator_{\lp -\infty, g_k(x,\theta_k) \rbk }:
(x,\theta_k)\in \mr\times\Theta_k \rbr \text{ for every }k.
\end{align*}
Define $\widehat{\mathcal{X}}_{n_x}$, $\widehat{\mathcal{Y}}_{n_k}$, $\cx$, and $\mathcal{Y}_k$ as 
\begin{align*}
\widehat{\mathcal{X}}_{n_x} (f)=\frac{1}{n_x}\sum_{i=1}^{n_x} f \lp X_i \rp, 
\widehat{\mathcal{Y}}_{n_k} (f)=\frac{1}{n_k}\sum_{i=1}^{n_k} f \lp Y_{ki} \rp, 
\cx(f)=\E\lbk f\lp X_i \rp \rbk, \text{ and }
\cy_k(f)=\E \lbk f\lp Y_{ki} \rp \rbk
\end{align*}
for all measurable $f$. Let $\{W_{xi}\}_{i=1}^{n_x},\{W_{1i}\}_{i=1}^{n_1},\ldots,\{W_{Ki}\}_{i=1}^{n_K}$ be jointly independent random vectors of multinomial weights that are independent of $\{X_i\}_{i=1}^{n_x},\{Y_{1i}\}_{i=1}^{n_1},\ldots,\{Y_{Ki}\}_{i=1}^{n_K} $. Define $\widehat{\mathcal{X}}_{n_x}^*$ and $\widehat{\mathcal{Y}}_{n_k}^*$ to be the bootstrap versions
of $\widehat{\mathcal{X}}_{n_x}$ and $\widehat{\mathcal{Y}}_{n_k}$, respectively, with
\begin{align*}
\widehat{\mathcal{X}}_{n_x}^* (f)=\frac{1}{n_x}\sum_{i=1}^{n_x} f \lp X^*_i \rp =\frac{1}{n_x}\sum_{i=1}^{n_x} W_{xi}f \lp X_i \rp
\text{ and }
\widehat{\mathcal{Y}}_{n_k}^* (f)=\frac{1}{n_k}\sum_{i=1}^{n_k} f \lp Y^*_{ki} \rp=\frac{1}{n_k}\sum_{i=1}^{n_k} W_{ki}f \lp Y_{ki} \rp
\end{align*}
for every measurable $f$.
By Example 2.5.4 of \citet{van1996weak}, the class $\mathcal{F}$ is
Donsker. Because $\mathcal{G}_k\subset\mathcal{F}$ for every $k$, by Theorem 2.10.1
of \citet{van1996weak}, the class $\mathcal{G}_k$ is also Donsker.
Therefore, \begin{align*}
\sqrt{n_x}\lp \widehat{\mathcal{X}}_{n_x}-\cx \rp \convd \bx  \text{ in } \ell^\infty(\mathcal{F})
\text{ and }
\sqrt{n_k}\lp\widehat{\mathcal{Y}}_{n_k}-\cy_k \rp \convd \by_k \text{ in } \ell^\infty(\mathcal{G}_k)
\end{align*}
as $n\to\infty$, where $\bx,\by_1,\ldots,\by_K$ are jointly independent centered Gaussian processes.
Moreover, because $\mathcal{F}$ and $\mathcal{G}_k$ are classes of indicator
functions, we have that
\begin{align*}
\cx \lbk \sup_{f\in\mathcal{F}} \lp f-\cx(f) \rp^2 \rbk \le 1 
\text{ and }
\cy_k \lbk \sup_{h\in\mathcal{G}_k} \lp h-\cy_k(h) \rp^2 \rbk \le 1.
\end{align*}
By Theorem 2.7 of \citet{kosorok2008introduction}, it follows that
\begin{align*}
\sqrt{n_x}\lp \widehat{\mathcal{X}}_{n_x}^*-\widehat{\mathcal{X}}_{n_x} \rp \overset{\text{a.s.}}{\leadsto} \mathbb{X} \text{ and } 
\sqrt{n_k}\lp \widehat{\mathcal{Y}}_{n_k}^*-\widehat{\mathcal{Y}}_{n_k} \rp \overset{\text{a.s.}}{\leadsto} \mathbb{Y}_k
\end{align*}
as $n\to\infty$. 

It is easy to show that 
\begin{align*}
&\widehat{F}_{n_x}(x)=\widehat{\mathcal{X}}_{n_x}\lp \indicator_{\lp-\infty, x\rbk} \rp , 
\lp \widehat{G}_{n_k}\circ g_k \rp (x,\theta_k)=\widehat{\mathcal{Y}}_{n_k}\lp 
\indicator_{\lp -\infty, g_k(x,\theta_k) \rbk } \rp, \\
&\widehat{F}_{n_x}^*(x)=\widehat{\mathcal{X}}_{n_x}^*\lp \indicator_{\lp-\infty, x\rbk} \rp , \text{ and }
\lp \widehat{G}_{n_k}^*\circ g_k \rp (x,\theta_k)=\widehat{\mathcal{Y}}_{n_k}^*
\lp \indicator_{\lp -\infty, g_k(x,\theta_k) \rbk } \rp
\end{align*}
for every $x\in\mr$, every $\theta_k\in\Theta_k$, and every $k$.
Define $W_F(x)=\bx (  \indicator_{\lp -\infty, x \rbk } )$ and
$W_k(x,\theta_k)=\by_k (  \indicator_{\lp -\infty, g_k(x,\theta_k) \rbk } )$
for every $x\in\mr$ and every $\theta_k\in\Theta_k$. By Lemma \ref{lemma:conditional weak convergence of transformed index}, we have that
\begin{align}\label{eq:almost sure weak convergence F G multi}
\sqrt{n_x} \lp \widehat{F}_{n_x}^*-\widehat{F}_{n_x} \rp \overset{\text{a.s.}}{\leadsto} {W}_F \text{ and }  \sqrt{n_k} \lp \widehat{G}_{n_k}^*\circ g_k-\widehat{G}_{n_k}\circ g_k \rp \overset{\text{a.s.}}{\leadsto} W_k.
\end{align}
For simplicity, let $\mathcal{Z}_n=\{ \lbr X_i \rbr_{i=1}^{n_x},
\lbr Y_{1i} \rbr_{i=1}^{n_1},\ldots,\lbr Y_{Ki} \rbr_{i=1}^{n_K} \}$, $\mathbb{A}=\ell^\infty (\mr)\times \prod_{k=1}^K\ell^\infty	(\mr\times\Theta_k)$, and $\mathbb{B}=\prod_{k=1}^K\ell^\infty	(\mr\times\Theta_k)$. Define norms $\Vert\cdot\Vert_{\mathbb{A}}$ and $\Vert\cdot\Vert_{\mathbb{B}}$ on $\mathbb{A}$ and $\mathbb{B}$, respectively, such that for every $(f,h)\in\mathbb{A}$ with $h=(h_1,\ldots,h_K)$ and every $w\in\mathbb{B}$ with  $w=(w_1,\ldots,w_K)$, 
\begin{align*}
\Vert (f,h) \Vert_{\mathbb{A}}=\Vert f\Vert_{\infty}+\sum_{k=1}^K\Vert h_k\Vert_{\infty} \text{ and } \Vert w \Vert_{\mathbb{B}}=\sum_{k=1}^K\Vert w_k\Vert_{\infty}.
\end{align*}
By the joint independence of the weight vectors, we have that for all bounded, nonnegative, Lipschitz functions $\Gamma_x$ on $ \ell^\infty(\mr) $ and $\Gamma_k$ on $ \ell^\infty(\mr\times\Theta_k)$, 
\begin{align*}
&\E\lbk  \Gamma_x \lp \sqrt{n_x} \lp \widehat{F}_{n_x}^*-\widehat{F}_{n_x} \rp \rp \prod_{k=1}^K \Gamma_k \lp \sqrt{n_k} \lp \widehat{G}_{n_k}^*\circ g_k-\widehat{G}_{n_k}\circ g_k \rp \rp \big|
\mathcal{Z}_n \rbk \\
=&\, \E\lbk  \Gamma_x \lp \sqrt{n_x} \lp \widehat{F}_{n_x}^*-\widehat{F}_{n_x} \rp \rp \big|
\mathcal{Z}_n \rbk \cdot\prod_{k=1}^K \E\lbk  \Gamma_k \lp \sqrt{n_k} \lp \widehat{G}_{n_k}^*\circ g_k-\widehat{G}_{n_k}\circ g_k \rp \rp \big|
\mathcal{Z}_n \rbk.
\end{align*}
Let $\lambda_{-x}=\prod_{k=1}^K\lambda_k$ and $\lambda_{-k}=(\lambda_x\cdot\prod_{j=1}^K\lambda_j)/\lambda_k$.
Then with the joint independence of the random elements $\{{W}_F,W_1,\ldots,W_K\}$, by Example 1.4.6 of
\citet{van1996weak} and Assumption \ref{ass:ratio lambda multi} of this paper, 
\begin{align*}
\sup_{\Gamma\in \mathrm{BL}_1\lp  \mathbb{A} \rp}
\lv \E \lbk \left. \Gamma \lp  
\begin{bmatrix}
\sqrt{T_n} \lp \widehat{F}_{n_x}^*-\widehat{F}_{n_x} \rp \\
\sqrt{T_n} \lp \widehat{G}_{n_1}^*\circ g_1 -\widehat{G}_{n_1} \circ g_1 \rp\\
\vdots\\
\sqrt{T_n} \lp \widehat{G}_{n_K}^*\circ g_K -\widehat{G}_{n_K} \circ g_K \rp
\end{bmatrix}
\rp
\rv  \mathcal{Z}_n 
\rbk-
\E \lbk \Gamma \lp \begin{bmatrix}
\sqrt{\lambda_{-x}} W_F \\ 
\sqrt{\lambda_{-1}}W_1\\
\vdots\\
\sqrt{\lambda_{-K}}W_K
\end{bmatrix} \rp \rbk \rv \convas 0
\end{align*}
as $n\to\infty$.

Define a map $\mathcal{I}: \mathbb{A}\to \mathbb{B}$, such
that 
\begin{align*}
\mathcal{I}\lp f, h \rp = (f -h_1 ,\ldots,f -h_K )
\end{align*}
for every $\lp f, h \rp\in \mathbb{A}$ with $h=(h_1,\ldots,h_K)$. It is easy to show the Lipschitz continuity of $\mathcal{I}$.
By the proof similar to that of Proposition 10.7(ii) of \citet{kosorok2008introduction}, we can show that 
\begin{align*}
\sup_{\Gamma\in\mathrm{BL}_1\lp \prod_{k=1}^K\ell^\infty(\mr\times\Theta_k) \rp}
\lv \E \lbk \left. \Gamma \lp \sqrt{T_n} \lp \phihat^*-\phihat \rp \rp
\rv \mathcal{Z}_n \rbk-
\E \lbk \Gamma \lp \tilde{\bG}_0 \rp \rbk \rv \convas 0
\end{align*}
as $n\to\infty$, where $\tilde{\mathbb{G}}_0=\mathcal{I}(\sqrt{\lambda_{-x}}{W}_F,\sqrt{\lambda_{-1}}W_{1},\ldots,\sqrt{\lambda_{-K}}W_{K})$. By the properties of the random elements $\{W_F,W_1,\ldots,W_K\}$, it can be verified that $\tilde{\mathbb{G}}_0$ is equivalent to ${\mathbb{G}}_0$ in law. 
The desired result follows from Lemma 1.9.2(i) of \citet{van1996weak}.

Because $\mathcal{F}$ and $\mathcal{G}_k$ are Donsker, by Theorem 2.6 of \citet{kosorok2008introduction}, $\sqrt{n_x} ( \widehat{\mathcal{X}}_{n_x}^*-\widehat{\mathcal{X}}_{n_x} )$ and
$\sqrt{n_k} ( \widehat{\mathcal{Y}}_{n_k}^*-\widehat{\mathcal{Y}}_{n_k} )$ (for every $k$) are asymptotically measurable. 
By Lemma \ref{lemma:conditional weak convergence of transformed index}, $\sqrt{n_x} ( \widehat{F}_{n_x}^*-\widehat{F}_{n_x} )$ and
$\sqrt{n_k} ( \widehat{G}_{n_k}^*\circ g_k-\widehat{G}_{n_k}\circ g_k )$ are asymptotically measurable. By \eqref{eq:almost sure weak convergence F G multi} and the asymptotic measurability of $\sqrt{n_x} ( \widehat{F}_{n_x}^*-\widehat{F}_{n_x} )$ and
$\sqrt{n_k} ( \widehat{G}_{n_k}^*\circ g_k-\widehat{G}_{n_k}\circ g_k )$, we can show that $\sqrt{n_x} ( \widehat{F}_{n_x}^*-\widehat{F}_{n_x} )$ and
$\sqrt{n_k} ( \widehat{G}_{n_k}^*\circ g_k-\widehat{G}_{n_k}\circ g_k )$ are asymptotically tight. Then by Lemmas 1.4.3 and 1.4.4 of \citet{van1996weak}, 
\begin{align*}
(\sqrt{n_x} ( \widehat{F}_{n_x}^*-\widehat{F}_{n_x} ),\sqrt{n_1} ( \widehat{G}_{n_1}^*\circ g_1-\widehat{G}_{n_1}\circ g_1 ),\ldots,\sqrt{n_K} ( \widehat{G}_{n_K}^*\circ g_K-\widehat{G}_{n_K}\circ g_K ))	
\end{align*}
is asymptotically measurable. The asymptotic measurability of $\sqrt{T_n} ( \phihat^*-\phihat )$ follows from the continuity of $\mathcal{I}$.
\end{pf}

\begin{pf}{ of Proposition \ref{thry:consistency of approximation of test statistic distribution multi}}
Note that both $\prod_{k=1}^K\ell^\infty(\mr\times\Theta_k)$ and $\mr$ are normed spaces,
and by Lemma \ref{prop:second order Hadamard of L multi}, the map
$\mL$ is second order Hadamard directionally differentiable at $\phi\in\bD_0$
tangentially to $\bD_{\mL0}$. Lemma \ref{prop:Donsker of phihat multi}
shows that $\sqrt{T_n}( \phihat-\phi )\convd \bG_0$ in $\prod_{k=1}^K\ell^\infty(\mr\times\Theta_k)$
as $n\to\infty$ and $\bG_0$ is tight with $\bG_0\in\bD_{\mL0}$ almost surely.
By Lemma \ref{lemma:properties of DL multi}, $\bD_{\mL0}$ is closed under
vector addition, that is, $\varphi_1+\varphi_2\in \bD_{\mL0}$ whenever
$\varphi_1,\varphi_2\in \bD_{\mL0}$. By construction, the random weights used to construct the bootstrap samples are independent of the data set, and $f(\sqrt{T_n}( \phihat^*-\phihat ))$ is a measurable function of the random weights for every continuous and bounded $f:\ell^\infty(\mr\times\Theta)\to\mathbb{R}$ given every sample. By Lemma
\ref{prop:weak convergence of bootstrap multi}, 
\begin{align*}
&\sup_{\Gamma\in\mathrm{BL}_1\lp \prod_{k=1}^K\ell^\infty(\mr\times\Theta_k) \rp}
\lv \E \lbk \left. \Gamma \lp \sqrt{T_n} \lp \phihat^*-\phihat \rp \rp
\rv \lbr X_i \rbr_{i=1}^{n_x}, \lbr Y_{1i} \rbr_{i=1}^{n_1},\ldots,\lbr Y_{Ki} \rbr_{i=1}^{n_K} \rbk-
\E \lbk \Gamma \lp \bG_0 \rp \rbk \rv \\
&\convp 0,
\end{align*}
and $\sqrt{T_n}( \phihat^*-\phihat )$ is asymptotically
measurable as $n\to \infty$.
Lemma \ref{prop:consistency of second derivative estimator multi} establishes
the consistency of $\mLhat$ for $\mathcal{L}^{\prime\prime}_{\phi}$. Therefore, Assumptions 2.1(i), 2.1(ii), 2.2, 3.1, 3.2, and 3.4 of \citet{chen2019inference} are satisfied, and the result follows from Theorem 3.3 of \citet{chen2019inference}.
\end{pf}

\begin{pf}{ of Theorem \ref{thry:size and power multi}}
Under Assumptions \ref{ass:properties of g multi}--\ref{ass:rate of tau multi}, with Propositions  \ref{thry:asymptotic distribution of test stat multi} and \ref{thry:consistency of approximation of test statistic distribution multi}, the desired results can be proved by arguments similar to those in the proof of Theorem \ref{thry:para size and power}.   
\end{pf}

\section{Additional Simulation Results}\label{sec: additional simulations}

In this section, we present simulation results for Case 1 with different choices of $\nu$ and larger sample sizes, and for Cases 2 and 3, as discussed in Section \ref{sec:simulation}. We also conduct additional Monte Carlo experiments to demonstrate the performance of the proposed test in testing symmetry, goodness of fit, and location transformation.

\subsection{Results for Section \ref{sec:simulation}}

\begin{table}[H]
\small\centering\setstretch{1.3}
\caption{Size and power for Case 1 with i.i.d.\ data ($\nu=\mathcal{N}(0,1)$, $\alpha=0.05$)}
\label{tab:case1 sig1}
\begin{tabular*}{15cm}{@{\extracolsep{\fill}}cccccccc}
\hline\hline
\multirow{2}{*}{DGP} & \multirow{2}{*}{$n$} & \multicolumn{6}{c}{$\tau_n$} \\
\cline{3-8}
& & $\sqrt{\ln(n)/n}$ & $n^{-2/5}$ & $ n^{-1/3}$ & $ n^{-1/4}$ & $ n^{-1/5}$ & $ n^{-1/6}$ \\
\hline
\multirow{4}{*}{DGP (0)}&$	100	$&$	0.044	$&$	0.029	$&$	0.044	$&$	0.052	$&$	0.052	$&$	0.052	$\\
&$	200	$&$	0.046	$&$	0.038	$&$	0.047	$&$	0.047	$&$	0.046	$&$	0.046	$\\
&$	400	$&$	0.059	$&$	0.047	$&$	0.063	$&$	0.081	$&$	0.078	$&$	0.078	$\\
&$	800	$&$	0.059	$&$	0.055	$&$	0.063	$&$	0.080	$&$	0.082	$&$	0.085	$\\
\hline
\multirow{4}{*}{DGP (1)}
&$	100	$&$	0.235	$&$	0.179	$&$	0.235	$&$	0.287	$&$	0.311	$&$	0.334	$\\
&$	200	$&$	0.392	$&$	0.329	$&$	0.405	$&$	0.524	$&$	0.569	$&$	0.581	$\\
&$	400	$&$	0.641	$&$	0.519	$&$	0.674	$&$	0.778	$&$	0.818	$&$	0.829	$\\
&$	800	$&$	0.846	$&$	0.759	$&$	0.886	$&$	0.966	$&$	0.978	$&$	0.983	$\\
\hline
\multirow{4}{*}{DGP (2)}
&$	100	$&$	0.810	$&$	0.706	$&$	0.812	$&$	0.890	$&$	0.916	$&$	0.932	$\\
&$	200	$&$	0.983	$&$	0.944	$&$	0.988	$&$	0.997	$&$	0.998	$&$	0.999	$\\
&$	400	$&$	1.000	$&$	1.000	$&$	1.000	$&$	1.000	$&$	1.000	$&$	1.000	$\\
&$	800	$&$	1.000	$&$	1.000	$&$	1.000	$&$	1.000	$&$	1.000	$&$	1.000	$\\
\hline
\multirow{4}{*}{DGP (3)}
&$	100	$&$	0.976	$&$	0.938	$&$	0.977	$&$	0.991	$&$	0.996	$&$	0.997	$\\
&$	200	$&$	1.000	$&$	1.000	$&$	1.000	$&$	1.000	$&$	1.000	$&$	1.000	$\\
&$	400	$&$	1.000	$&$	1.000	$&$	1.000	$&$	1.000	$&$	1.000	$&$	1.000	$\\
&$	800	$&$	1.000	$&$	1.000	$&$	1.000	$&$	1.000	$&$	1.000	$&$	1.000	$\\
\hline\hline
\end{tabular*}
\end{table}

\begin{table}[H]
\small\centering\setstretch{1.3}
\caption{Size for Case 1 with dependent data ($\nu=\mathcal{N}(0,1)$, $\alpha=0.05$)}
\label{tab:case1dep null sig1}
\begin{tabular*}{15cm}{@{\extracolsep{\fill}}cccccccc}
\hline\hline
\multirow{2}{*}{$b(n)$} & \multirow{2}{*}{$n$} & \multicolumn{6}{c}{$\tau_n$} \\
\cline{3-8}
& & $\sqrt{\ln(n)/n}$ & $n^{-2/5}$ & $ n^{-1/3}$ & $ n^{-1/4}$ & $ n^{-1/5}$ & $ n^{-1/6}$ \\
\hline
\multirow{4}{*}{$n^{1/6}$}
&$	100	$&$	0.032	$&$	0.022	$&$	0.032	$&$	0.046	$&$	0.048	$&$	0.048	$\\
&$	200	$&$	0.047	$&$	0.038	$&$	0.049	$&$	0.052	$&$	0.060	$&$	0.062	$\\
&$	400	$&$	0.068	$&$	0.063	$&$	0.070	$&$	0.088	$&$	0.098	$&$	0.098	$\\
&$	800	$&$	0.055	$&$	0.049	$&$	0.056	$&$	0.062	$&$	0.065	$&$	0.067	$\\
\hline
\multirow{4}{*}{$n^{1/5}$}
&$	100	$&$	0.042	$&$	0.033	$&$	0.042	$&$	0.046	$&$	0.048	$&$	0.048	$\\
&$	200	$&$	0.033	$&$	0.030	$&$	0.034	$&$	0.038	$&$	0.040	$&$	0.041	$\\
&$	400	$&$	0.068	$&$	0.063	$&$	0.070	$&$	0.088	$&$	0.098	$&$	0.098	$\\
&$	800	$&$	0.053	$&$	0.046	$&$	0.062	$&$	0.064	$&$	0.074	$&$	0.082	$\\
\hline
\multirow{4}{*}{$n^{1/4}$}
&$	100	$&$	0.042	$&$	0.033	$&$	0.042	$&$	0.046	$&$	0.048	$&$	0.048	$\\
&$	200	$&$	0.040	$&$	0.035	$&$	0.046	$&$	0.057	$&$	0.064	$&$	0.068	$\\
&$	400	$&$	0.070	$&$	0.060	$&$	0.079	$&$	0.087	$&$	0.084	$&$	0.079	$\\
&$	800	$&$	0.074	$&$	0.063	$&$	0.076	$&$	0.082	$&$	0.075	$&$	0.084	$\\
\hline
\multirow{4}{*}{$n^{1/3}$}
&$	100	$&$	0.048	$&$	0.042	$&$	0.048	$&$	0.066	$&$	0.066	$&$	0.066	$\\
&$	200	$&$	0.039	$&$	0.030	$&$	0.040	$&$	0.050	$&$	0.053	$&$	0.060	$\\
&$	400	$&$	0.067	$&$	0.057	$&$	0.068	$&$	0.087	$&$	0.082	$&$	0.084	$\\
&$	800	$&$	0.064	$&$	0.054	$&$	0.065	$&$	0.086	$&$	0.103	$&$	0.109	$\\
\hline\hline
\end{tabular*}
\end{table}

\begin{table}[H]
\small\centering\setstretch{1.3}
\caption{Power for DGP (1) of Case 1 with dependent data ($\nu=\mathcal{N}(0,1)$, $\alpha=0.05$)}
\label{tab:case1dep DGP (1) sig1}
\begin{tabular*}{15cm}{@{\extracolsep{\fill}}cccccccc}
\hline\hline
\multirow{2}{*}{$b(n)$} & \multirow{2}{*}{$n$} & \multicolumn{6}{c}{$\tau_n$} \\
\cline{3-8}
& & $\sqrt{\ln(n)/n}$ & $n^{-2/5}$ & $ n^{-1/3}$ & $ n^{-1/4}$ & $ n^{-1/5}$ & $ n^{-1/6}$ \\
\hline
\multirow{4}{*}{$n^{1/6}$}
&$	100	$&$	0.283	$&$	0.245	$&$	0.283	$&$	0.324	$&$	0.352	$&$	0.376	$\\
&$	200	$&$	0.493	$&$	0.412	$&$	0.510	$&$	0.613	$&$	0.690	$&$	0.701	$\\
&$	400	$&$	0.750	$&$	0.637	$&$	0.783	$&$	0.881	$&$	0.908	$&$	0.921	$\\
&$	800	$&$	0.985	$&$	0.960	$&$	0.993	$&$	0.997	$&$	0.998	$&$	0.998	$\\
\hline
\multirow{4}{*}{$n^{1/5}$}
&$	100	$&$	0.242	$&$	0.164	$&$	0.242	$&$	0.278	$&$	0.304	$&$	0.318	$\\
&$	200	$&$	0.484	$&$	0.380	$&$	0.497	$&$	0.607	$&$	0.659	$&$	0.671	$\\
&$	400	$&$	0.750	$&$	0.637	$&$	0.783	$&$	0.881	$&$	0.908	$&$	0.921	$\\
&$	800	$&$	0.986	$&$	0.961	$&$	0.993	$&$	0.997	$&$	0.998	$&$	0.998	$\\
\hline
\multirow{4}{*}{$n^{1/4}$}
&$	100	$&$	0.242	$&$	0.164	$&$	0.242	$&$	0.278	$&$	0.304	$&$	0.318	$\\
&$	200	$&$	0.510	$&$	0.410	$&$	0.528	$&$	0.631	$&$	0.668	$&$	0.696	$\\
&$	400	$&$	0.768	$&$	0.647	$&$	0.790	$&$	0.874	$&$	0.900	$&$	0.908	$\\
&$	800	$&$	0.983	$&$	0.957	$&$	0.991	$&$	0.997	$&$	0.998	$&$	0.998	$\\
\hline
\multirow{4}{*}{$n^{1/3}$}
&$	100	$&$	0.223	$&$	0.148	$&$	0.223	$&$	0.264	$&$	0.287	$&$	0.289	$\\
&$	200	$&$	0.447	$&$	0.344	$&$	0.451	$&$	0.576	$&$	0.596	$&$	0.613	$\\
&$	400	$&$	0.695	$&$	0.588	$&$	0.738	$&$	0.848	$&$	0.886	$&$	0.897	$\\
&$	800	$&$	0.976	$&$	0.942	$&$	0.986	$&$	0.997	$&$	0.998	$&$	0.998	$\\
\hline\hline
\end{tabular*}
\end{table}

\begin{table}[H]
\small\centering\setstretch{1.3}
\caption{Power for DGP (2) of Case 1 with dependent data ($\nu=\mathcal{N}(0,1)$, $\alpha=0.05$)}
\label{tab:case1dep DGP (2) sig1}
\begin{tabular*}{15cm}{@{\extracolsep{\fill}}cccccccc}
\hline\hline
\multirow{2}{*}{$b(n)$} & \multirow{2}{*}{$n$} & \multicolumn{6}{c}{$\tau_n$} \\
\cline{3-8}
& & $\sqrt{\ln(n)/n}$ & $n^{-2/5}$ & $ n^{-1/3}$ & $ n^{-1/4}$ & $ n^{-1/5}$ & $ n^{-1/6}$ \\
\hline
\multirow{4}{*}{$n^{1/6}$}
&$	100	$&$	0.915	$&$	0.840	$&$	0.916	$&$	0.966	$&$	0.974	$&$	0.979	$\\
&$	200	$&$	0.995	$&$	0.991	$&$	0.995	$&$	0.999	$&$	1.000	$&$	1.000	$\\
&$	400	$&$	1.000	$&$	1.000	$&$	1.000	$&$	1.000	$&$	1.000	$&$	1.000	$\\
&$	800	$&$	1.000	$&$	1.000	$&$	1.000	$&$	1.000	$&$	1.000	$&$	1.000	$\\
\hline
\multirow{4}{*}{$n^{1/5}$}
&$	100	$&$	0.894	$&$	0.801	$&$	0.894	$&$	0.954	$&$	0.971	$&$	0.977	$\\
&$	200	$&$	0.995	$&$	0.991	$&$	0.995	$&$	0.999	$&$	1.000	$&$	1.000	$\\
&$	400	$&$	1.000	$&$	1.000	$&$	1.000	$&$	1.000	$&$	1.000	$&$	1.000	$\\
&$	800	$&$	1.000	$&$	1.000	$&$	1.000	$&$	1.000	$&$	1.000	$&$	1.000	$\\
\hline
\multirow{4}{*}{$n^{1/4}$}
&$	100	$&$	0.894	$&$	0.801	$&$	0.894	$&$	0.954	$&$	0.971	$&$	0.977	$\\
&$	200	$&$	0.995	$&$	0.991	$&$	0.995	$&$	1.000	$&$	1.000	$&$	1.000	$\\
&$	400	$&$	1.000	$&$	1.000	$&$	1.000	$&$	1.000	$&$	1.000	$&$	1.000	$\\
&$	800	$&$	1.000	$&$	1.000	$&$	1.000	$&$	1.000	$&$	1.000	$&$	1.000	$\\
\hline
\multirow{4}{*}{$n^{1/3}$}
&$	100	$&$	0.897	$&$	0.810	$&$	0.899	$&$	0.949	$&$	0.968	$&$	0.974	$\\
&$	200	$&$	0.995	$&$	0.991	$&$	0.995	$&$	0.999	$&$	1.000	$&$	1.000	$\\
&$	400	$&$	1.000	$&$	1.000	$&$	1.000	$&$	1.000	$&$	1.000	$&$	1.000	$\\
&$	800	$&$	1.000	$&$	1.000	$&$	1.000	$&$	1.000	$&$	1.000	$&$	1.000	$\\
\hline\hline
\end{tabular*}
\end{table}

\begin{table}[H]
\small\centering\setstretch{1.3}
\caption{Power for DGP (3) of Case 1 with dependent data ($\nu=\mathcal{N}(0,1)$, $\alpha=0.05$)}
\label{tab:case1dep DGP (3) sig1}
\begin{tabular*}{15cm}{@{\extracolsep{\fill}}cccccccc}
\hline\hline
\multirow{2}{*}{$b(n)$} & \multirow{2}{*}{$n$} & \multicolumn{6}{c}{$\tau_n$} \\
\cline{3-8}
& & $\sqrt{\ln(n)/n}$ & $n^{-2/5}$ & $ n^{-1/3}$ & $ n^{-1/4}$ & $ n^{-1/5}$ & $ n^{-1/6}$ \\
\hline
\multirow{4}{*}{$n^{1/6}$}
&$	100	$&$	0.986	$&$	0.957	$&$	0.986	$&$	0.995	$&$	0.999	$&$	0.999	$\\
&$	200	$&$	1.000	$&$	1.000	$&$	1.000	$&$	1.000	$&$	1.000	$&$	1.000	$\\
&$	400	$&$	1.000	$&$	1.000	$&$	1.000	$&$	1.000	$&$	1.000	$&$	1.000	$\\
&$	800	$&$	1.000	$&$	1.000	$&$	1.000	$&$	1.000	$&$	1.000	$&$	1.000	$\\
\hline
\multirow{4}{*}{$n^{1/5}$}
&$	100	$&$	0.979	$&$	0.947	$&$	0.979	$&$	0.993	$&$	0.996	$&$	0.998	$\\
&$	200	$&$	1.000	$&$	1.000	$&$	1.000	$&$	1.000	$&$	1.000	$&$	1.000	$\\
&$	400	$&$	1.000	$&$	1.000	$&$	1.000	$&$	1.000	$&$	1.000	$&$	1.000	$\\
&$	800	$&$	1.000	$&$	1.000	$&$	1.000	$&$	1.000	$&$	1.000	$&$	1.000	$\\
\hline
\multirow{4}{*}{$n^{1/4}$}
&$	100	$&$	0.979	$&$	0.947	$&$	0.979	$&$	0.993	$&$	0.996	$&$	0.998	$\\
&$	200	$&$	1.000	$&$	1.000	$&$	1.000	$&$	1.000	$&$	1.000	$&$	1.000	$\\
&$	400	$&$	1.000	$&$	1.000	$&$	1.000	$&$	1.000	$&$	1.000	$&$	1.000	$\\
&$	800	$&$	1.000	$&$	1.000	$&$	1.000	$&$	1.000	$&$	1.000	$&$	1.000	$\\
\hline
\multirow{4}{*}{$n^{1/3}$}
&$	100	$&$	0.986	$&$	0.957	$&$	0.986	$&$	0.995	$&$	0.998	$&$	0.999	$\\
&$	200	$&$	1.000	$&$	1.000	$&$	1.000	$&$	1.000	$&$	1.000	$&$	1.000	$\\
&$	400	$&$	1.000	$&$	1.000	$&$	1.000	$&$	1.000	$&$	1.000	$&$	1.000	$\\
&$	800	$&$	1.000	$&$	1.000	$&$	1.000	$&$	1.000	$&$	1.000	$&$	1.000	$\\
\hline\hline
\end{tabular*}
\end{table}

\begin{table}[H]
\small\centering\setstretch{1.3}
\caption{Size and power for Case 1 with i.i.d.\ data ($\nu=\mathcal{N}(0,5^2)$, $\alpha=0.05$)}
\label{tab:case1 sig5}
\begin{tabular*}{15cm}{@{\extracolsep{\fill}}cccccccc}
\hline\hline
\multirow{2}{*}{DGP} & \multirow{2}{*}{$n$} & \multicolumn{6}{c}{$\tau_n$} \\
\cline{3-8}
& & $\sqrt{\ln(n)/n}$ & $n^{-2/5}$ & $ n^{-1/3}$ & $ n^{-1/4}$ & $ n^{-1/5}$ & $ n^{-1/6}$ \\
\hline
\multirow{4}{*}{DGP (0)}&$	100	$&$	0.043	$&$	0.037	$&$	0.043	$&$	0.051	$&$	0.054	$&$	0.056	$\\
&$	200	$&$	0.041	$&$	0.034	$&$	0.041	$&$	0.046	$&$	0.048	$&$	0.049	$\\
&$	400	$&$	0.059	$&$	0.045	$&$	0.068	$&$	0.069	$&$	0.061	$&$	0.067	$\\
&$	800	$&$	0.051	$&$	0.045	$&$	0.051	$&$	0.069	$&$	0.073	$&$	0.074	$\\
\hline
\multirow{4}{*}{DGP (1)}
&$	100	$&$	0.247	$&$	0.185	$&$	0.248	$&$	0.316	$&$	0.348	$&$	0.373	$\\
&$	200	$&$	0.438	$&$	0.360	$&$	0.455	$&$	0.569	$&$	0.620	$&$	0.637	$\\
&$	400	$&$	0.677	$&$	0.583	$&$	0.706	$&$	0.814	$&$	0.849	$&$	0.860	$\\
&$	800	$&$	0.887	$&$	0.822	$&$	0.921	$&$	0.976	$&$	0.990	$&$	0.992	$\\
\hline
\multirow{4}{*}{DGP (2)}
&$	100	$&$	0.861	$&$	0.793	$&$	0.863	$&$	0.923	$&$	0.948	$&$	0.956	$\\
&$	200	$&$	0.997	$&$	0.982	$&$	0.997	$&$	1.000	$&$	1.000	$&$	1.000	$\\
&$	400	$&$	1.000	$&$	1.000	$&$	1.000	$&$	1.000	$&$	1.000	$&$	1.000	$\\
&$	800	$&$	1.000	$&$	1.000	$&$	1.000	$&$	1.000	$&$	1.000	$&$	1.000	$\\
\hline
\multirow{4}{*}{DGP (3)}
&$	100	$&$	0.992	$&$	0.983	$&$	0.992	$&$	0.998	$&$	0.999	$&$	0.999	$\\
&$	200	$&$	1.000	$&$	1.000	$&$	1.000	$&$	1.000	$&$	1.000	$&$	1.000	$\\
&$	400	$&$	1.000	$&$	1.000	$&$	1.000	$&$	1.000	$&$	1.000	$&$	1.000	$\\
&$	800	$&$	1.000	$&$	1.000	$&$	1.000	$&$	1.000	$&$	1.000	$&$	1.000	$\\
\hline\hline
\end{tabular*}
\end{table}

\begin{table}[H]
\small\centering\setstretch{1.3}
\caption{Size for Case 1 with dependent data ($\nu=\mathcal{N}(0,5^2)$, $\alpha=0.05$)}
\label{tab:case1dep null sig5}
\begin{tabular*}{15cm}{@{\extracolsep{\fill}}cccccccc}
\hline\hline
\multirow{2}{*}{$b(n)$} & \multirow{2}{*}{$n$} & \multicolumn{6}{c}{$\tau_n$} \\
\cline{3-8}
& & $\sqrt{\ln(n)/n}$ & $n^{-2/5}$ & $ n^{-1/3}$ & $ n^{-1/4}$ & $ n^{-1/5}$ & $ n^{-1/6}$ \\
\hline
\multirow{4}{*}{$n^{1/6}$}
&$	100	$&$	0.037	$&$	0.028	$&$	0.038	$&$	0.046	$&$	0.052	$&$	0.052	$\\
&$	200	$&$	0.053	$&$	0.038	$&$	0.052	$&$	0.059	$&$	0.067	$&$	0.069	$\\
&$	400	$&$	0.071	$&$	0.066	$&$	0.073	$&$	0.075	$&$	0.081	$&$	0.080	$\\
&$	800	$&$	0.062	$&$	0.051	$&$	0.070	$&$	0.077	$&$	0.083	$&$	0.083	$\\
\hline
\multirow{4}{*}{$n^{1/5}$}
&$	100	$&$	0.038	$&$	0.030	$&$	0.038	$&$	0.046	$&$	0.046	$&$	0.046	$\\
&$	200	$&$	0.037	$&$	0.029	$&$	0.037	$&$	0.040	$&$	0.048	$&$	0.050	$\\
&$	400	$&$	0.071	$&$	0.066	$&$	0.073	$&$	0.075	$&$	0.081	$&$	0.080	$\\
&$	800	$&$	0.046	$&$	0.050	$&$	0.064	$&$	0.077	$&$	0.083	$&$	0.083	$\\
\hline
\multirow{4}{*}{$n^{1/4}$}
&$	100	$&$	0.038	$&$	0.030	$&$	0.038	$&$	0.046	$&$	0.046	$&$	0.046	$\\
&$	200	$&$	0.040	$&$	0.033	$&$	0.042	$&$	0.055	$&$	0.058	$&$	0.061	$\\
&$	400	$&$	0.067	$&$	0.059	$&$	0.070	$&$	0.073	$&$	0.072	$&$	0.072	$\\
&$	800	$&$	0.083	$&$	0.072	$&$	0.088	$&$	0.097	$&$	0.097	$&$	0.100	$\\
\hline
\multirow{4}{*}{$n^{1/3}$}
&$	100	$&$	0.056	$&$	0.046	$&$	0.057	$&$	0.066	$&$	0.070	$&$	0.072	$\\
&$	200	$&$	0.046	$&$	0.037	$&$	0.047	$&$	0.055	$&$	0.059	$&$	0.067	$\\
&$	400	$&$	0.066	$&$	0.058	$&$	0.067	$&$	0.070	$&$	0.074	$&$	0.075	$\\
&$	800	$&$	0.060	$&$	0.038	$&$	0.072	$&$	0.081	$&$	0.086	$&$	0.087	$\\
\hline\hline
\end{tabular*}
\end{table}

\begin{table}[H]
\small\centering\setstretch{1.3}
\caption{Power for DGP (1) of Case 1 with dependent data ($\nu=\mathcal{N}(0,5^2)$, $\alpha=0.05$)}
\label{tab:case1dep DGP (1) sig5}
\begin{tabular*}{15cm}{@{\extracolsep{\fill}}cccccccc}
\hline\hline
\multirow{2}{*}{$b(n)$} & \multirow{2}{*}{$n$} & \multicolumn{6}{c}{$\tau_n$} \\
\cline{3-8}
& & $\sqrt{\ln(n)/n}$ & $n^{-2/5}$ & $ n^{-1/3}$ & $ n^{-1/4}$ & $ n^{-1/5}$ & $ n^{-1/6}$ \\
\hline
\multirow{4}{*}{$n^{1/6}$}
&$	100	$&$	0.331	$&$	0.249	$&$	0.331	$&$	0.393	$&$	0.416	$&$	0.436	$\\
&$	200	$&$	0.517	$&$	0.393	$&$	0.552	$&$	0.654	$&$	0.683	$&$	0.704	$\\
&$	400	$&$	0.758	$&$	0.671	$&$	0.802	$&$	0.898	$&$	0.916	$&$	0.925	$\\
&$	800	$&$	0.988	$&$	0.965	$&$	0.992	$&$	1.000	$&$	1.000	$&$	1.000	$\\
\hline
\multirow{4}{*}{$n^{1/5}$}
&$	100	$&$	0.255	$&$	0.206	$&$	0.255	$&$	0.334	$&$	0.355	$&$	0.375	$\\
&$	200	$&$	0.495	$&$	0.372	$&$	0.510	$&$	0.625	$&$	0.677	$&$	0.688	$\\
&$	400	$&$	0.758	$&$	0.671	$&$	0.802	$&$	0.898	$&$	0.916	$&$	0.925	$\\
&$	800	$&$	0.990	$&$	0.969	$&$	0.992	$&$	1.000	$&$	1.000	$&$	1.000	$\\
\hline
\multirow{4}{*}{$n^{1/4}$}
&$	100	$&$	0.255	$&$	0.206	$&$	0.255	$&$	0.334	$&$	0.355	$&$	0.375	$\\
&$	200	$&$	0.552	$&$	0.423	$&$	0.576	$&$	0.683	$&$	0.690	$&$	0.705	$\\
&$	400	$&$	0.758	$&$	0.652	$&$	0.799	$&$	0.894	$&$	0.920	$&$	0.925	$\\
&$	800	$&$	0.988	$&$	0.962	$&$	0.992	$&$	1.000	$&$	1.000	$&$	1.000	$\\
\hline
\multirow{4}{*}{$n^{1/3}$}
&$	100	$&$	0.261	$&$	0.184	$&$	0.262	$&$	0.332	$&$	0.361	$&$	0.367	$\\
&$	200	$&$	0.483	$&$	0.374	$&$	0.504	$&$	0.622	$&$	0.669	$&$	0.688	$\\
&$	400	$&$	0.746	$&$	0.642	$&$	0.776	$&$	0.884	$&$	0.916	$&$	0.922	$\\
&$	800	$&$	0.977	$&$	0.950	$&$	0.989	$&$	1.000	$&$	1.000	$&$	1.000	$\\
\hline\hline
\end{tabular*}
\end{table}

\begin{table}[H]
\small\centering\setstretch{1.3}
\caption{Power for DGP (2) of Case 1 with dependent data ($\nu=\mathcal{N}(0,5^2)$, $\alpha=0.05$)}
\label{tab:case1dep DGP (2) sig5}
\begin{tabular*}{15cm}{@{\extracolsep{\fill}}cccccccc}
\hline\hline
\multirow{2}{*}{$b(n)$} & \multirow{2}{*}{$n$} & \multicolumn{6}{c}{$\tau_n$} \\
\cline{3-8}
& & $\sqrt{\ln(n)/n}$ & $n^{-2/5}$ & $ n^{-1/3}$ & $ n^{-1/4}$ & $ n^{-1/5}$ & $ n^{-1/6}$ \\
\hline
\multirow{4}{*}{$n^{1/6}$}
&$	100	$&$	0.976	$&$	0.920	$&$	0.976	$&$	0.990	$&$	0.992	$&$	0.993	$\\
&$	200	$&$	1.000	$&$	0.998	$&$	1.000	$&$	1.000	$&$	1.000	$&$	1.000	$\\
&$	400	$&$	1.000	$&$	1.000	$&$	1.000	$&$	1.000	$&$	1.000	$&$	1.000	$\\
&$	800	$&$	1.000	$&$	1.000	$&$	1.000	$&$	1.000	$&$	1.000	$&$	1.000	$\\
\hline
\multirow{4}{*}{$n^{1/5}$}
&$	100	$&$	0.966	$&$	0.920	$&$	0.966	$&$	0.990	$&$	0.992	$&$	0.993	$\\
&$	200	$&$	1.000	$&$	0.999	$&$	1.000	$&$	1.000	$&$	1.000	$&$	1.000	$\\
&$	400	$&$	1.000	$&$	1.000	$&$	1.000	$&$	1.000	$&$	1.000	$&$	1.000	$\\
&$	800	$&$	1.000	$&$	1.000	$&$	1.000	$&$	1.000	$&$	1.000	$&$	1.000	$\\
\hline
\multirow{4}{*}{$n^{1/4}$}
&$	100	$&$	0.966	$&$	0.920	$&$	0.966	$&$	0.990	$&$	0.992	$&$	0.993	$\\
&$	200	$&$	1.000	$&$	0.999	$&$	1.000	$&$	1.000	$&$	1.000	$&$	1.000	$\\
&$	400	$&$	1.000	$&$	1.000	$&$	1.000	$&$	1.000	$&$	1.000	$&$	1.000	$\\
&$	800	$&$	1.000	$&$	1.000	$&$	1.000	$&$	1.000	$&$	1.000	$&$	1.000	$\\
\hline
\multirow{4}{*}{$n^{1/3}$}
&$	100	$&$	0.961	$&$	0.912	$&$	0.961	$&$	0.984	$&$	0.991	$&$	0.992	$\\
&$	200	$&$	1.000	$&$	0.999	$&$	1.000	$&$	1.000	$&$	1.000	$&$	1.000	$\\
&$	400	$&$	1.000	$&$	1.000	$&$	1.000	$&$	1.000	$&$	1.000	$&$	1.000	$\\
&$	800	$&$	1.000	$&$	1.000	$&$	1.000	$&$	1.000	$&$	1.000	$&$	1.000	$\\
\hline\hline
\end{tabular*}
\end{table}

\begin{table}[H]
\small\centering\setstretch{1.3}
\caption{Power for DGP (3) of Case 1 with dependent data ($\nu=\mathcal{N}(0,5^2)$, $\alpha=0.05$)}
\label{tab:case1dep DGP (3) sig5}
\begin{tabular*}{15cm}{@{\extracolsep{\fill}}cccccccc}
\hline\hline
\multirow{2}{*}{$b(n)$} & \multirow{2}{*}{$n$} & \multicolumn{6}{c}{$\tau_n$} \\
\cline{3-8}
& & $\sqrt{\ln(n)/n}$ & $n^{-2/5}$ & $ n^{-1/3}$ & $ n^{-1/4}$ & $ n^{-1/5}$ & $ n^{-1/6}$ \\
\hline
\multirow{4}{*}{$n^{1/6}$}
&$	100	$&$	0.999	$&$	0.995	$&$	0.999	$&$	1.000	$&$	1.000	$&$	1.000	$\\
&$	200	$&$	1.000	$&$	1.000	$&$	1.000	$&$	1.000	$&$	1.000	$&$	1.000	$\\
&$	400	$&$	1.000	$&$	1.000	$&$	1.000	$&$	1.000	$&$	1.000	$&$	1.000	$\\
&$	800	$&$	1.000	$&$	1.000	$&$	1.000	$&$	1.000	$&$	1.000	$&$	1.000	$\\
\hline
\multirow{4}{*}{$n^{1/5}$}
&$	100	$&$	0.999	$&$	0.994	$&$	0.999	$&$	1.000	$&$	1.000	$&$	1.000	$\\
&$	200	$&$	1.000	$&$	1.000	$&$	1.000	$&$	1.000	$&$	1.000	$&$	1.000	$\\
&$	400	$&$	1.000	$&$	1.000	$&$	1.000	$&$	1.000	$&$	1.000	$&$	1.000	$\\
&$	800	$&$	1.000	$&$	1.000	$&$	1.000	$&$	1.000	$&$	1.000	$&$	1.000	$\\
\hline
\multirow{4}{*}{$n^{1/4}$}
&$	100	$&$	0.999	$&$	0.994	$&$	0.999	$&$	1.000	$&$	1.000	$&$	1.000	$\\
&$	200	$&$	1.000	$&$	1.000	$&$	1.000	$&$	1.000	$&$	1.000	$&$	1.000	$\\
&$	400	$&$	1.000	$&$	1.000	$&$	1.000	$&$	1.000	$&$	1.000	$&$	1.000	$\\
&$	800	$&$	1.000	$&$	1.000	$&$	1.000	$&$	1.000	$&$	1.000	$&$	1.000	$\\
\hline
\multirow{4}{*}{$n^{1/3}$}
&$	100	$&$	0.999	$&$	0.995	$&$	0.999	$&$	1.000	$&$	1.000	$&$	1.000	$\\
&$	200	$&$	1.000	$&$	1.000	$&$	1.000	$&$	1.000	$&$	1.000	$&$	1.000	$\\
&$	400	$&$	1.000	$&$	1.000	$&$	1.000	$&$	1.000	$&$	1.000	$&$	1.000	$\\
&$	800	$&$	1.000	$&$	1.000	$&$	1.000	$&$	1.000	$&$	1.000	$&$	1.000	$\\
\hline\hline
\end{tabular*}
\end{table}

\begin{table}[H]
\small\centering\setstretch{1.3}
\caption{Size for Case 1 with dependent data and larger samples ($\alpha=0.05$)}
\label{tab:case1dep null large samples}
\begin{tabular*}{15cm}{@{\extracolsep{\fill}}cccccccc}
\hline\hline
\multirow{2}{*}{$b(n)$} & \multirow{2}{*}{$n$} & \multicolumn{6}{c}{$\tau_n$} \\
\cline{3-8}
& & $\sqrt{\ln(n)/n}$ & $n^{-2/5}$ & $ n^{-1/3}$ & $ n^{-1/4}$ & $ n^{-1/5}$ & $ n^{-1/6}$ \\
\hline
\multirow{2}{*}{$n^{1/6}$}&$	1600	$&$	0.045	$&$	0.037	$&$	0.045	$&$	0.049	$&$	0.059	$&$	0.069	$\\
&$	3200	$&$	0.044	$&$	0.034	$&$	0.057	$&$	0.070	$&$	0.079	$&$	0.084	$\\
\hline
\multirow{2}{*}{$n^{1/5}$}
&$	1600	$&$	0.050	$&$	0.045	$&$	0.059	$&$	0.073	$&$	0.079	$&$	0.082	$\\
&$	3200	$&$	0.037	$&$	0.032	$&$	0.049	$&$	0.071	$&$	0.072	$&$	0.078	$\\
\hline
\multirow{2}{*}{$n^{1/4}$}
&$	1600	$&$	0.048	$&$	0.045	$&$	0.049	$&$	0.058	$&$	0.065	$&$	0.071	$\\
&$	3200	$&$	0.036	$&$	0.034	$&$	0.049	$&$	0.070	$&$	0.078	$&$	0.076	$\\
\hline
\multirow{2}{*}{$n^{1/3}$}
&$	1600	$&$	0.051	$&$	0.048	$&$	0.061	$&$	0.073	$&$	0.076	$&$	0.086	$\\
&$	3200	$&$	0.045	$&$	0.037	$&$	0.050	$&$	0.072	$&$	0.070	$&$	0.071	$\\
\hline\hline
\end{tabular*}
\end{table}

\begin{table}[H]
\small\centering\setstretch{1.3}
\caption{Size for Case 2 with i.i.d.\ data}
\label{tab:case2 null}
\begin{tabular*}{15cm}{@{\extracolsep{\fill}}cccccccc}
\hline\hline
\multirow{2}{*}{$\alpha$} & \multirow{2}{*}{$n$} & \multicolumn{6}{c}{$\tau_n$} \\
\cline{3-8}
& & $\sqrt{\ln(n)/n}$ & $n^{-2/5}$ & $ n^{-1/3}$ & $ n^{-1/4}$ & $ n^{-1/5}$ & $ n^{-1/6}$ \\
\hline
\multirow{4}{*}{$0.01$}
&$	100	$&$	0.004	$&$	0.004	$&$	0.004	$&$	0.004	$&$	0.004	$&$	0.004	$\\
&$	200	$&$	0.007	$&$	0.006	$&$	0.007	$&$	0.009	$&$	0.008	$&$	0.009	$\\
&$	400	$&$	0.006	$&$	0.004	$&$	0.007	$&$	0.009	$&$	0.009	$&$	0.009	$\\
&$	800	$&$	0.001	$&$	0.001	$&$	0.001	$&$	0.001	$&$	0.001	$&$	0.001	$\\
\hline
\multirow{4}{*}{$0.025$}
&$	100	$&$	0.016	$&$	0.012	$&$	0.016	$&$	0.017	$&$	0.017	$&$	0.017	$\\
&$	200	$&$	0.017	$&$	0.010	$&$	0.017	$&$	0.023	$&$	0.026	$&$	0.026	$\\
&$	400	$&$	0.012	$&$	0.012	$&$	0.012	$&$	0.016	$&$	0.017	$&$	0.017	$\\
&$	800	$&$	0.019	$&$	0.013	$&$	0.020	$&$	0.027	$&$	0.030	$&$	0.032	$\\
\hline
\multirow{4}{*}{$0.05$}
&$	100	$&$	0.025	$&$	0.021	$&$	0.025	$&$	0.034	$&$	0.042	$&$	0.043	$\\
&$	200	$&$	0.043	$&$	0.040	$&$	0.043	$&$	0.049	$&$	0.051	$&$	0.052	$\\
&$	400	$&$	0.031	$&$	0.030	$&$	0.031	$&$	0.035	$&$	0.038	$&$	0.038	$\\
&$	800	$&$	0.048	$&$	0.047	$&$	0.048	$&$	0.057	$&$	0.057	$&$	0.059	$\\
\hline
\multirow{4}{*}{$0.1$}
&$	100	$&$	0.063	$&$	0.054	$&$	0.063	$&$	0.074	$&$	0.077	$&$	0.082	$\\
&$	200	$&$	0.099	$&$	0.088	$&$	0.100	$&$	0.109	$&$	0.113	$&$	0.116	$\\
&$	400	$&$	0.082	$&$	0.074	$&$	0.083	$&$	0.089	$&$	0.092	$&$	0.089	$\\
&$	800	$&$	0.093	$&$	0.084	$&$	0.096	$&$	0.104	$&$	0.105	$&$	0.110	$\\
\hline
\multirow{4}{*}{$0.2$}
&$	100	$&$	0.154	$&$	0.150	$&$	0.154	$&$	0.167	$&$	0.170	$&$	0.171	$\\
&$	200	$&$	0.229	$&$	0.210	$&$	0.233	$&$	0.233	$&$	0.238	$&$	0.239	$\\
&$	400	$&$	0.172	$&$	0.155	$&$	0.172	$&$	0.175	$&$	0.173	$&$	0.178	$\\
&$	800	$&$	0.215	$&$	0.204	$&$	0.216	$&$	0.218	$&$	0.215	$&$	0.215	$\\
\hline\hline
\end{tabular*}
\end{table}

\begin{table}[H]
\small\centering\setstretch{1.3}
\caption{Power for Case 2 with i.i.d.\ data ($\alpha=0.05$)}
\label{tab:case2 alt}
\begin{tabular*}{15cm}{@{\extracolsep{\fill}}cccccccc}
\hline\hline
\multirow{2}{*}{DGP} & \multirow{2}{*}{$n$} & \multicolumn{6}{c}{$\tau_n$} \\
\cline{3-8}
& & $\sqrt{\ln(n)/n}$ & $n^{-2/5}$ & $ n^{-1/3}$ & $ n^{-1/4}$ & $ n^{-1/5}$ & $ n^{-1/6}$ \\
\hline
\multirow{4}{*}{DGP (1)}
&$	100	$&$	0.177	$&$	0.137	$&$	0.177	$&$	0.216	$&$	0.230	$&$	0.233	$\\
&$	200	$&$	0.332	$&$	0.255	$&$	0.345	$&$	0.425	$&$	0.464	$&$	0.479	$\\
&$	400	$&$	0.615	$&$	0.536	$&$	0.634	$&$	0.708	$&$	0.728	$&$	0.738	$\\
&$	800	$&$	0.767	$&$	0.716	$&$	0.791	$&$	0.860	$&$	0.880	$&$	0.887	$\\
\hline
\multirow{4}{*}{DGP (2)}
&$	100	$&$	0.769	$&$	0.684	$&$	0.771	$&$	0.829	$&$	0.843	$&$	0.856	$\\
&$	200	$&$	0.915	$&$	0.876	$&$	0.920	$&$	0.957	$&$	0.967	$&$	0.972	$\\
&$	400	$&$	0.997	$&$	0.990	$&$	0.997	$&$	0.999	$&$	0.999	$&$	0.999	$\\
&$	800	$&$	1.000	$&$	1.000	$&$	1.000	$&$	1.000	$&$	1.000	$&$	1.000	$\\
\hline
\multirow{4}{*}{DGP (3)}
&$	100	$&$	0.935	$&$	0.889	$&$	0.935	$&$	0.974	$&$	0.983	$&$	0.985	$\\
&$	200	$&$	0.997	$&$	0.994	$&$	0.998	$&$	1.000	$&$	1.000	$&$	1.000	$\\
&$	400	$&$	1.000	$&$	1.000	$&$	1.000	$&$	1.000	$&$	1.000	$&$	1.000	$\\
&$	800	$&$	1.000	$&$	1.000	$&$	1.000	$&$	1.000	$&$	1.000	$&$	1.000	$\\
\hline\hline
\end{tabular*}
\end{table}

\begin{table}[H]
\small\centering\setstretch{1.3}
\caption{Size for Case 2 with dependent data ($\alpha=0.05$)}
\label{tab:case2dep null}
\begin{tabular*}{15cm}{@{\extracolsep{\fill}}cccccccc}
\hline\hline
\multirow{2}{*}{$b(n)$} & \multirow{2}{*}{$n$} & \multicolumn{6}{c}{$\tau_n$} \\
\cline{3-8}
& & $\sqrt{\ln(n)/n}$ & $n^{-2/5}$ & $ n^{-1/3}$ & $ n^{-1/4}$ & $ n^{-1/5}$ & $ n^{-1/6}$ \\
\hline
\multirow{4}{*}{$n^{1/6}$}
&$	100	$&$	0.030	$&$	0.029	$&$	0.030	$&$	0.037	$&$	0.039	$&$	0.039	$\\
&$	200	$&$	0.040	$&$	0.036	$&$	0.040	$&$	0.052	$&$	0.059	$&$	0.057	$\\
&$	400	$&$	0.030	$&$	0.024	$&$	0.034	$&$	0.040	$&$	0.046	$&$	0.049	$\\
&$	800	$&$	0.036	$&$	0.034	$&$	0.039	$&$	0.046	$&$	0.047	$&$	0.047	$\\
\hline
\multirow{4}{*}{$n^{1/5}$}
&$	100	$&$	0.041	$&$	0.030	$&$	0.041	$&$	0.044	$&$	0.048	$&$	0.050	$\\
&$	200	$&$	0.048	$&$	0.038	$&$	0.048	$&$	0.056	$&$	0.056	$&$	0.060	$\\
&$	400	$&$	0.030	$&$	0.024	$&$	0.034	$&$	0.040	$&$	0.046	$&$	0.049	$\\
&$	800	$&$	0.045	$&$	0.039	$&$	0.045	$&$	0.044	$&$	0.044	$&$	0.045	$\\
\hline
\multirow{4}{*}{$n^{1/4}$}
&$	100	$&$	0.041	$&$	0.030	$&$	0.041	$&$	0.044	$&$	0.048	$&$	0.050	$\\
&$	200	$&$	0.052	$&$	0.042	$&$	0.053	$&$	0.057	$&$	0.060	$&$	0.060	$\\
&$	400	$&$	0.032	$&$	0.024	$&$	0.034	$&$	0.046	$&$	0.049	$&$	0.053	$\\
&$	800	$&$	0.046	$&$	0.039	$&$	0.046	$&$	0.046	$&$	0.046	$&$	0.046	$\\
\hline
\multirow{4}{*}{$n^{1/3}$}
&$	100	$&$	0.029	$&$	0.027	$&$	0.029	$&$	0.033	$&$	0.036	$&$	0.039	$\\
&$	200	$&$	0.047	$&$	0.038	$&$	0.048	$&$	0.054	$&$	0.056	$&$	0.057	$\\
&$	400	$&$	0.037	$&$	0.028	$&$	0.038	$&$	0.055	$&$	0.055	$&$	0.055	$\\
&$	800	$&$	0.032	$&$	0.025	$&$	0.033	$&$	0.039	$&$	0.042	$&$	0.044	$\\
\hline\hline
\end{tabular*}
\end{table}

\begin{table}[H]
\small\centering\setstretch{1.3}
\caption{Power for DGP (1) of Case 2 with dependent data ($\alpha=0.05$)}
\label{tab:case2dep DGP (1)}
\begin{tabular*}{15cm}{@{\extracolsep{\fill}}cccccccc}
\hline\hline
\multirow{2}{*}{$b(n)$} & \multirow{2}{*}{$n$} & \multicolumn{6}{c}{$\tau_n$} \\
\cline{3-8}
& & $\sqrt{\ln(n)/n}$ & $n^{-2/5}$ & $ n^{-1/3}$ & $ n^{-1/4}$ & $ n^{-1/5}$ & $ n^{-1/6}$ \\
\hline
\multirow{4}{*}{$n^{1/6}$}
&$	100	$&$	0.175	$&$	0.129	$&$	0.175	$&$	0.210	$&$	0.231	$&$	0.249	$\\
&$	200	$&$	0.283	$&$	0.223	$&$	0.287	$&$	0.383	$&$	0.414	$&$	0.431	$\\
&$	400	$&$	0.589	$&$	0.505	$&$	0.617	$&$	0.684	$&$	0.712	$&$	0.719	$\\
&$	800	$&$	0.761	$&$	0.692	$&$	0.787	$&$	0.859	$&$	0.872	$&$	0.880	$\\
\hline
\multirow{4}{*}{$n^{1/5}$}
&$	100	$&$	0.158	$&$	0.126	$&$	0.159	$&$	0.206	$&$	0.222	$&$	0.227	$\\
&$	200	$&$	0.320	$&$	0.248	$&$	0.327	$&$	0.413	$&$	0.445	$&$	0.460	$\\
&$	400	$&$	0.589	$&$	0.505	$&$	0.617	$&$	0.684	$&$	0.712	$&$	0.719	$\\
&$	800	$&$	0.764	$&$	0.704	$&$	0.789	$&$	0.865	$&$	0.880	$&$	0.886	$\\
\hline
\multirow{4}{*}{$n^{1/4}$}
&$	100	$&$	0.158	$&$	0.126	$&$	0.159	$&$	0.206	$&$	0.222	$&$	0.227	$\\
&$	200	$&$	0.320	$&$	0.248	$&$	0.325	$&$	0.413	$&$	0.444	$&$	0.465	$\\
&$	400	$&$	0.558	$&$	0.465	$&$	0.587	$&$	0.667	$&$	0.697	$&$	0.711	$\\
&$	800	$&$	0.797	$&$	0.752	$&$	0.829	$&$	0.879	$&$	0.901	$&$	0.911	$\\
\hline
\multirow{4}{*}{$n^{1/3}$}
&$	100	$&$	0.153	$&$	0.120	$&$	0.154	$&$	0.183	$&$	0.211	$&$	0.222	$\\
&$	200	$&$	0.307	$&$	0.248	$&$	0.314	$&$	0.406	$&$	0.431	$&$	0.444	$\\
&$	400	$&$	0.547	$&$	0.455	$&$	0.572	$&$	0.657	$&$	0.677	$&$	0.700	$\\
&$	800	$&$	0.796	$&$	0.738	$&$	0.823	$&$	0.878	$&$	0.898	$&$	0.911	$\\
\hline\hline
\end{tabular*}
\end{table}

\begin{table}[H]
\small\centering\setstretch{1.3}
\caption{Power for DGP (2) of Case 2 with dependent data ($\alpha=0.05$)}
\label{tab:case2dep DGP (2)}
\begin{tabular*}{15cm}{@{\extracolsep{\fill}}cccccccc}
\hline\hline
\multirow{2}{*}{$b(n)$} & \multirow{2}{*}{$n$} & \multicolumn{6}{c}{$\tau_n$} \\
\cline{3-8}
& & $\sqrt{\ln(n)/n}$ & $n^{-2/5}$ & $ n^{-1/3}$ & $ n^{-1/4}$ & $ n^{-1/5}$ & $ n^{-1/6}$ \\
\hline
\multirow{4}{*}{$n^{1/6}$}
&$	100	$&$	0.714	$&$	0.607	$&$	0.715	$&$	0.783	$&$	0.814	$&$	0.830	$\\
&$	200	$&$	0.914	$&$	0.858	$&$	0.921	$&$	0.948	$&$	0.960	$&$	0.970	$\\
&$	400	$&$	0.993	$&$	0.987	$&$	0.996	$&$	0.999	$&$	0.999	$&$	0.999	$\\
&$	800	$&$	1.000	$&$	1.000	$&$	1.000	$&$	1.000	$&$	1.000	$&$	1.000	$\\
\hline
\multirow{4}{*}{$n^{1/5}$}
&$	100	$&$	0.742	$&$	0.662	$&$	0.744	$&$	0.809	$&$	0.830	$&$	0.842	$\\
&$	200	$&$	0.911	$&$	0.857	$&$	0.915	$&$	0.946	$&$	0.960	$&$	0.966	$\\
&$	400	$&$	0.993	$&$	0.987	$&$	0.996	$&$	0.999	$&$	0.999	$&$	0.999	$\\
&$	800	$&$	1.000	$&$	1.000	$&$	1.000	$&$	1.000	$&$	1.000	$&$	1.000	$\\
\hline
\multirow{4}{*}{$n^{1/4}$}
&$	100	$&$	0.742	$&$	0.662	$&$	0.744	$&$	0.809	$&$	0.830	$&$	0.842	$\\
&$	200	$&$	0.898	$&$	0.842	$&$	0.906	$&$	0.942	$&$	0.955	$&$	0.960	$\\
&$	400	$&$	0.990	$&$	0.984	$&$	0.993	$&$	0.999	$&$	0.999	$&$	0.999	$\\
&$	800	$&$	1.000	$&$	1.000	$&$	1.000	$&$	1.000	$&$	1.000	$&$	1.000	$\\
\hline
\multirow{4}{*}{$n^{1/3}$}
&$	100	$&$	0.745	$&$	0.671	$&$	0.746	$&$	0.810	$&$	0.833	$&$	0.845	$\\
&$	200	$&$	0.919	$&$	0.866	$&$	0.922	$&$	0.950	$&$	0.962	$&$	0.970	$\\
&$	400	$&$	0.991	$&$	0.985	$&$	0.993	$&$	0.999	$&$	0.999	$&$	0.999	$\\
&$	800	$&$	1.000	$&$	1.000	$&$	1.000	$&$	1.000	$&$	1.000	$&$	1.000	$\\
\hline\hline
\end{tabular*}
\end{table}

\begin{table}[H]
\small\centering\setstretch{1.3}
\caption{Power for DGP (3) of Case 2 with dependent data ($\alpha=0.05$)}
\label{tab:case2dep DGP (3)}
\begin{tabular*}{15cm}{@{\extracolsep{\fill}}cccccccc}
\hline\hline
\multirow{2}{*}{$b(n)$} & \multirow{2}{*}{$n$} & \multicolumn{6}{c}{$\tau_n$} \\
\cline{3-8}
& & $\sqrt{\ln(n)/n}$ & $n^{-2/5}$ & $ n^{-1/3}$ & $ n^{-1/4}$ & $ n^{-1/5}$ & $ n^{-1/6}$ \\
\hline
\multirow{4}{*}{$n^{1/6}$}
&$	100	$&$	0.926	$&$	0.872	$&$	0.927	$&$	0.962	$&$	0.972	$&$	0.977	$\\
&$	200	$&$	0.999	$&$	0.994	$&$	0.999	$&$	1.000	$&$	1.000	$&$	1.000	$\\
&$	400	$&$	1.000	$&$	1.000	$&$	1.000	$&$	1.000	$&$	1.000	$&$	1.000	$\\
&$	800	$&$	1.000	$&$	1.000	$&$	1.000	$&$	1.000	$&$	1.000	$&$	1.000	$\\
\hline
\multirow{4}{*}{$n^{1/5}$}
&$	100	$&$	0.918	$&$	0.864	$&$	0.918	$&$	0.957	$&$	0.970	$&$	0.973	$\\
&$	200	$&$	0.999	$&$	0.993	$&$	0.999	$&$	1.000	$&$	1.000	$&$	1.000	$\\
&$	400	$&$	1.000	$&$	1.000	$&$	1.000	$&$	1.000	$&$	1.000	$&$	1.000	$\\
&$	800	$&$	1.000	$&$	1.000	$&$	1.000	$&$	1.000	$&$	1.000	$&$	1.000	$\\
\hline
\multirow{4}{*}{$n^{1/4}$}
&$	100	$&$	0.918	$&$	0.864	$&$	0.918	$&$	0.957	$&$	0.970	$&$	0.973	$\\
&$	200	$&$	0.999	$&$	0.994	$&$	0.999	$&$	1.000	$&$	1.000	$&$	1.000	$\\
&$	400	$&$	1.000	$&$	1.000	$&$	1.000	$&$	1.000	$&$	1.000	$&$	1.000	$\\
&$	800	$&$	1.000	$&$	1.000	$&$	1.000	$&$	1.000	$&$	1.000	$&$	1.000	$\\
\hline
\multirow{4}{*}{$n^{1/3}$}
&$	100	$&$	0.926	$&$	0.874	$&$	0.926	$&$	0.960	$&$	0.972	$&$	0.976	$\\
&$	200	$&$	0.999	$&$	0.996	$&$	0.999	$&$	1.000	$&$	1.000	$&$	1.000	$\\
&$	400	$&$	1.000	$&$	1.000	$&$	1.000	$&$	1.000	$&$	1.000	$&$	1.000	$\\
&$	800	$&$	1.000	$&$	1.000	$&$	1.000	$&$	1.000	$&$	1.000	$&$	1.000	$\\
\hline\hline
\end{tabular*}
\end{table}

\begin{table}[H]
\small\centering\setstretch{1.3}
\caption{Size and power for Case 3 with i.i.d.\ data ($\alpha=0.05$)}
\label{tab:case3}
\begin{tabular*}{15cm}{@{\extracolsep{\fill}}cccccccc}
\hline\hline
\multirow{2}{*}{DGP} & \multirow{2}{*}{$n$} & \multicolumn{6}{c}{$\tau_n$} \\
\cline{3-8}
& & $\sqrt{\ln(n)/n}$ & $n^{-2/5}$ & $ n^{-1/3}$ & $ n^{-1/4}$ & $ n^{-1/5}$ & $ n^{-1/6}$ \\
\hline
\multirow{4}{*}{DGP (0)}
&$	100	$&$	0.039	$&$	0.027	$&$	0.039	$&$	0.050	$&$	0.053	$&$	0.056	$\\
&$	200	$&$	0.054	$&$	0.040	$&$	0.055	$&$	0.058	$&$	0.058	$&$	0.061	$\\
&$	400	$&$	0.039	$&$	0.033	$&$	0.043	$&$	0.050	$&$	0.050	$&$	0.051	$\\
&$	800	$&$	0.039	$&$	0.037	$&$	0.044	$&$	0.044	$&$	0.046	$&$	0.044	$\\
\hline
\multirow{4}{*}{DGP (1)}
&$	100	$&$	0.136	$&$	0.104	$&$	0.137	$&$	0.160	$&$	0.162	$&$	0.169	$\\
&$	200	$&$	0.198	$&$	0.173	$&$	0.209	$&$	0.265	$&$	0.283	$&$	0.291	$\\
&$	400	$&$	0.408	$&$	0.325	$&$	0.439	$&$	0.516	$&$	0.536	$&$	0.553	$\\
&$	800	$&$	0.713	$&$	0.616	$&$	0.748	$&$	0.811	$&$	0.830	$&$	0.847	$\\
\hline
\multirow{4}{*}{DGP (2)}
&$	100	$&$	0.631	$&$	0.514	$&$	0.632	$&$	0.737	$&$	0.788	$&$	0.811	$\\
&$	200	$&$	0.860	$&$	0.782	$&$	0.868	$&$	0.941	$&$	0.961	$&$	0.966	$\\
&$	400	$&$	0.997	$&$	0.987	$&$	0.998	$&$	1.000	$&$	1.000	$&$	1.000	$\\
&$	800	$&$	1.000	$&$	1.000	$&$	1.000	$&$	1.000	$&$	1.000	$&$	1.000	$\\
\hline
\multirow{4}{*}{DGP (3)}
&$	100	$&$	0.906	$&$	0.823	$&$	0.906	$&$	0.949	$&$	0.972	$&$	0.976	$\\
&$	200	$&$	0.998	$&$	0.995	$&$	0.998	$&$	0.999	$&$	1.000	$&$	1.000	$\\
&$	400	$&$	1.000	$&$	1.000	$&$	1.000	$&$	1.000	$&$	1.000	$&$	1.000	$\\
&$	800	$&$	1.000	$&$	1.000	$&$	1.000	$&$	1.000	$&$	1.000	$&$	1.000	$\\
\hline\hline
\end{tabular*}
\end{table}

\begin{table}[H]
\small\centering\setstretch{1.3}
\caption{Size for Case 3 with dependent data ($\alpha=0.05$)}
\label{tab:case3dep null}
\begin{tabular*}{15cm}{@{\extracolsep{\fill}}cccccccc}
\hline\hline
\multirow{2}{*}{$b(n)$} & \multirow{2}{*}{$n$} & \multicolumn{6}{c}{$\tau_n$} \\
\cline{3-8}
& & $\sqrt{\ln(n)/n}$ & $n^{-2/5}$ & $ n^{-1/3}$ & $ n^{-1/4}$ & $ n^{-1/5}$ & $ n^{-1/6}$ \\
\hline
\multirow{4}{*}{$n^{1/6}$}
&$	100	$&$	0.050	$&$	0.040	$&$	0.050	$&$	0.060	$&$	0.056	$&$	0.057	$\\
&$	200	$&$	0.038	$&$	0.031	$&$	0.038	$&$	0.039	$&$	0.043	$&$	0.042	$\\
&$	400	$&$	0.058	$&$	0.050	$&$	0.058	$&$	0.059	$&$	0.060	$&$	0.060	$\\
&$	800	$&$	0.044	$&$	0.040	$&$	0.046	$&$	0.054	$&$	0.058	$&$	0.059	$\\
\hline
\multirow{4}{*}{$n^{1/5}$}
&$	100	$&$	0.034	$&$	0.025	$&$	0.034	$&$	0.047	$&$	0.050	$&$	0.050	$\\
&$	200	$&$	0.036	$&$	0.030	$&$	0.037	$&$	0.040	$&$	0.040	$&$	0.043	$\\
&$	400	$&$	0.058	$&$	0.050	$&$	0.058	$&$	0.059	$&$	0.060	$&$	0.060	$\\
&$	800	$&$	0.027	$&$	0.021	$&$	0.028	$&$	0.040	$&$	0.044	$&$	0.044	$\\
\hline
\multirow{4}{*}{$n^{1/4}$}
&$	100	$&$	0.034	$&$	0.025	$&$	0.034	$&$	0.047	$&$	0.050	$&$	0.050	$\\
&$	200	$&$	0.038	$&$	0.032	$&$	0.039	$&$	0.040	$&$	0.040	$&$	0.040	$\\
&$	400	$&$	0.059	$&$	0.051	$&$	0.059	$&$	0.061	$&$	0.060	$&$	0.060	$\\
&$	800	$&$	0.034	$&$	0.028	$&$	0.037	$&$	0.048	$&$	0.054	$&$	0.054	$\\
\hline
\multirow{4}{*}{$n^{1/3}$}
&$	100	$&$	0.034	$&$	0.025	$&$	0.035	$&$	0.053	$&$	0.058	$&$	0.059	$\\
&$	200	$&$	0.038	$&$	0.033	$&$	0.039	$&$	0.048	$&$	0.052	$&$	0.053	$\\
&$	400	$&$	0.042	$&$	0.034	$&$	0.045	$&$	0.059	$&$	0.059	$&$	0.065	$\\
&$	800	$&$	0.041	$&$	0.032	$&$	0.044	$&$	0.052	$&$	0.054	$&$	0.054	$\\
\hline\hline
\end{tabular*}
\end{table}

\begin{table}[H]
\small\centering\setstretch{1.3}
\caption{Power for DGP (1) of Case 3 with dependent data ($\alpha=0.05$)}
\label{tab:case3dep DGP (1)}
\begin{tabular*}{15cm}{@{\extracolsep{\fill}}cccccccc}
\hline\hline
\multirow{2}{*}{$b(n)$} & \multirow{2}{*}{$n$} & \multicolumn{6}{c}{$\tau_n$} \\
\cline{3-8}
& & $\sqrt{\ln(n)/n}$ & $n^{-2/5}$ & $ n^{-1/3}$ & $ n^{-1/4}$ & $ n^{-1/5}$ & $ n^{-1/6}$ \\
\hline
\multirow{4}{*}{$n^{1/6}$}
&$	100	$&$	0.165	$&$	0.146	$&$	0.165	$&$	0.198	$&$	0.221	$&$	0.224	$\\
&$	200	$&$	0.272	$&$	0.223	$&$	0.286	$&$	0.309	$&$	0.337	$&$	0.343	$\\
&$	400	$&$	0.429	$&$	0.355	$&$	0.453	$&$	0.519	$&$	0.534	$&$	0.549	$\\
&$	800	$&$	0.645	$&$	0.538	$&$	0.675	$&$	0.759	$&$	0.791	$&$	0.809	$\\
\hline
\multirow{4}{*}{$n^{1/5}$}
&$	100	$&$	0.165	$&$	0.136	$&$	0.165	$&$	0.187	$&$	0.188	$&$	0.193	$\\
&$	200	$&$	0.240	$&$	0.192	$&$	0.246	$&$	0.294	$&$	0.319	$&$	0.330	$\\
&$	400	$&$	0.429	$&$	0.355	$&$	0.453	$&$	0.519	$&$	0.534	$&$	0.549	$\\
&$	800	$&$	0.669	$&$	0.573	$&$	0.707	$&$	0.788	$&$	0.824	$&$	0.824	$\\
\hline
\multirow{4}{*}{$n^{1/4}$}
&$	100	$&$	0.165	$&$	0.136	$&$	0.165	$&$	0.187	$&$	0.188	$&$	0.193	$\\
&$	200	$&$	0.214	$&$	0.198	$&$	0.222	$&$	0.287	$&$	0.306	$&$	0.309	$\\
&$	400	$&$	0.417	$&$	0.351	$&$	0.441	$&$	0.510	$&$	0.528	$&$	0.525	$\\
&$	800	$&$	0.637	$&$	0.533	$&$	0.675	$&$	0.774	$&$	0.802	$&$	0.826	$\\
\hline
\multirow{4}{*}{$n^{1/3}$}
&$	100	$&$	0.150	$&$	0.137	$&$	0.151	$&$	0.176	$&$	0.188	$&$	0.199	$\\
&$	200	$&$	0.232	$&$	0.175	$&$	0.241	$&$	0.309	$&$	0.332	$&$	0.343	$\\
&$	400	$&$	0.417	$&$	0.342	$&$	0.433	$&$	0.482	$&$	0.503	$&$	0.521	$\\
&$	800	$&$	0.697	$&$	0.627	$&$	0.733	$&$	0.799	$&$	0.826	$&$	0.831	$\\
\hline\hline
\end{tabular*}
\end{table}

\begin{table}[H]
\small\centering\setstretch{1.3}
\caption{Power for DGP (2) of Case 3 with dependent data ($\alpha=0.05$)}
\label{tab:case3dep DGP (2)}
\begin{tabular*}{15cm}{@{\extracolsep{\fill}}cccccccc}
\hline\hline
\multirow{2}{*}{$b(n)$} & \multirow{2}{*}{$n$} & \multicolumn{6}{c}{$\tau_n$} \\
\cline{3-8}
& & $\sqrt{\ln(n)/n}$ & $n^{-2/5}$ & $ n^{-1/3}$ & $ n^{-1/4}$ & $ n^{-1/5}$ & $ n^{-1/6}$ \\
\hline
\multirow{4}{*}{$n^{1/6}$}
&$	100	$&$	0.606	$&$	0.521	$&$	0.609	$&$	0.718	$&$	0.760	$&$	0.788	$\\
&$	200	$&$	0.889	$&$	0.821	$&$	0.900	$&$	0.951	$&$	0.964	$&$	0.970	$\\
&$	400	$&$	0.993	$&$	0.981	$&$	0.994	$&$	0.999	$&$	0.999	$&$	1.000	$\\
&$	800	$&$	1.000	$&$	1.000	$&$	1.000	$&$	1.000	$&$	1.000	$&$	1.000	$\\
\hline
\multirow{4}{*}{$n^{1/5}$}
&$	100	$&$	0.680	$&$	0.579	$&$	0.683	$&$	0.755	$&$	0.785	$&$	0.809	$\\
&$	200	$&$	0.890	$&$	0.821	$&$	0.901	$&$	0.952	$&$	0.964	$&$	0.970	$\\
&$	400	$&$	0.993	$&$	0.981	$&$	0.994	$&$	0.999	$&$	0.999	$&$	1.000	$\\
&$	800	$&$	1.000	$&$	1.000	$&$	1.000	$&$	1.000	$&$	1.000	$&$	1.000	$\\
\hline
\multirow{4}{*}{$n^{1/4}$}
&$	100	$&$	0.680	$&$	0.579	$&$	0.683	$&$	0.755	$&$	0.785	$&$	0.809	$\\
&$	200	$&$	0.889	$&$	0.814	$&$	0.899	$&$	0.952	$&$	0.966	$&$	0.970	$\\
&$	400	$&$	0.992	$&$	0.975	$&$	0.993	$&$	0.999	$&$	0.999	$&$	0.999	$\\
&$	800	$&$	1.000	$&$	1.000	$&$	1.000	$&$	1.000	$&$	1.000	$&$	1.000	$\\
\hline
\multirow{4}{*}{$n^{1/3}$}
&$	100	$&$	0.628	$&$	0.526	$&$	0.628	$&$	0.726	$&$	0.767	$&$	0.782	$\\
&$	200	$&$	0.879	$&$	0.808	$&$	0.889	$&$	0.942	$&$	0.959	$&$	0.969	$\\
&$	400	$&$	0.993	$&$	0.981	$&$	0.994	$&$	0.999	$&$	0.999	$&$	0.999	$\\
&$	800	$&$	1.000	$&$	1.000	$&$	1.000	$&$	1.000	$&$	1.000	$&$	1.000	$\\
\hline\hline
\end{tabular*}
\end{table}

\begin{table}[H]
\small\centering\setstretch{1.3}
\caption{Power for DGP (3) of Case 3 with dependent data ($\alpha=0.05$)}
\label{tab:case3dep DGP (3)}
\begin{tabular*}{15cm}{@{\extracolsep{\fill}}cccccccc}
\hline\hline
\multirow{2}{*}{$b(n)$} & \multirow{2}{*}{$n$} & \multicolumn{6}{c}{$\tau_n$} \\
\cline{3-8}
& & $\sqrt{\ln(n)/n}$ & $n^{-2/5}$ & $ n^{-1/3}$ & $ n^{-1/4}$ & $ n^{-1/5}$ & $ n^{-1/6}$ \\
\hline
\multirow{4}{*}{$n^{1/6}$}
&$	100	$&$	0.943	$&$	0.883	$&$	0.943	$&$	0.970	$&$	0.979	$&$	0.987	$\\
&$	200	$&$	0.997	$&$	0.995	$&$	0.997	$&$	0.999	$&$	1.000	$&$	1.000	$\\
&$	400	$&$	1.000	$&$	1.000	$&$	1.000	$&$	1.000	$&$	1.000	$&$	1.000	$\\
&$	800	$&$	1.000	$&$	1.000	$&$	1.000	$&$	1.000	$&$	1.000	$&$	1.000	$\\
\hline
\multirow{4}{*}{$n^{1/5}$}
&$	100	$&$	0.944	$&$	0.883	$&$	0.944	$&$	0.973	$&$	0.984	$&$	0.991	$\\
&$	200	$&$	0.997	$&$	0.995	$&$	0.997	$&$	0.999	$&$	1.000	$&$	1.000	$\\
&$	400	$&$	1.000	$&$	1.000	$&$	1.000	$&$	1.000	$&$	1.000	$&$	1.000	$\\
&$	800	$&$	1.000	$&$	1.000	$&$	1.000	$&$	1.000	$&$	1.000	$&$	1.000	$\\
\hline
\multirow{4}{*}{$n^{1/4}$}
&$	100	$&$	0.944	$&$	0.883	$&$	0.944	$&$	0.973	$&$	0.984	$&$	0.991	$\\
&$	200	$&$	0.997	$&$	0.991	$&$	0.997	$&$	0.999	$&$	0.999	$&$	1.000	$\\
&$	400	$&$	1.000	$&$	1.000	$&$	1.000	$&$	1.000	$&$	1.000	$&$	1.000	$\\
&$	800	$&$	1.000	$&$	1.000	$&$	1.000	$&$	1.000	$&$	1.000	$&$	1.000	$\\
\hline
\multirow{4}{*}{$n^{1/3}$}
&$	100	$&$	0.929	$&$	0.865	$&$	0.929	$&$	0.962	$&$	0.976	$&$	0.981	$\\
&$	200	$&$	0.997	$&$	0.997	$&$	0.997	$&$	0.999	$&$	1.000	$&$	1.000	$\\
&$	400	$&$	1.000	$&$	1.000	$&$	1.000	$&$	1.000	$&$	1.000	$&$	1.000	$\\
&$	800	$&$	1.000	$&$	1.000	$&$	1.000	$&$	1.000	$&$	1.000	$&$	1.000	$\\
\hline\hline
\end{tabular*}
\end{table}

\subsection{Symmetry}

We test the symmetry of the distribution of $Z$, as discussed in Example \ref{exam:symmetry.prob}. The DGPs are constructed based on those of \citet{psaradakis2022using}, and we consider i.i.d. samples. We let $Z_1, \ldots, Z_n$ be independently and identically drawn from the generalized lambda distribution $\mathrm{GL}(\lambda_1, \lambda_2, \lambda_3, \lambda_4)$ with quantile function (inverse distribution function) $F^{-1}(u)=\lambda_1+(1/\lambda_2)[u^{\lambda_3}-(1-u)^{\lambda_4}], u\in(0,1)$. By choosing different values of the parameters $(\lambda_1, \lambda_2, \lambda_3, \lambda_4)$, we may allow the distribution of $Z_i$ to exhibit various degrees of skewness as summarized in Table \ref{tab:Summary of DGPs}. Specifically, DGP (0) satisfies the null hypothesis, and DGP (1) to DGP (3) satisfy the alternative hypothesis. The grid for $\theta$ is $\{-0.3, -0.298, -0.296, \ldots, 0.3\}$. The choices of the tuning parameters and other implementation details follow those elaborated in Section \ref{sec:simulation}.

\begin{table}[ht]
    \small\centering\setstretch{1.3}
    \caption{Summary of DGPs}
    \label{tab:Summary of DGPs}
    \begin{tabular}{cccccc}
    \hline\hline
     & $\lambda_1$ & $\lambda_2$ & $\lambda_3$ & $\lambda_4$ & Skewness \\
    \hline
    DGP (0) & $0$ & $-0.397912$ & $-0.16$ & $-0.16$ & $0$ \\
    DGP (1) & $0$ & $-1$ & $-0.0075$ & $-0.03$ & $1.5$ \\
    DGP (2) & $0$ & $-1$ & $-0.1009$ & $-0.1802$ & $2.0$ \\
    DGP (3) & $0$ & $-1$ & $-0.001$ & $-0.13$ & $3.2$ \\
    \hline\hline
    \end{tabular}
\end{table}

Table \ref{tab:symmetry simulation} displays the rejection rates in these Monte Carlo experiments. As the sample sizes increase, the rejection rates under DGP (0) (i.e., empirical size) approach the significance level $\alpha$, while the rejection rates under DGP (1)--DGP (3) (i.e., empirical power) approach $1$. These simulation results show the good empirical properties of the test.

\begin{table}[H]
\small\centering\setstretch{1.3}
\caption{Size and power for testing symmetry ($\alpha=0.05$)}
\label{tab:symmetry simulation}
\begin{tabular*}{15cm}{@{\extracolsep{\fill}}cccccccc}
\hline\hline
\multirow{2}{*}{DGP} & \multirow{2}{*}{$n$} & \multicolumn{6}{c}{$\tau_n$} \\
\cline{3-8}
& & $\sqrt{\ln(n)/n}$ & $n^{-2/5}$ & $ n^{-1/3}$ & $ n^{-1/4}$ & $ n^{-1/5}$ & $ n^{-1/6}$ \\
\hline
\multirow{6}{*}{DGP (0)}
&$	100	$&$	0.019	$&$	0.024	$&$	0.019	$&$	0.008	$&$	0.004	$&$	0.004	$\\
&$	200	$&$	0.042	$&$	0.033	$&$	0.043	$&$	0.030	$&$	0.017	$&$	0.013	$\\
&$	400	$&$	0.035	$&$	0.034	$&$	0.036	$&$	0.030	$&$	0.016	$&$	0.007	$\\
&$	800	$&$	0.027	$&$	0.026	$&$	0.027	$&$	0.024	$&$	0.017	$&$	0.010	$\\
&$	1600	$&$	0.044	$&$	0.039	$&$	0.047	$&$	0.050	$&$	0.035	$&$	0.024	$\\
&$	3200	$&$	0.045	$&$	0.035	$&$	0.054	$&$	0.065	$&$	0.063	$&$	0.035	$\\
\hline
\multirow{4}{*}{DGP (1)}
&$	100	$&$	0.784	$&$	0.668	$&$	0.785	$&$	0.875	$&$	0.917	$&$	0.941	$\\
&$	200	$&$	0.978	$&$	0.953	$&$	0.982	$&$	0.997	$&$	0.997	$&$	0.999	$\\
&$	400	$&$	1.000	$&$	1.000	$&$	1.000	$&$	1.000	$&$	1.000	$&$	1.000	$\\
&$	800	$&$	1.000	$&$	1.000	$&$	1.000	$&$	1.000	$&$	1.000	$&$	1.000	$\\
\hline
\multirow{4}{*}{DGP (2)}
&$	100	$&$	0.348	$&$	0.257	$&$	0.349	$&$	0.428	$&$	0.483	$&$	0.489	$\\
&$	200	$&$	0.642	$&$	0.495	$&$	0.655	$&$	0.747	$&$	0.787	$&$	0.814	$\\
&$	400	$&$	0.887	$&$	0.807	$&$	0.916	$&$	0.975	$&$	0.982	$&$	0.982	$\\
&$	800	$&$	0.998	$&$	0.991	$&$	1.000	$&$	1.000	$&$	1.000	$&$	1.000	$\\
\hline
\multirow{4}{*}{DGP (3)}
&$	100	$&$	0.994	$&$	0.978	$&$	0.994	$&$	1.000	$&$	1.000	$&$	1.000	$\\
&$	200	$&$	1.000	$&$	1.000	$&$	1.000	$&$	1.000	$&$	1.000	$&$	1.000	$\\
&$	400	$&$	1.000	$&$	1.000	$&$	1.000	$&$	1.000	$&$	1.000	$&$	1.000	$\\
&$	800	$&$	1.000	$&$	1.000	$&$	1.000	$&$	1.000	$&$	1.000	$&$	1.000	$\\
\hline\hline
\end{tabular*}
\end{table}

\subsection{Goodness of Fit}

For Example \ref{exam:fit.prob}, we test whether the distribution of $Z$ belongs to the normal family $\{\mathcal{N}(\theta,1):\theta\in\Theta\subset \mr\}$. We let $U_1, \ldots, U_n$ be i.i.d.\ from $\mathrm{Unif}[0,1]$, and $V_1,\ldots, V_n$ be i.i.d.\ from $\mathcal{N}(0,1)$. We consider the following four DGPs. Specifically, DGP (0) satisfies the null hypothesis, and DGP (1) to DGP (3) satisfy the alternative hypothesis. In addition, the grid for $\theta$ is $\{-0.3, -0.298, -0.296, \ldots, 0.3\}$. The choices of the tuning parameters and other implementation details follow those elaborated in Section \ref{sec:simulation}.

\begin{itemize}
\item DGP (0): $Z_i=V_i$.
\item DGP (1): $Z_i=0.2 U_i+ 0.8 V_i$.
\item DGP (2): $Z_i=0.6 U_i+ 0.4 V_i$.
\item DGP (3): $Z_i=U_i$.
\end{itemize}

Table \ref{tab:GoF simulation} shows the rejection rates for the DGPs above, which illustrate the good empirical properties of the test.

\begin{table}[H]
\small\centering\setstretch{1.3}
\caption{Size and power for testing goodness of fit ($\alpha=0.05$)}
\label{tab:GoF simulation}
\begin{tabular*}{15cm}{@{\extracolsep{\fill}}cccccccc}
\hline\hline
\multirow{2}{*}{DGP} & \multirow{2}{*}{$n$} & \multicolumn{6}{c}{$\tau_n$} \\
\cline{3-8}
& & $\sqrt{\ln(n)/n}$ & $n^{-2/5}$ & $ n^{-1/3}$ & $ n^{-1/4}$ & $ n^{-1/5}$ & $ n^{-1/6}$ \\
\hline
\multirow{6}{*}{DGP (0)}
&$	100	$&$	0.018	$&$	0.017	$&$	0.018	$&$	0.014	$&$	0.008	$&$	0.004	$\\
&$	200	$&$	0.016	$&$	0.014	$&$	0.018	$&$	0.008	$&$	0.007	$&$	0.004	$\\
&$	400	$&$	0.028	$&$	0.024	$&$	0.030	$&$	0.025	$&$	0.019	$&$	0.008	$\\
&$	800	$&$	0.039	$&$	0.035	$&$	0.039	$&$	0.036	$&$	0.022	$&$	0.015	$\\
&$	1600	$&$	0.042	$&$	0.036	$&$	0.046	$&$	0.042	$&$	0.027	$&$	0.018	$\\
&$	3200	$&$	0.050	$&$	0.041	$&$	0.058	$&$	0.058	$&$	0.044	$&$	0.030	$\\
\hline
\multirow{4}{*}{DGP (1)}
&$	100	$&$	0.566	$&$	0.501	$&$	0.568	$&$	0.627	$&$	0.621	$&$	0.601	$\\
&$	200	$&$	0.852	$&$	0.760	$&$	0.854	$&$	0.891	$&$	0.891	$&$	0.873	$\\
&$	400	$&$	0.992	$&$	0.980	$&$	0.994	$&$	0.998	$&$	0.998	$&$	0.998	$\\
&$	800	$&$	1.000	$&$	1.000	$&$	1.000	$&$	1.000	$&$	1.000	$&$	1.000	$\\
\hline
\multirow{4}{*}{DGP (2)}
&$	100	$&$	1.000	$&$	1.000	$&$	1.000	$&$	1.000	$&$	1.000	$&$	1.000	$\\
&$	200	$&$	1.000	$&$	1.000	$&$	1.000	$&$	1.000	$&$	1.000	$&$	1.000	$\\
&$	400	$&$	1.000	$&$	1.000	$&$	1.000	$&$	1.000	$&$	1.000	$&$	1.000	$\\
&$	800	$&$	1.000	$&$	1.000	$&$	1.000	$&$	1.000	$&$	1.000	$&$	1.000	$\\
\hline
\multirow{4}{*}{DGP (3)}
&$	100	$&$	1.000	$&$	1.000	$&$	1.000	$&$	1.000	$&$	1.000	$&$	1.000	$\\
&$	200	$&$	1.000	$&$	1.000	$&$	1.000	$&$	1.000	$&$	1.000	$&$	1.000	$\\
&$	400	$&$	1.000	$&$	1.000	$&$	1.000	$&$	1.000	$&$	1.000	$&$	1.000	$\\
&$	800	$&$	1.000	$&$	1.000	$&$	1.000	$&$	1.000	$&$	1.000	$&$	1.000	$\\
\hline\hline
\end{tabular*}
\end{table}

\subsection{Location Transformation}

For random variables $X$ and $Y$ with cumulative distribution functions $F$ and $G$, we want to test whether there exists $\theta\in\Theta\subset\mr$ such that $F(x)=G(x-\theta)$ for all $x\in\mr$. We let $X_1,\ldots, X_n$ be i.i.d.\ from $\mathcal{N}(0,1)$, $U_1, \ldots, U_n$ be i.i.d.\ from $\mathrm{Unif}[0,1]$, and $V_1,\ldots, V_n$ be i.i.d.\ from $\mathcal{N}(-1,1)$. We consider the following four DGPs, where DGP (0) satisfies the null hypothesis, and DGP (1) to DGP (3) satisfy the alternative hypothesis. The choices of the tuning parameters and other implementation details are as elaborated in Section \ref{sec:simulation}.

\begin{itemize}
\item DGP (0): $Y_i=V_i$.
\item DGP (1): $Y_i=0.2 U_i+ 0.8 V_i$.
\item DGP (2): $Y_i=0.6 U_i+ 0.4 V_i$.
\item DGP (3): $Y_i=U_i$.
\end{itemize}

Table \ref{tab:location simulation} presents the rejection rates in these Monte Carlo simulations. The results show that the test is slightly conservative for some choices of $\tau_n$, while it has a good empirical power property in finite samples.

\begin{table}[H]
\small\centering\setstretch{1.3}
\caption{Size and power for testing location transformation ($\alpha=0.05$)}
\label{tab:location simulation}
\begin{tabular*}{15cm}{@{\extracolsep{\fill}}cccccccc}
\hline\hline
\multirow{2}{*}{DGP} & \multirow{2}{*}{$n$} & \multicolumn{6}{c}{$\tau_n$} \\
\cline{3-8}
& & $\sqrt{\ln(n)/n}$ & $n^{-2/5}$ & $ n^{-1/3}$ & $ n^{-1/4}$ & $ n^{-1/5}$ & $ n^{-1/6}$ \\
\hline
\multirow{6}{*}{DGP (0)}
&$	100	$&$	0.012	$&$	0.016	$&$	0.012	$&$	0.005	$&$	0.002	$&$	0.002	$\\
&$	200	$&$	0.014	$&$	0.014	$&$	0.014	$&$	0.006	$&$	0.004	$&$	0.002	$\\
&$	400	$&$	0.028	$&$	0.027	$&$	0.027	$&$	0.012	$&$	0.008	$&$	0.004	$\\
&$	800	$&$	0.035	$&$	0.027	$&$	0.035	$&$	0.019	$&$	0.009	$&$	0.004	$\\
&$	1600	$&$	0.040	$&$	0.038	$&$	0.042	$&$	0.026	$&$	0.017	$&$	0.015	$\\
&$	3200	$&$	0.034	$&$	0.032	$&$	0.040	$&$	0.034	$&$	0.023	$&$	0.015	$\\
\hline
\multirow{4}{*}{DGP (1)}
&$	100	$&$	0.094	$&$	0.073	$&$	0.094	$&$	0.135	$&$	0.146	$&$	0.146	$\\
&$	200	$&$	0.278	$&$	0.199	$&$	0.299	$&$	0.357	$&$	0.364	$&$	0.374	$\\
&$	400	$&$	0.584	$&$	0.545	$&$	0.615	$&$	0.716	$&$	0.743	$&$	0.745	$\\
&$	800	$&$	0.966	$&$	0.946	$&$	0.980	$&$	0.991	$&$	0.996	$&$	0.997	$\\
\hline
\multirow{4}{*}{DGP (2)}
&$	100	$&$	1.000	$&$	1.000	$&$	1.000	$&$	1.000	$&$	1.000	$&$	1.000	$\\
&$	200	$&$	1.000	$&$	1.000	$&$	1.000	$&$	1.000	$&$	1.000	$&$	1.000	$\\
&$	400	$&$	1.000	$&$	1.000	$&$	1.000	$&$	1.000	$&$	1.000	$&$	1.000	$\\
&$	800	$&$	1.000	$&$	1.000	$&$	1.000	$&$	1.000	$&$	1.000	$&$	1.000	$\\
\hline
\multirow{4}{*}{DGP (3)}
&$	100	$&$	1.000	$&$	1.000	$&$	1.000	$&$	1.000	$&$	1.000	$&$	1.000	$\\
&$	200	$&$	1.000	$&$	1.000	$&$	1.000	$&$	1.000	$&$	1.000	$&$	1.000	$\\
&$	400	$&$	1.000	$&$	1.000	$&$	1.000	$&$	1.000	$&$	1.000	$&$	1.000	$\\
&$	800	$&$	1.000	$&$	1.000	$&$	1.000	$&$	1.000	$&$	1.000	$&$	1.000	$\\
\hline\hline
\end{tabular*}
\end{table}

\end{appendices}
\bibliographystyle{apalike}
\bibliography{references}		

\end{document}